\documentclass[preprint,11pt,superscriptaddress,nofootinbib]{revtex4}

\usepackage{graphicx}
\usepackage{dcolumn}
\usepackage{amsmath}
\usepackage[colorlinks=true, citecolor=blue, urlcolor = blue, linkcolor= blue, bookmarks=true]{hyperref}

\begin{document}
%opening
\title{Ergosphere and shadow of a rotating regular black hole}

\author{Sushant~G.~Ghosh} 
\email{sghosh2@jmi.ac.in, sgghosh@gmail.com} 
\affiliation{ Centre for Theoretical Physics, Jamia Millia Islamia, 
New Delhi 110025, India}
\affiliation{Multidisciplinary Centre for Advanced Research and Studies 
(MCARS),\\ 
Jamia Millia Islamia, New Delhi 110025, India}
\affiliation{Astrophysics and Cosmology Research Unit, School of 
Mathematics, Statistics and Computer Science, University of KwaZulu-Natal, 
Private Bag X54001, Durban 4000, South Africa}

\author{Muhammed Amir}
\email{amirctp12@gmail.com}
\affiliation{Astrophysics and Cosmology Research Unit, School of Mathematics,
Statistics and Computer Science, University of KwaZulu-Natal,
Private Bag X54001, Durban 4000, South Africa}

\author{Sunil D. Maharaj}
\email{maharaj@ukzn.ac.za}
\affiliation{Astrophysics and Cosmology Research Unit, School of Mathematics, 
Statistics and Computer Science, University of KwaZulu-Natal, 
Private Bag X54001, Durban 4000, South Africa}
 
\date{\today}

\begin{abstract}
The spacetime singularities in classical general relativity predicted by 
the celebrated singularity theorems are formed at the end of gravitational 
collapse. Quantum gravity is the expected theory to resolve the singularity 
problem, but we are now far from it. Therefore attention has shifted to models 
of regular black holes free from the singularities. A spherically symmetric 
regular toy model was obtained by Dymnikova (1992) which we demonstrate as an 
exact solution of Einstein's field equations coupled to nonlinear electrodynamics 
for a Lagrangian with parameter $b$ related to magnetic charge. We construct 
rotating counterpart of this solution which encompasses the Kerr black hole 
as a special case when charge is switched off ($b=0$). Event Horizon 
Telescope has released the first image of supermassive black hole M87$^*$, 
revealing the structure near black hole horizon. The rotating regular black 
hole's shadow may be useful to determine strong field regime. We investigate 
ergosphere and black hole shadow of rotating regular black hole to infer 
that their sizes are sensitive to charge $b$ and have a richer chaotic 
structure. In particular, rotating regular black hole possess larger size, 
but less distorted  shadows when compared with Kerr black holes. We find one 
to one correspondence between ergosphere and shadow of the black hole.
\end{abstract}

%\pacs{04.50.Kd, 04.20.Jb, 04.40.Nr, 04.70.Bw}

\maketitle

\section{Introduction}
The elegant theorems of Hawking and Penrose imply  that spacetime singularities are pervasive features 
of general relativity, so that the theory itself predicts its own failure to describe the physics of 
these extreme situations. The spacetime singularities that arise in gravitational collapse are always 
hidden inside black holes, which is the essence of the weak cosmic censorship conjecture, put forward 
40 years ago by Penrose \cite{Penrose:1969pc}. This signature is still one of the most important open 
questions in general relativity. We are far away from any robust and reliable quantum theory of gravity 
capable of resolving the singularities in the interior of black holes. Hence, there is significant 
attention towards models of black hole solutions without the central singularity. The earliest idea of 
regular models dates back to the pioneering work of Sakharov \cite{Sakharov:1966} and Gliner 
\cite{Gliner:1966} where they proposed that the spacetime in the highly dense central region of a 
black hole should be de Sitter-like at $r \simeq 0$. Thus spacetime filled with vacuum could provide 
a proper discrimination at the final stage of gravitational collapse, replacing future singularity 
\cite{Gliner:1966}. The prototype of these regular black holes is the 
Bardeen metric \cite{Bardeen:1968}, which can be formally obtained by coupling Einstein's gravity to a 
nonlinear electrodynamic field \cite{AyonBeato:2000zs}, thought to be an alteration of the 
Reissner-Nordstr{\"o}m solution. There has been enormous attentions to obtain regular black holes 
\cite{Dymnikova:2004zc,Bronnikov:2000vy,Shankaranarayanan:2003qm,Hayward:2005gi,Ansoldi:2008jw,Culetu:2014lca,Balart:2014jia,Balart:2014cga,Xiang:2013sza}; most of these regular black holes were driven by 
the Bardeen idea \cite{Bardeen:1968}. However, these are only nonrotating black holes, which can  be 
hardly tested by observations, as black hole spin is an important parameter in astrophysical processes. 

Further, a study of rotating regular solutions is important as the astronomical observations predict 
that the astrophysical black hole might be a Kerr black hole \cite{Bambi:2011mj}. However, the real 
nature of these astrophysical black holes is still to be verified \cite{Bambi:2011mj}. Analyzing the 
iron $k {\alpha}$ lines and the continuum fitting method are possible procedures, which can probe the 
geometry of spacetime around astrophysical black holes candidates \cite{Bambi:2014nta}. The rotating 
solutions of the Bardeen metrics have been obtained in Ref.~\cite{Bambi:2011mj}. There exists a number 
of rotating regular black holes \cite{Bambi:2013ufa,Neves:2014aba,Toshmatov:2014nya,Ghosh:2014hea,Toshmatov:2017zpr} which 
were discovered with the help of the Newman-Janis algorithm \cite{Newman:1965tw}, and also by using other 
techniques \cite{Azreg-Ainou:2014aqa,Azreg-Ainou:2014nra,Azreg-Ainou:2014pra,Ghosh:2014pba}. 
The authors in Refs.~\cite{Ghosh:2014mea,Amir:2015pja,Ghosh:2015pra,Amir:2016nti,Ahmed:2018fge} 
demonstrated that the regular black holes can be considered as the particle accelerator. 
The quasinormal modes of test fields around the regular 
black holes has been discussed explicitly in \cite{Toshmatov:2015wga}. An 
interesting study on electromagnetic perturbation of the regular black 
holes has been discussed thoroughly in 
\cite{Toshmatov:2018tyo,Toshmatov:2018ell,Toshmatov:2019gxg}. 

The black hole shadow is a dark zone in the sky and the shadow is a useful tool for measuring black 
hole parameters because its shape and size carry impression of the geometry surrounding the black hole.
Observing the black hole shadow is a possible method to determine the spin of the black hole, and this 
subject is popular nowadays. This triggered theoretical investigation of the black holes for wide 
variety of spacetimes
\cite{Chandrasekhar92,Falcke:1999pj,Takahashi04,Takahashi:2005hy,Bambi:2008jg,Hioki:2009na,Bambi:2010hf,Amarilla:2010zq,Amarilla:2011fxx,Abdujabbarov:2012bnn,Yumoto:2012kz,Li:2013jra,Amarilla:2013sj,Atamurotov:2013dpa,Atamurotov:2013sca,Wei:2013kza,Grenzebach:2014fha,Papnoi:2014,Abdujabbarov:2015xqa,Johannsen:2015qca,Goddi:2016jrs,Abdujabbarov:2016hnw,Younsi:2016azx,Amir:2016cen,Grenzebach:2016,Zakharov,Singh:2017vfr,Amir:2017slq,Kumar:2017vuh}. 
Apart from black holes, the theoretical investigation of shadow of wormholes has also been discussed 
in \cite{Nedkova:2013msa,Bambi:2013nla,AzregAinou:2014dwa,Ohgami:2015nra,Ohgami:2016iqm,Abdujabbarov:2016efm,Shaikh:2018kfv,Amir:2018szm,Amir:2018pcu}. 
Hioki and Maeda \cite{Hioki:2009na} discuss a simple relation between the shape of shadow, inclination 
angle and spin parameter for the Kerr black hole. A coordinate-independent method was proposed to 
distinguish the shape of black hole shadow from other rotating black holes \cite{Abdujabbarov:2015xqa}. 
From the observational point of view, the study of shadow is significant because of the Event Horizon 
Telescope\footnote{https://eventhorizontelescope.org}. It was setup with an aim to image the event 
horizon of supermassive black holes Sgr A$^*$ and M87$^*$. They have successfully released the first image 
of M87$^*$ which is in accordance with the Kerr black hole predicted by general relativity 
\cite{Akiyama:2019cqa,Akiyama:2019fyp,Akiyama:2019eap}. The photon ring and shadow have been observed which 
open a new window to test general relativity and modified theories of gravity in strong field regime. 

The purpose of this paper is to construct the rotating counterpart or Kerr-like regular black hole from 
the spherically symmetric regular black hole proposed by Dymnikova \cite{Dymnikova:1992ux}. 
The black hole is singularity free or regular black hole with an additional parameter 
$b$, and for definiteness, we can call it the rotating regular black hole. We investigate several 
properties including the shadow of this black hole to explicitly bring out the effect of parameter $b$, 
and compare with the Kerr black hole. It turns out that the parameter $b$ leaves a significant imprint 
on the black hole shadow, and also affects the other properties. 

The paper is organized as follows. In Sec.~\ref{spctm}, we show that the spherically symmetric 
regular black hole obtained by Dymnikova \cite{Dymnikova:1992ux} is an exact solution of 
general relativity coupled to nonlinear electrodynamics. We introduce a line element of the 
rotating regular black hole spacetime and discuss associated sources with it in Sec.~\ref{rrb}.
The weak energy conditions is the subject of Sec.~\ref{energy}. 
In Sec.~\ref{properties}, we discuss some basic properties of the rotating regular black 
hole. Shadow of rotating regular black hole is the subject of Sec.~\ref{shadow} where we 
derive analytical formulae for the shadow and we conclude in the Sec.~\ref{conclusion}. We 
consider the signature convention ($-,+,+,+$) for the spacetime metric and use the geometric 
unit $G=c=1$ throughout the paper.

\section{Nonlinear electrodynamics and exact regular black hole solution}
\label{spctm}
The spherically symmetric metric proposed by Dymnikova \cite{Dymnikova:1992ux} 
represents a regular black hole with de Sitter core instead of a singularity. 
The stress-energy tensor responsible for the geometry describes a smooth 
transition from the standard vacuum state at infinity to isotropic vacuum 
state through anisotropic vacuum state in the intermediate region. 
The spacetime metric of the Dymnikova solution reads
\begin{eqnarray}
    ds^2 &=& -\left[1-\frac{2 M (1-e^{-r^3/b^3})}{r}\right] dt^2 
    + \left[1-\frac{2M (1-e^{-r^3/b^3})}{r}\right]^{-1}dr^2 
    + r^2 d\Omega^2, \label{dymn}
\end{eqnarray}
where $d \Omega^2 = d \theta^2 + \sin^2 \theta d \phi^2$, $M$ is the black 
hole mass, and the parameter $b$ is given by $b^3=r_0^2M $. The corresponding 
energy-momentum tensor of metric \eqref{dymn} is \cite{Dymnikova:1992ux}
\begin{eqnarray}
    T_0^0 &=& T_1^1=\frac{3}{r_0^2}e^{-r^3/b^3}, \nonumber\\
    T_2^2 &=& T_3^3=\frac{3}{r_0^2}\left(1-\frac{3r^3}{2b^3}\right)
    e^{-r^3/b^3}. \label{set}
\end{eqnarray}
The metric \eqref{dymn} admits two horizons, i.e., the event horizon and 
the Cauchy horizon, but there is no singularity. We know that the horizons 
are zeros of $g^{rr}=0$, which are given \cite{Dymnikova:1992ux} by
$$r_+ \approx M \left[1-\mathcal{O} \left(\exp(-\frac{M^2}{r_0^2})\right)\right],$$
$$r_- \approx r_0 \left[1-\mathcal{O} \left(\exp(-\frac{r_0}{4M})\right)\right].$$
The metric \eqref{dymn} is regular, everywhere including at $r=0$, which is 
evident from the behavior of the curvature invariants. Thus the  metric 
(\ref{dymn}) that for large $r$ coincides with the Schwarzschild solution, 
for small $r$ behaves like the de Sitter spacetime and describes a 
spherically symmetric regular black hole. In what follows, we show the 
metric \eqref{dymn} as an exact solution of general relativity 
coupled to an appropriate nonlinear electrodynamics. 

We start with action of general relativity coupled to the nonlinear 
electrodynamics which is given by
\begin{equation}
    S = \frac{1}{16 \pi} \int d^4 x \sqrt{-g} \left[R 
    - \mathcal{L} (F)\right], \label{action}
\end{equation}
where $R$ is the Ricci scalar and $\mathcal{L}(F)$ is an arbitrary 
function of $F = \frac{1}{4} F_{\mu \nu} F^{\mu \nu}$ with 
$F_{\mu \nu} = 2 \nabla_{[\mu} A_{\nu]}$, and $A_{\mu}$ denotes the 
electromagnetic potential. The Einstein field equations are derived 
from the action \eqref{action} which read simply
\begin{eqnarray}
    G_{\mu \nu} &=& 2 \left(\frac{\partial \mathcal{L}}{\partial F} 
    F_{\mu \alpha} F^{\alpha}_{\nu} - g_{\mu \nu} \mathcal{L} \right), 
    \label{eqe1} \\
    && \nabla_{\mu} \left(\frac{\partial \mathcal{L}}{\partial F} 
    F^{\mu \nu} \right) =0. \label{eqe2}
\end{eqnarray}
We further consider the spherically symmetric line element of the 
following form
\begin{eqnarray}
    ds^2 &=& -\left[1-\frac{2M(r)}{r}\right]dt^2 
    + \left[1-\frac{2M (r)}{r}\right]^{-1}dr^2 + r^2 d\Omega^2.
    \label{ansatz}
\end{eqnarray} 
For the spherically symmetric spacetime the only nonzero components of 
$F_{\mu \nu}$ appropriate to magnetic charge are $F_{\theta \phi}$ and 
$F_{\phi \theta}$ with $F_{\theta \phi} =- F_{\phi \theta}$ such that 
$F_{\theta \phi} = 2 b \sin \theta$ \cite{Fernando:2016ksb}. Now it is 
straight forward to compute
\begin{equation}
    F = \frac{2 b^2}{r^4} = 2 B^2,
\end{equation}
where $B$ is the absolute value of magnetic field. The regular solution 
\cite{Dymnikova:1992ux} in which we are much interested comes from 
particular nonlinear electrodynamics source
\begin{equation}
    \mathcal{L}(F) = \frac{3}{s b^2} \exp \left[-\left(\frac{2}
    {b^2 F}\right)^{3/4}\right], \label{lf}
\end{equation}
where the parameter $s$ is given by $s = {|b|}/{M}$ with $b$ and $M$ are 
arbitrary constant can be related to the magnetic charge and the black hole 
mass, respectively. The $(t,t)$ component of the Einstein's field equations 
\eqref{eqe1} gives
\begin{equation}
    M'(r) = r^2 \mathcal{L}(F). \label{mprm}
\end{equation}
On substituting \eqref{lf} into \eqref{mprm} and then integrating it 
results in
\begin{equation}
    \int_r^{\infty} M'(r) dr = M - M(r) = M e^{-r^3/b^3},
\end{equation}
where the integration constant is purposely chosen as $M$, thus the above 
equation yields
\begin{equation}
    M(r) = M (1- e^{-r^3/b^3}). \label{mr}
\end{equation}
When we substitute \eqref{mr} in \eqref{ansatz} eventually we get the 
spherically symmetric Dymnikova regular spacetime \eqref{dymn}. Thus, we can 
say that the Dymnikova spacetime \eqref{dymn} can be obtained exactly with 
source from a particular nonlinear electrodynamics where the source is given 
by the Lagrangian \eqref{lf}.

\section{Rotating regular black holes}
\label{rrb}
In this section, we aim to study the rotating counterpart of the Dymnikova 
spacetime which is a generalization of the Kerr black hole. We start by 
constructing a rotating counterpart of the regular spacetime \eqref{dymn} 
which in the Boyer-Lindquist coordinates reads
\begin{eqnarray}
    ds^2 & = & - \left( 1- \frac{2M(r)r}{\Sigma} \right) dt^2  
    -\frac{4aM(r)r\sin^2 \theta}{\Sigma} dt d\phi
    +\frac{\Sigma}{\Delta}dr^2 +  \Sigma d \theta^2  \nonumber \\ && 
    + \left(r^2+ a^2 +  \frac{2 a^2 M(r) r  \sin^2 \theta}{\Sigma} \right) 
    \sin^2 \theta d\phi^2, \label{metric}
\end{eqnarray}
where 
\[\Sigma = r^2 + a^2 \cos^2\theta, \quad \Delta=r^2 + a^2 - 2 r M(r) .\] 
Here $a$ represents the rotation parameter and $M(r)$ is the mass function 
given in \eqref{mr}. Henceforth, we use the term rotating regular black hole 
for the spacetime metric \eqref{metric}. It is noticeable that in the 
absence of charge ($b=0$), we obtain the Kerr spacetime. The spacetime 
\eqref{metric} has independence on $t$ and $\phi$ coordinates that means $t$ 
and $\phi$ coordinates are cyclic coordinates in the Lagrangian which 
corresponding to two Killing vectors, namely, the time translation Killing 
vector $\chi^a$, and the azimuthal Killing vector $\zeta^{a}$.

Now we are going to check the validity of the rotating solution by computing 
the source nonlinear electrodynamics equations. The action and corresponding 
field equations are explicitly expressed in the previous section. The vector 
potential in case of spherically symmetric spacetime is given by 
$A_{\mu} = -b \cos \theta \delta^{\phi}_{\mu}$. The vector potential for 
the rotating spacetime gets modify \cite{Toshmatov:2017zpr} and given as 
follows
\begin{equation}
    A_{\mu} = \left[\frac{b a \cos \theta}{\Sigma}, 0, 0, 
    -\frac{b(r^2+a^2) \cos \theta}{\Sigma} \right].
\end{equation}
It is easy to compute the field strength $(F)$ for the rotating spacetime, 
which turns out to be in following form
\begin{equation}
    F = \frac{2b^2 [(r^2 -a^2 \cos^2 \theta)^2 
    -4a^2r^2 \cos^2 \theta]}{\Sigma^4}. \label{fs}
\end{equation}
We can immediately recover $F=2b^2/r^4$ while substitute $a=0$ in 
\eqref{fs}. In order to compute the source for the rotating spacetime, we 
solve the Einstein field equations, $G_{\mu \nu} =T_{\mu \nu}$ with respect 
to $\mathcal{L}$ and $\partial \mathcal{L}/\partial F$. Consequently, the 
Lagrangian density reads simply
\begin{eqnarray}
  \mathcal{L} = \frac{3 Mr^4 e^{-r^3/b^3}[r^4b^3 + 2a^2r^2(b^3 -3r^3)
  \cos^2 \theta +3a^4 (3b^3 -2r^3) \cos^4 \theta]} {b^6 \Sigma^4}, \label{rl}
\end{eqnarray}
and $\partial \mathcal{L}/\partial F$ has the form
\begin{eqnarray}
  \frac{\partial \mathcal{L}}{\partial F} = \frac{6 Mr^2 e^{-r^3/b^3}
  [3r^5 -a^2 (4b^3 -3r^3)\cos^2 \theta]}{8 b^8}. \label{rlf}
\end{eqnarray}
Interestingly, when we substitute $a=0$ into Eqs.~\eqref{rl} and \eqref{rlf} 
as a result we obtain
\begin{equation}
    \mathcal{L} = \frac{3 M e^{-r^3/b^3}}{b^3}, \quad
    \frac{\partial \mathcal{L}}{\partial F} 
    = \frac{9 M r^7 e^{-r^3/b^3}}{8 b^8}.
\end{equation}
These expressions are exactly similar to that of the spherically symmetric 
spacetime.

\section{Energy conditions}
\label{energy}
Now we are much interested in computing the orthonormal basis also known 
as tetrads of the spacetime \eqref{metric} so that we can analyze the 
matter associated with the rotating regular spacetime 
\cite{Neves:2014aba,Bardeen:1972fi}. The computation of the orthonormal 
basis gives the following form \cite{Neves:2014aba,Bardeen:1972fi}:
\begin{equation}
    e^{(a)}_{\mu}=\left(\begin{array}{cccc}
    \sqrt{\mp(g_{tt}-\Omega g_{t\phi})}& 0 & 0 & 0 \\
    0 & \sqrt{\pm g_{rr}} & 0 & 0 \\
    0 & 0 & \sqrt{g_{\theta \theta}} & 0 \\
    {g_{t\phi}}/{\sqrt{g_{\phi\phi}}} & 0 & 0 & \sqrt{g_{\phi\phi}}
    \end{array}\right), \label{Matrix}
\end{equation}
where $\Omega= g_{t\phi} /{g_{\phi\phi}}$ is the angular velocity of the 
rotating regular black hole. We further determine the components of the 
energy-momentum tensor by using the following relation
\cite{Neves:2014aba,Bardeen:1972fi}
\begin{equation}
    T^{(a)(b)} = e^{(a)}_{\mu} e^{(b)}_{\nu} G^{\mu \nu}. \nonumber
\end{equation}
When we compute the components of the energy-momentum tensor $T^{(a)(b)}$, 
then we realize that it has only diagonal components 
\cite{Neves:2014aba,Bardeen:1972fi}, i.e.,
\begin{equation}
    T^{(a)(b)} = \mbox{diag}(\rho, P_1,P_2,P_3), \label{t_ab}
\end{equation}
where $\rho$ and $P$ are the matter density and the pressure, respectively. 
In case of the rotating regular black hole, the quantities $\rho$ and $P$ 
that appearing in (\ref{t_ab}) have the forms as follows
\begin{eqnarray}\label{rset}
    \rho &=& \frac{6 M r^4 e^{-r^3/b^3}}{b^3 \Sigma^2} = -P_1, \nonumber \\
    P_2 &=& - \frac{3 M r^2 e^{-r^3/b^3} \left[-3 r^3 \Sigma 
    +2 b^3(r^2+2a^2 \cos^2 \theta)\right]}{ b^6\Sigma^2}  = P_3.
\end{eqnarray}
When we compare them with the components of static counterpart 
\eqref{set}, we find that the components of \eqref{rset} are different 
because of the black hole rotation. This also happens when we consider 
rotating Vaidya solution \cite{Carmeli:1975kg}, it has in addition null 
radiation some another source. Having the expressions of matter density 
$\rho$ and pressure $P_i$, we can verify the weak energy condition for the 
spacetime \eqref{metric} that requires the inequalities $\rho \geq 0$ and 
$\rho + P_i \geq 0$, ($i=1,2,3$) to be satisfied. 
\begin{figure*}
    \begin{tabular}{c c}
    \includegraphics[scale=0.6]{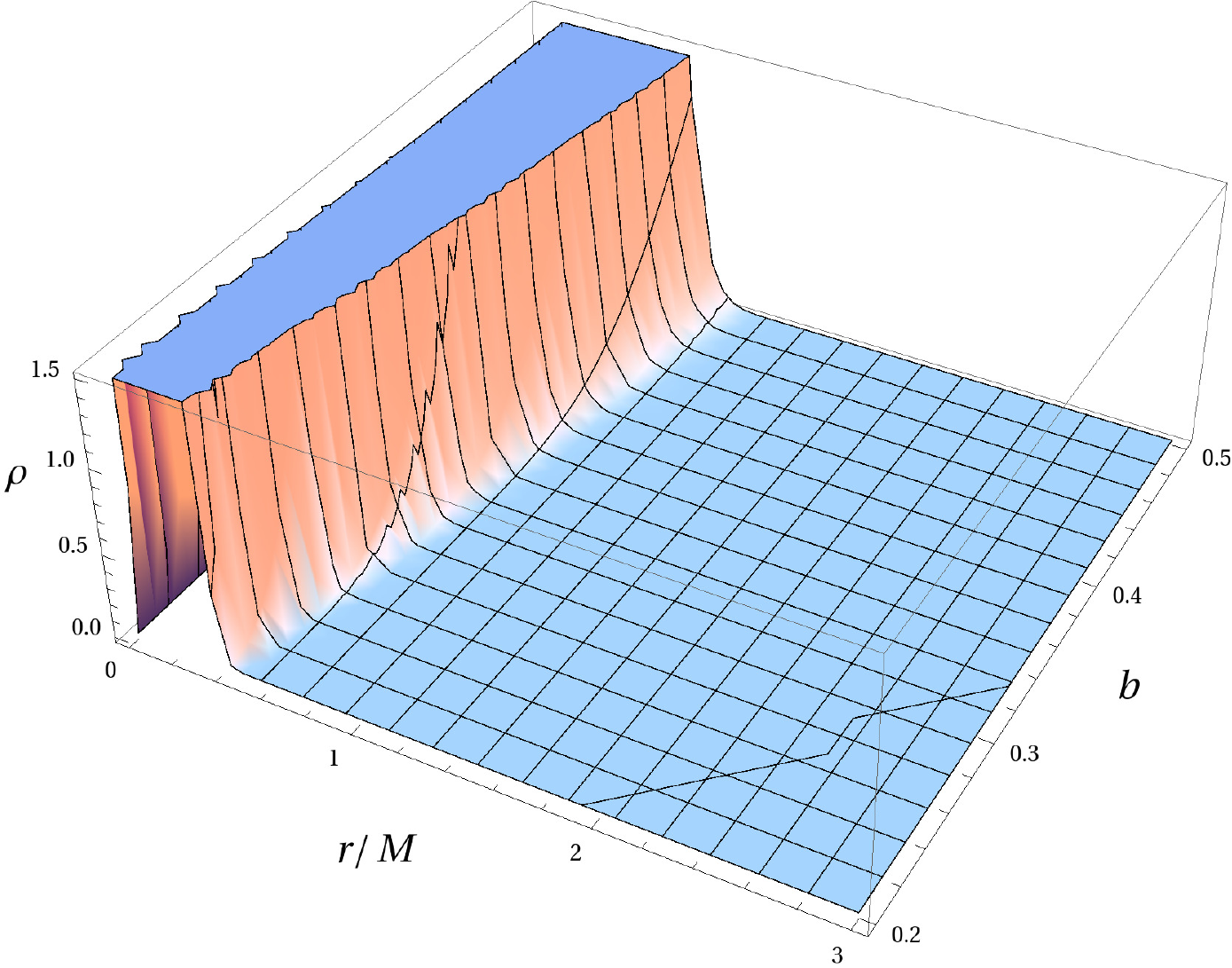} \quad
    \includegraphics[scale=0.6]{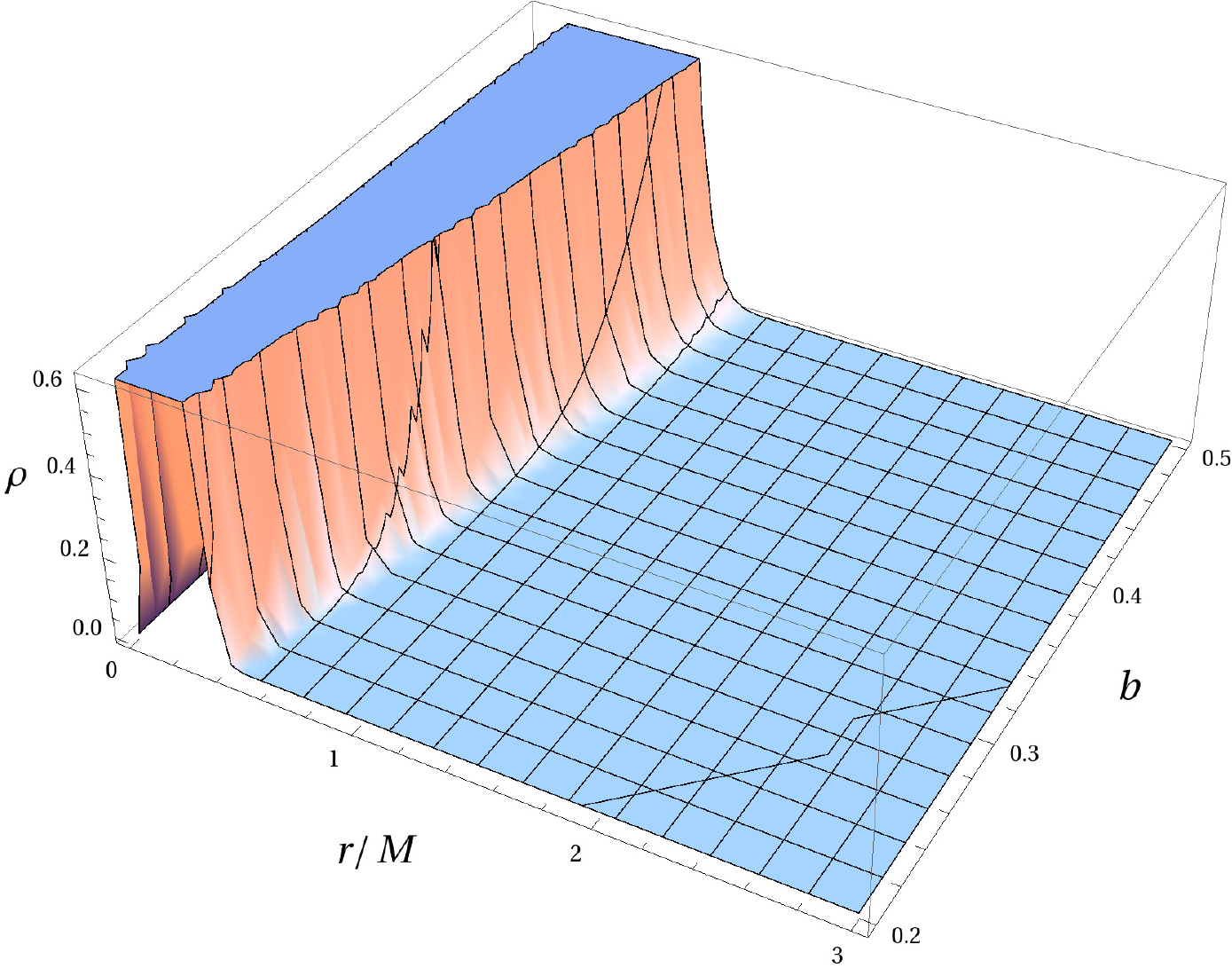}\\
    \includegraphics[scale=0.6]{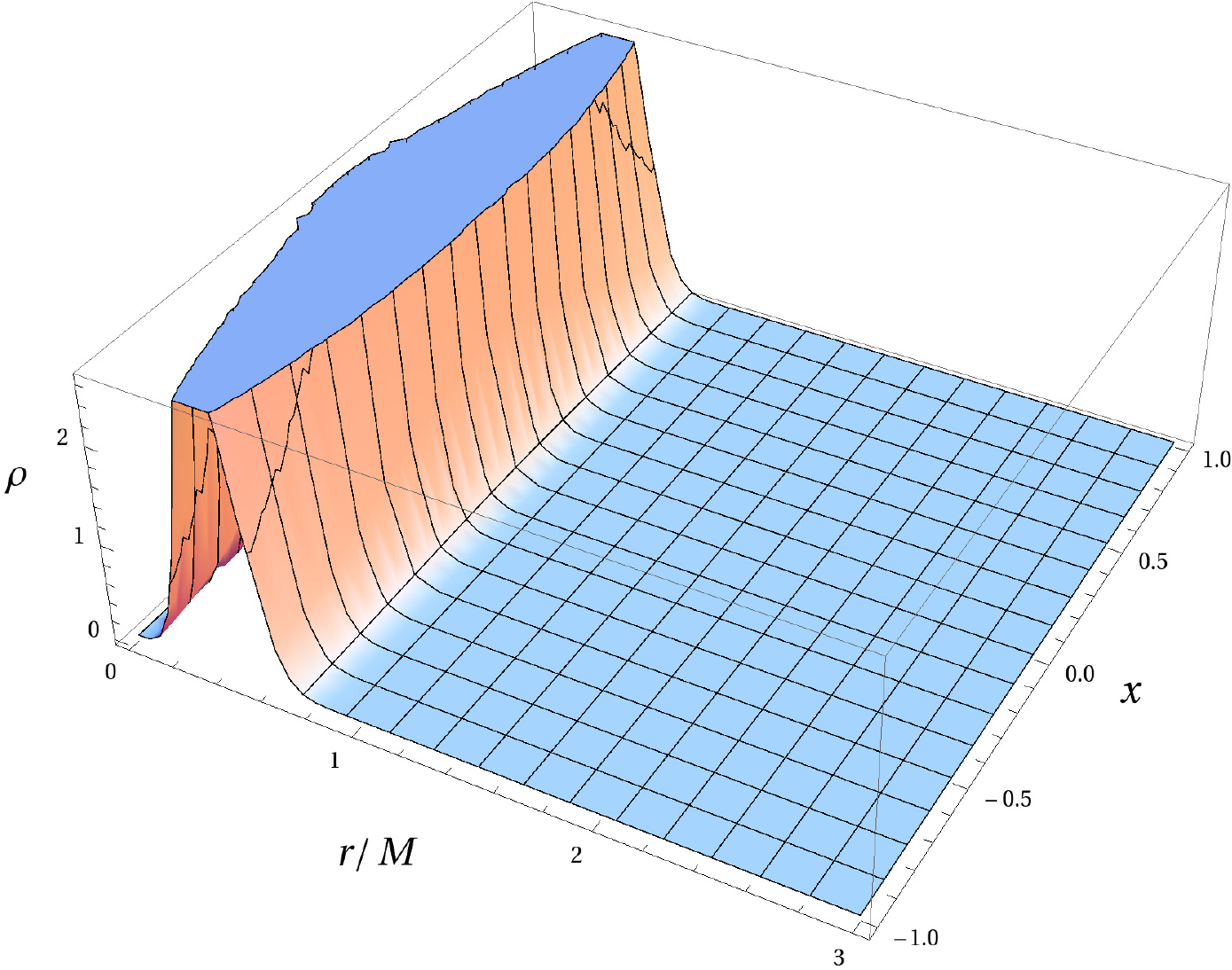} \quad
    \includegraphics[scale=0.6]{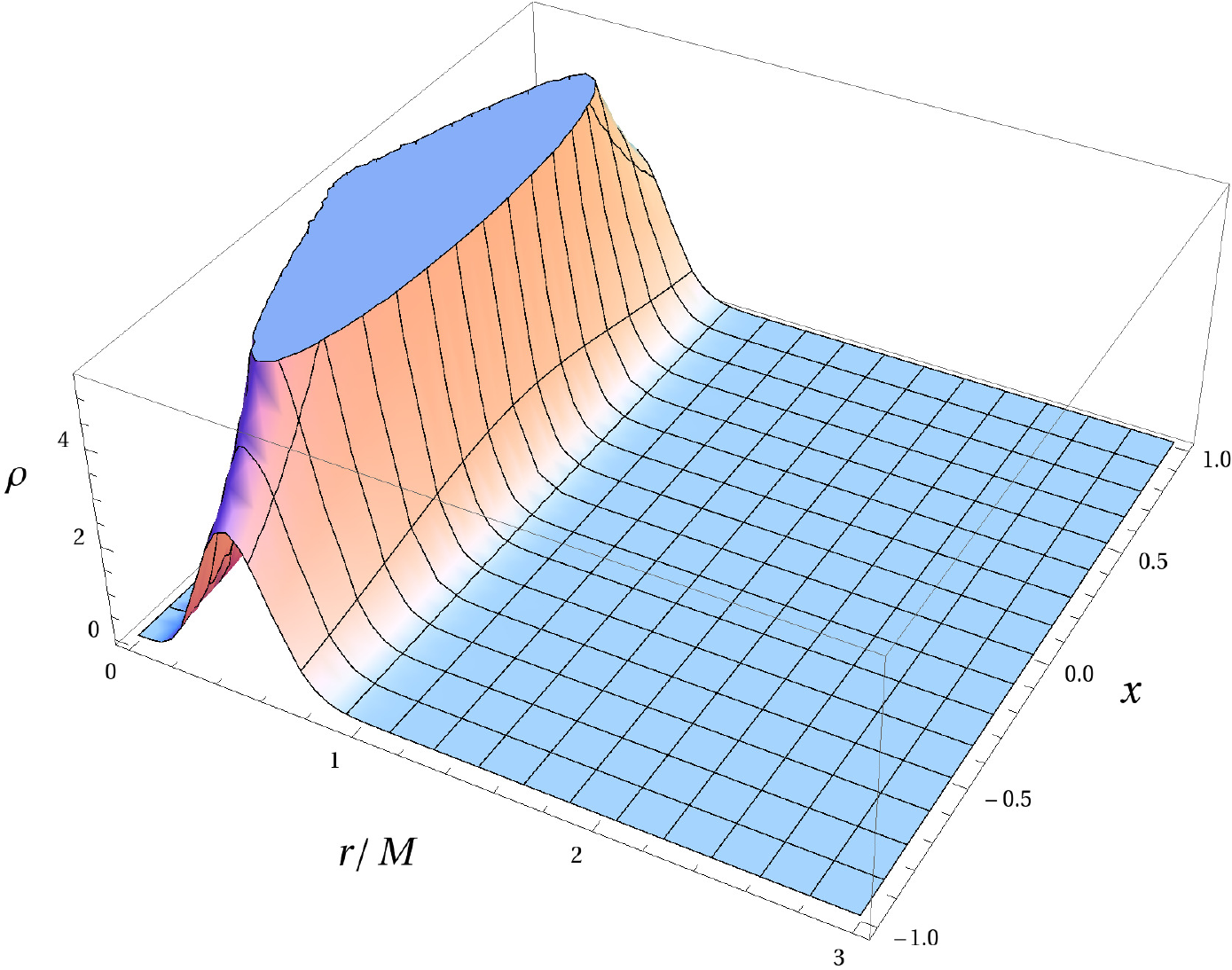}
    \end{tabular}
    \caption{\label{rho} Radial and angular dependence of matter density 
    $\rho$ for the rotating regular black hole, for the given values of the 
    rotation parameter $a=0.4, 0.6$, the parameter $b =0.4, 0.5$, and 
    $\theta =\pi/4$ ($ x=\cos \theta $).}
\end{figure*}
\begin{figure*}
    \begin{tabular}{c c}
    \includegraphics[scale=0.6]{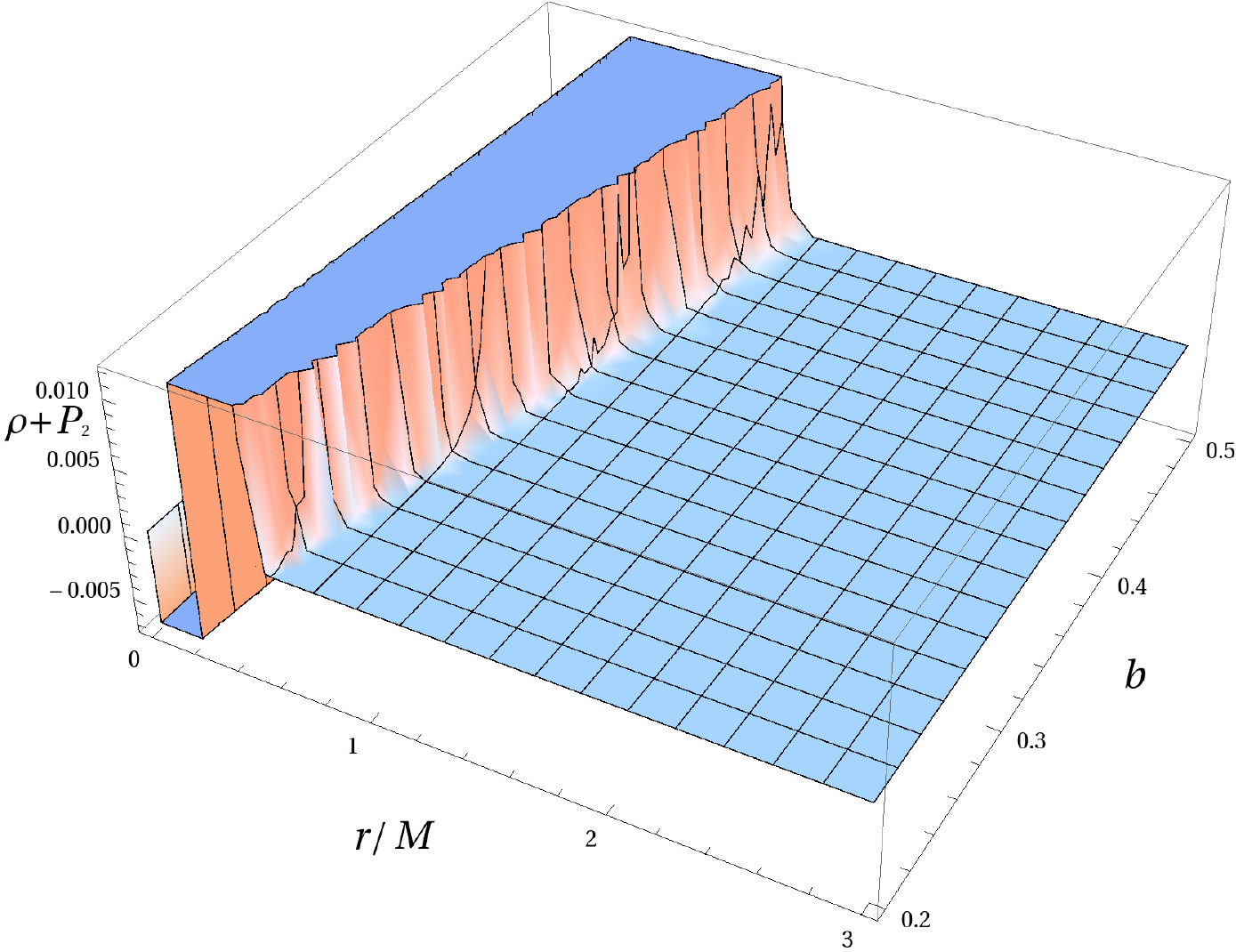} \quad
    \includegraphics[scale=0.6]{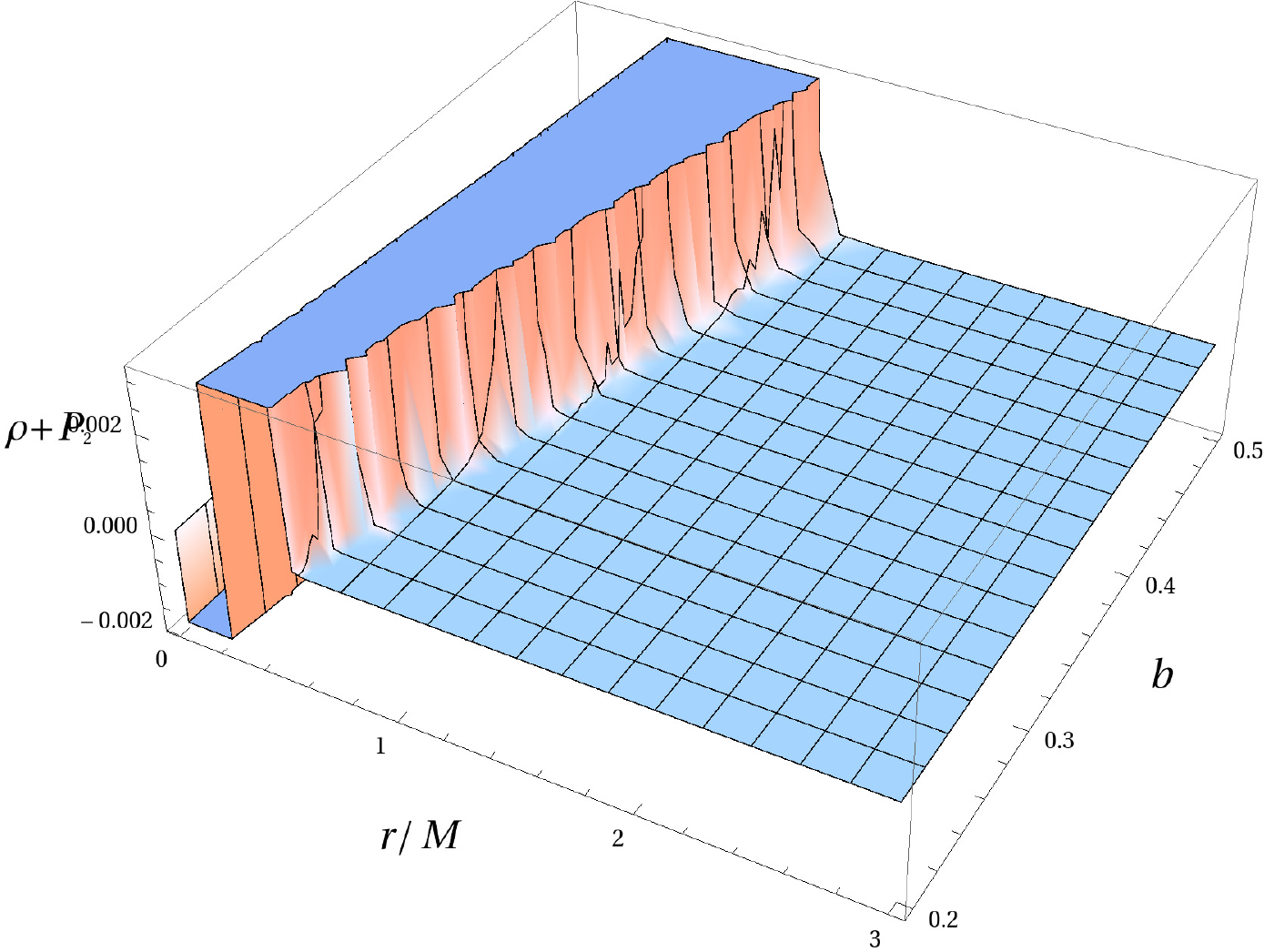}\\
    \includegraphics[scale=0.6]{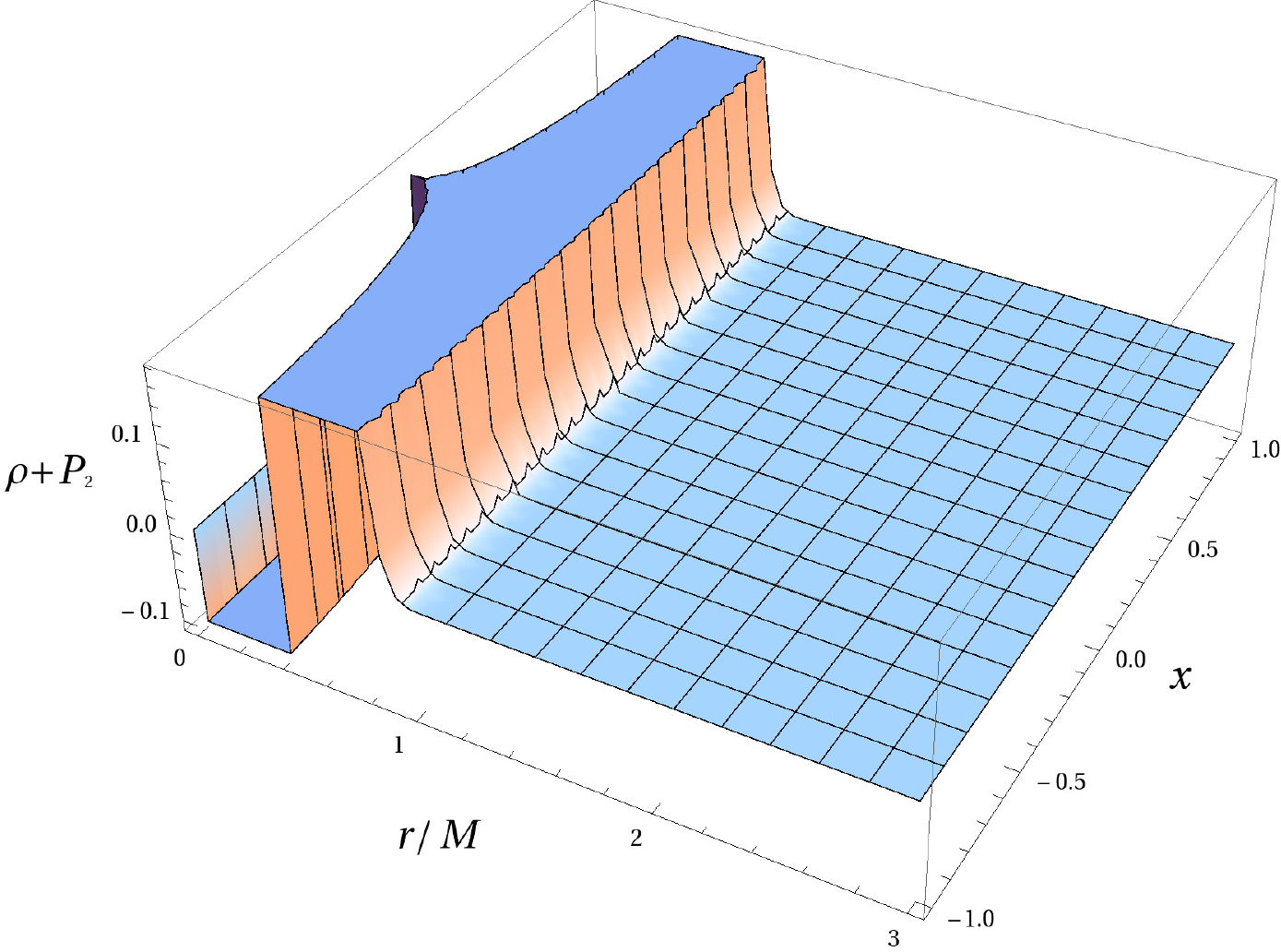} \quad
    \includegraphics[scale=0.6]{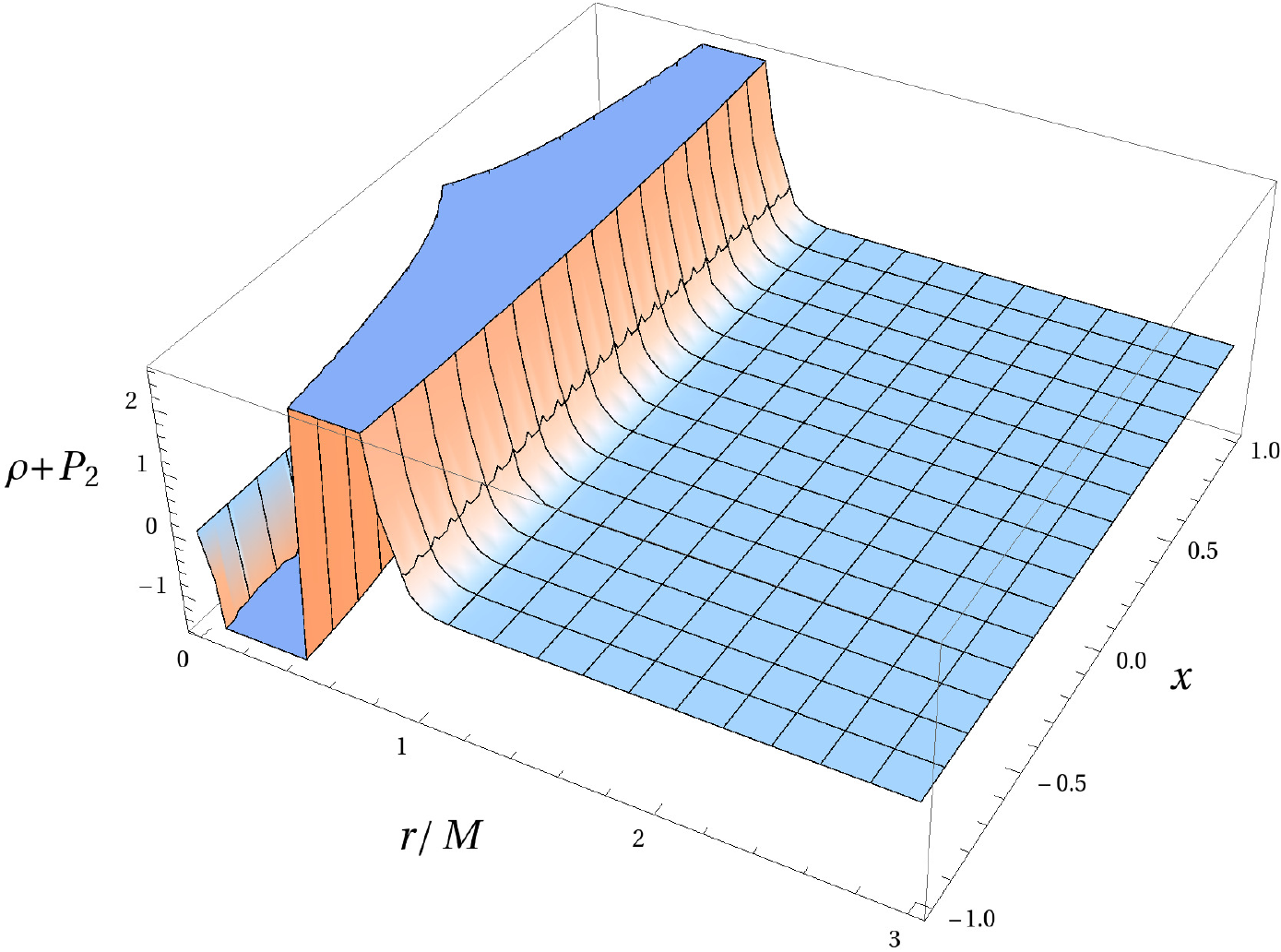}
    \end{tabular}
    \caption{\label{rhop} Radial and angular dependence of $\rho+P_2$ for the 
    rotating regular black hole, for the given values of the rotation parameter 
    $a=0.4, 0.6$, the parameter $b= 0.4, 0.5$, and $\theta =\pi/4$ 
    ($ x=\cos \theta $).}
\end{figure*}
Moreover, a straight forward calculation gives
\begin{equation}
    \rho+P_2 = \rho+P_3 = 
    \frac{3 M r^2 e^{-r^3/b^3} \left[3r^3 \Sigma -4a^2b^3 
    \cos^2 \theta\right]}{b^6 \Sigma^2}, \label{wec}
\end{equation}
which shows that the energy condition may be violated and it can be seen in 
the plots (cf. Fig.~\ref{rho} and \ref{rhop}). We find that the matter 
density is always positive for all values of $a$ and $b$ as can be seen 
from the Fig.~\ref{rho}. It turns out that the rotating regular black hole 
solution may violate energy conditions; however, this happens for all the 
rotating regular black holes in the literature (see, e.g., 
\cite{Neves:2014aba,Bambi:2013ufa}). Despite small violation of such 
solutions are important from phenomenologically and also they are important 
as astrophysical black holes are rotating.

\section{Properties of rotating regular spacetime}
\label{properties}
In this section, we will discuss some important properties of the rotating 
regular black hole solution, for instance, the curvature scalars, the 
horizons, and the ergospheres etc. It is important to study these basic 
properties of the spacetime from physical point of view.

\subsection{Curvature invariants}
\label{curv}
In order to check the curvature singularity within the spacetime, we 
compute the curvature invariants or scalars of the rotating regular black 
hole spacetime. We find that the curvature invariants of the spacetime 
\eqref{metric} have cumbersome forms, but in the limit 
$\theta \rightarrow \pi/2$, they reduce to the following expressions
\begin{eqnarray}
    \lim_{\theta \rightarrow \pi/2} \mathcal{R} &=&
    \frac{18M^2(9r^6 -12r^3b^3 +8b^6)e^{-2r^3/b^3}}{b^{12}}, \nonumber\\
    \lim_{\theta \rightarrow \pi/2} K & = & 
    \frac{12M^{2}e^{-2r^3/b^3}}{b^{12} r^6}
    \Big[4b^{12}(1+e^{2r^3/b^3}) - 4 (3r^6b^6 +2r^3b^9 +2b^{12})e^{r^3/b^3} 
    \nonumber\\ 
    && + 27r^{12} + 24r^6b^6 + 8r^3b^9 \Big], \nonumber\\
    \lim_{\theta \rightarrow \pi/2} R &=& 
    -\frac{6 M (3 r^3-4 b^3) e^{-r^3/b^3}}{b^6}. \label{limits}
\end{eqnarray}
We further consider the limit $r \rightarrow 0$ of \eqref{limits}, in this 
limit they turn out to be
\begin{eqnarray}
    \lim_{r \rightarrow 0}\lim_{\theta \rightarrow \pi/2} \mathcal{R} 
    &=& \frac{144M^2}{b^{6}}, \quad
    \lim_{r \rightarrow 0} \lim_{\theta \rightarrow \pi/2} K 
    = \frac{96M^{2}}{b^{6}}, \quad
    \lim_{r \rightarrow 0} \lim_{\theta \rightarrow \pi/2} R 
    = \frac{24 M}{b^3}. \label{limits1}
\end{eqnarray}
\begin{figure*}
    \begin{tabular}{c c}
        \includegraphics[scale=0.6]{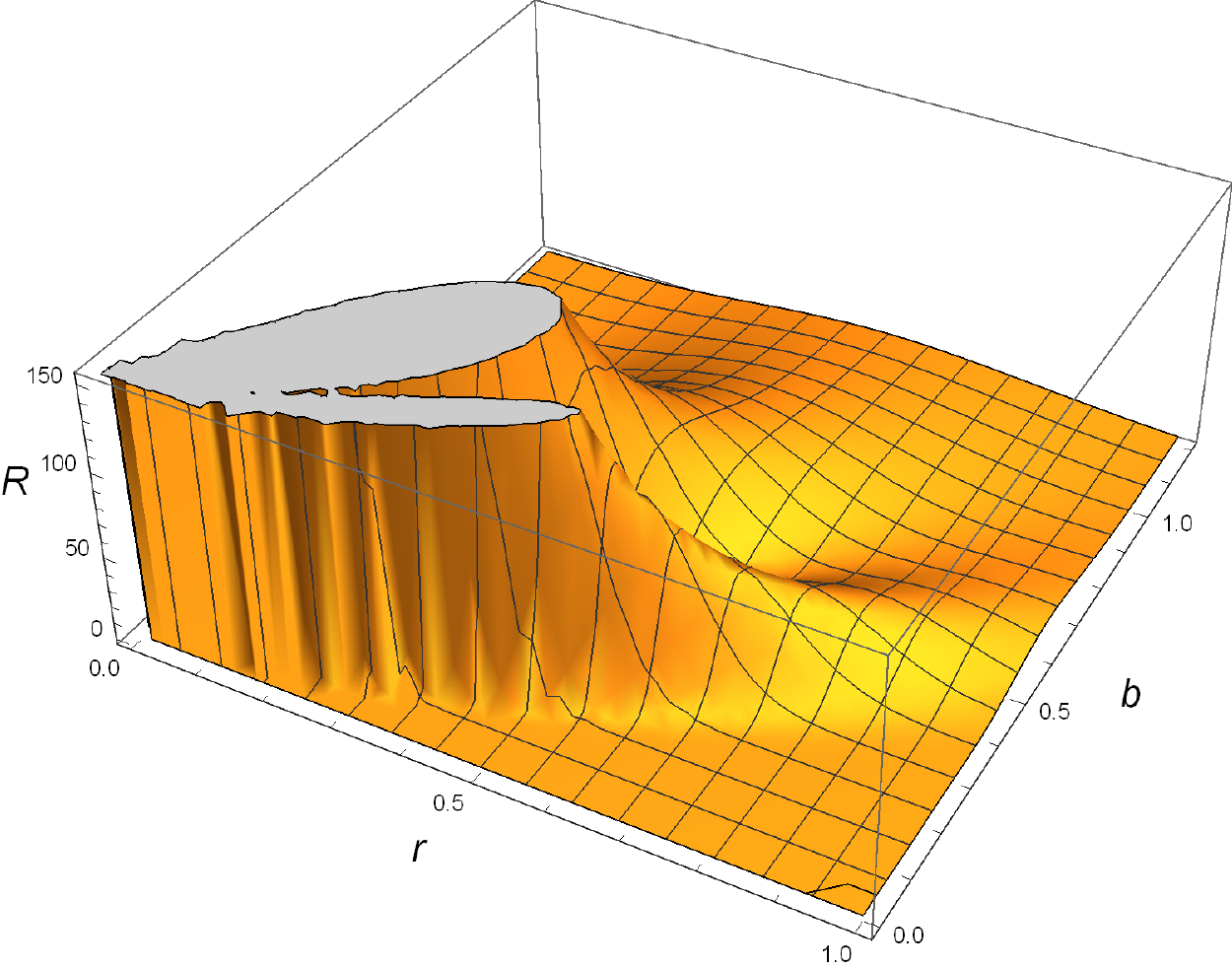} \quad
        \includegraphics[scale=0.6]{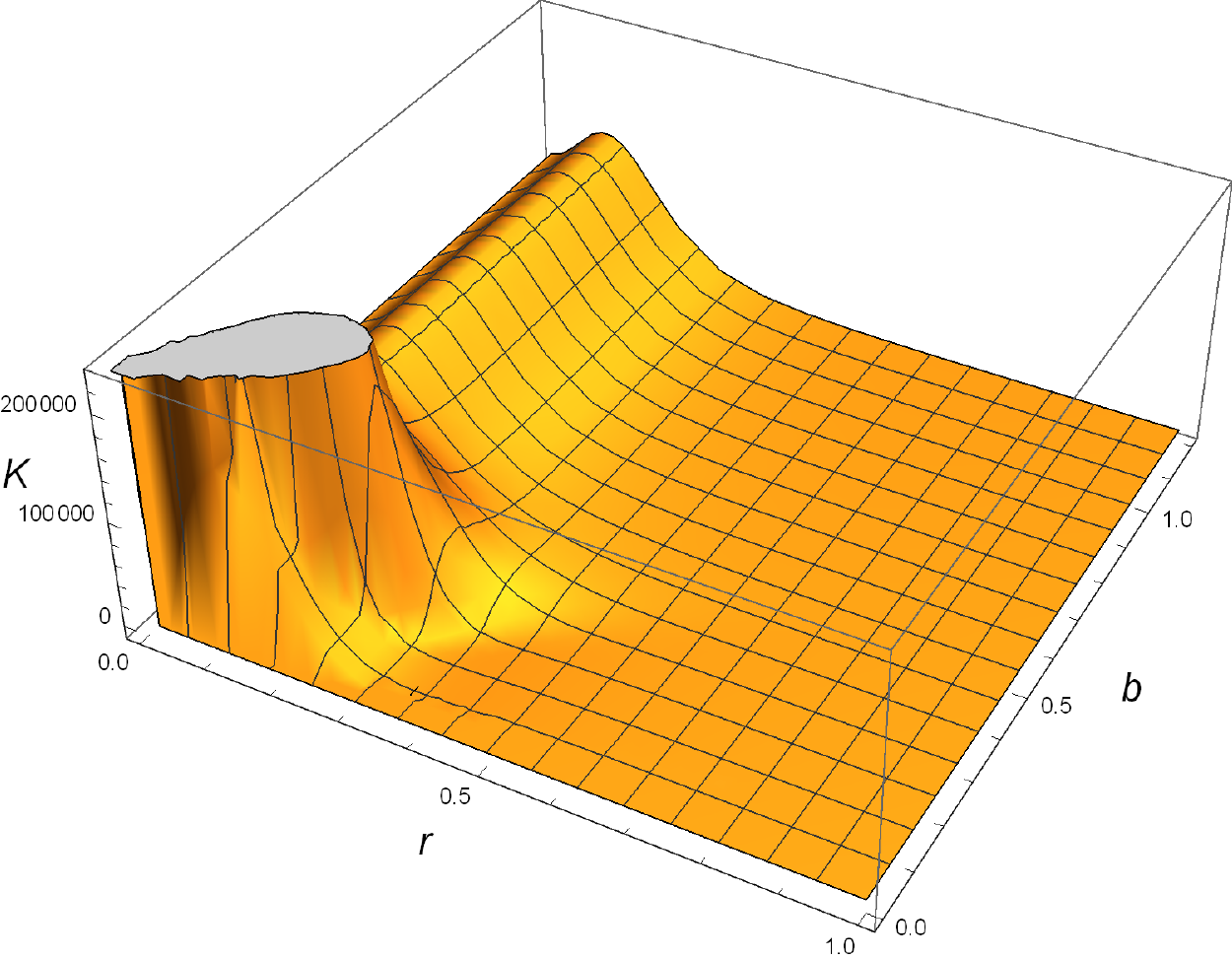}
    \end{tabular}
    \caption{\label{ks} The behavior of Ricci square and Riemann square 
    for the rotating regular black hole.}
\end{figure*}
The curvature invariants have finite values for $M,b\neq 0$, therefore, 
we can say that the metric \eqref{metric} is regular everywhere as can be 
seen from the Fig.~\ref{ks}. The figure shows that the Kretschmann scalar 
and the square of Ricci tensor are well behaving for different values of 
the parameters $b$ and $a$. It is noticeable that the presence of the 
exponential factor in the mass of the black hole \eqref{metric}, i.e., 
$e^{-r^3/b^3}$, makes the black hole singularity free.

\subsection{Horizons}
\label{hor}
\begin{figure}
    \includegraphics[scale=0.57]{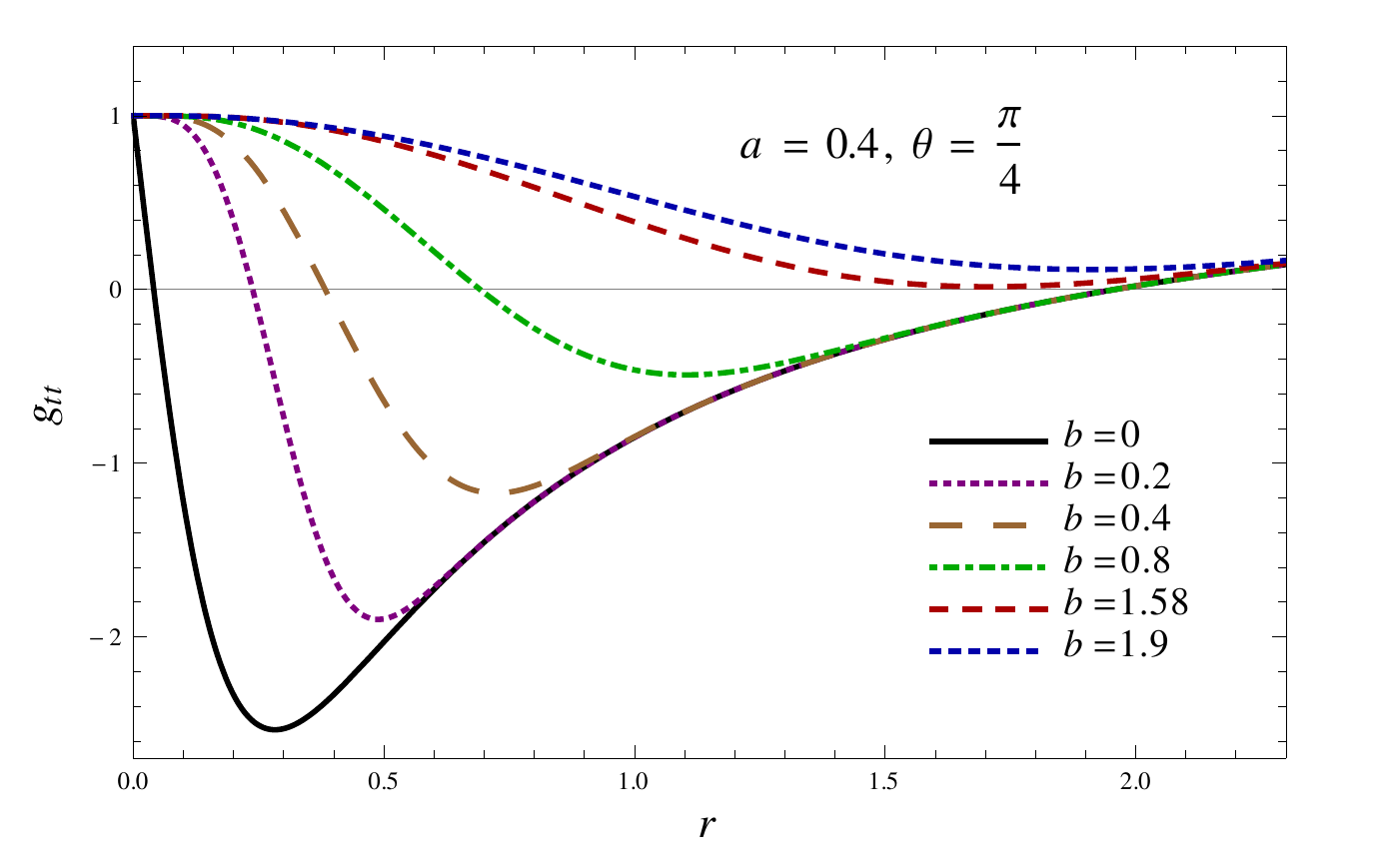}
    \includegraphics[scale=0.57]{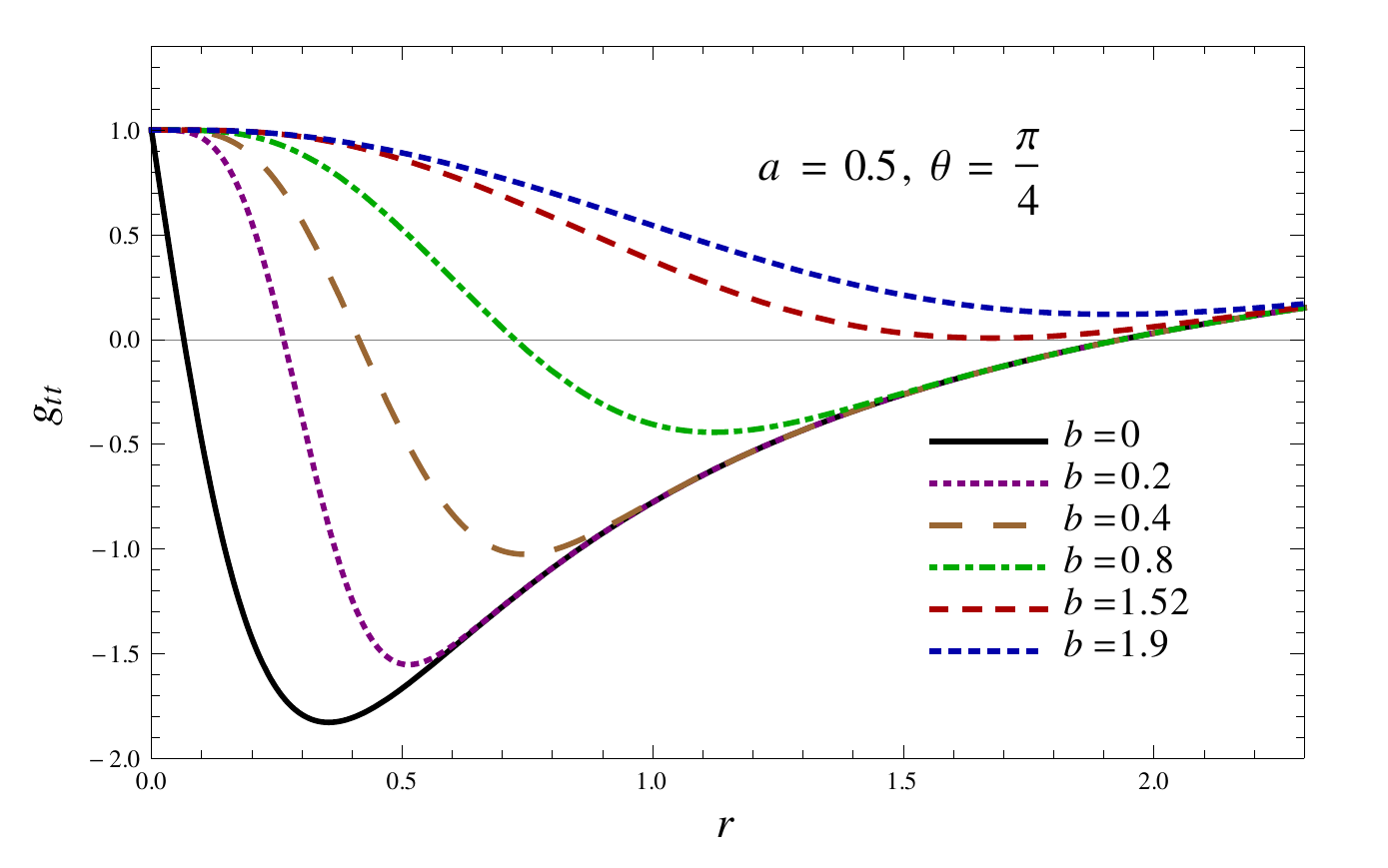}
    \caption{\label{slsf} Plots showing the variation of static limit 
    surface with radius $r$ for different values of the parameter $b$ 
    and the rotation parameter $a$.}
\end{figure}

Now we wish to discuss the effect of charge on the structure of horizons 
and ergosphere. It turns out that as like the Kerr black hole, the 
spacetime has two surfaces, viz., static limit surfaces and horizons. The 
static limit surface is a surface on which the time translation Killing 
vectors ($\chi^a$) becomes null or $\chi^a \chi_a =0$. It requires that 
the $g_{tt}$ component to be vanished,
\begin{equation}
    r^2 +a^2 \cos^2 \theta -2M(1-e^{-r^3/b^3})r =0. \label{sls}
\end{equation}
The typical behavior of the static limit surface is depicted in 
Fig.~\ref{slsf} for various values of the parameters 
(see also Table~\ref{table:ergo}). It turns out that the behavior 
is similar to that of the Kerr black holes. However, the radii of 
the static limit surface shrink with increasing values of charge $b$ 
(Table~\ref{table:ergo}).
\begin{table}
    \begin{center}
        \caption{Radius of EHs, SLSs and $\delta^{a}=r^{+}_{SLS}-r^{+}_{EH}$ 
        for different values of parameter $b$}\label{table:ergo}
    \resizebox{\textwidth}{!}{
        \begin{tabular}{| c | c c c | c c c | c c c | c c c |}
            \hline 
        &\multicolumn{3}{c}{$a=0.40$}  &\multicolumn{3}{c}{$a=0.45$}
        &\multicolumn{3}{c}{$a=0.50$} &  \multicolumn{3}{c|}{$a=0.55$}\\
            \hline
    $b$ & $r^{+}_{H}$ & $r^{+}_{sls}$ & $\delta^{0.40}$  & $r^{+}_{H}$ 
    & $r^{+}_{sls}$ & $\delta^{0.45}$ 
    & $r^{+}_{H}$ & $r^{+}_{sls}$  & $\delta^{0.50}$ & $r^{+}_{H}$ 
    & $r^{+}_{sls}$  & $\delta^{0.55}$ \\
            \hline
    0    & 1.91652  & 1.95917  & 0.04265  & 1.89303  & 1.94802  & 0.05499  
    & 1.86603  & 1.93541  & 0.06938 & 1.83516  & 1.92128  & 0.08612 \\
    0.20    & 1.91652  & 1.95917  & 0.04265  & 1.89303  & 1.94802  & 0.05499  
    & 1.86603  & 1.93541  & 0.06938 & 1.83516  & 1.92128  & 0.08612 \\
    0.40    & 1.91652  & 1.95917  & 0.04265  & 1.89303  & 1.94802  & 0.05499  
    & 1.86603  & 1.93541  & 0.06938 & 1.83516  & 1.92128  & 0.08612 \\
    0.80    & 1.91648  & 1.95915  & 0.04267  & 1.89298  & 1.94800  & 0.05502  
    & 1.86594  & 1.93539  & 0.06945 & 1.83502  & 1.92124  & 0.08622 \\
    1.10    & 1.90991  & 1.95491  & 0.04500  & 1.88464  & 1.94325  & 0.05861 
    & 1.85500  & 1.92997  & 0.07497 & 1.82009  & 1.91499  & 0.0949 \\
    1.20    & 1.89844  & 1.94703  & 0.04859  & 1.87049  & 1.93456  & 0.06407  
    & 1.83687  & 1.92027  & 0.08340 & 1.79566  & 1.90399  & 0.10833 \\
    1.30    & 1.81955  & 1.89605  & 0.07650  & 1.76301  & 1.87791  & 0.11490  
    & 1.79581  & 1.89955  & 0.10374 & 1.73422  & 1.88046  & 0.14624 \\	
            \hline 
        \end{tabular}
    }
    \end{center}
\end{table}

\begin{figure}
    \includegraphics[scale=0.56]{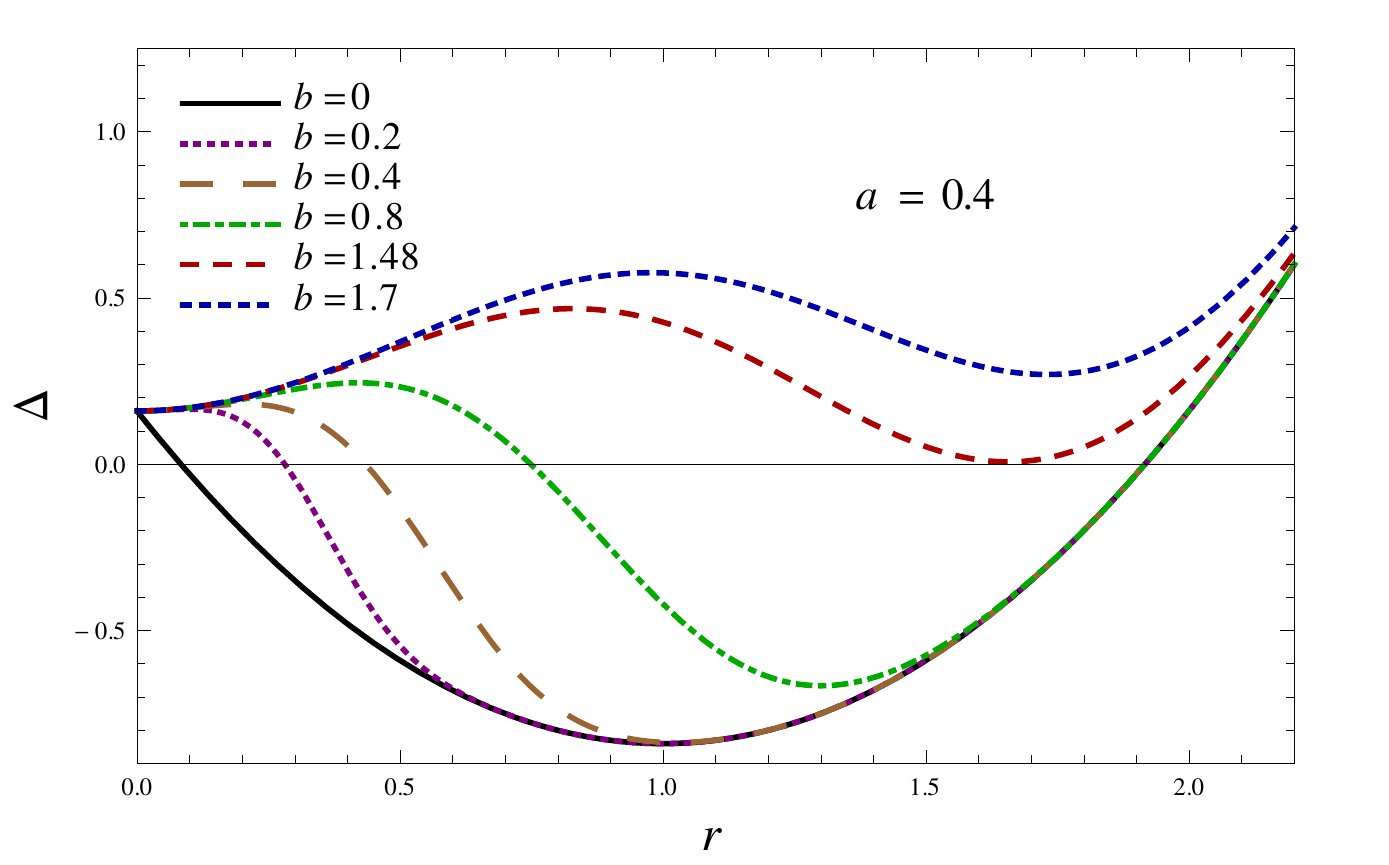} 
    \includegraphics[scale=0.56]{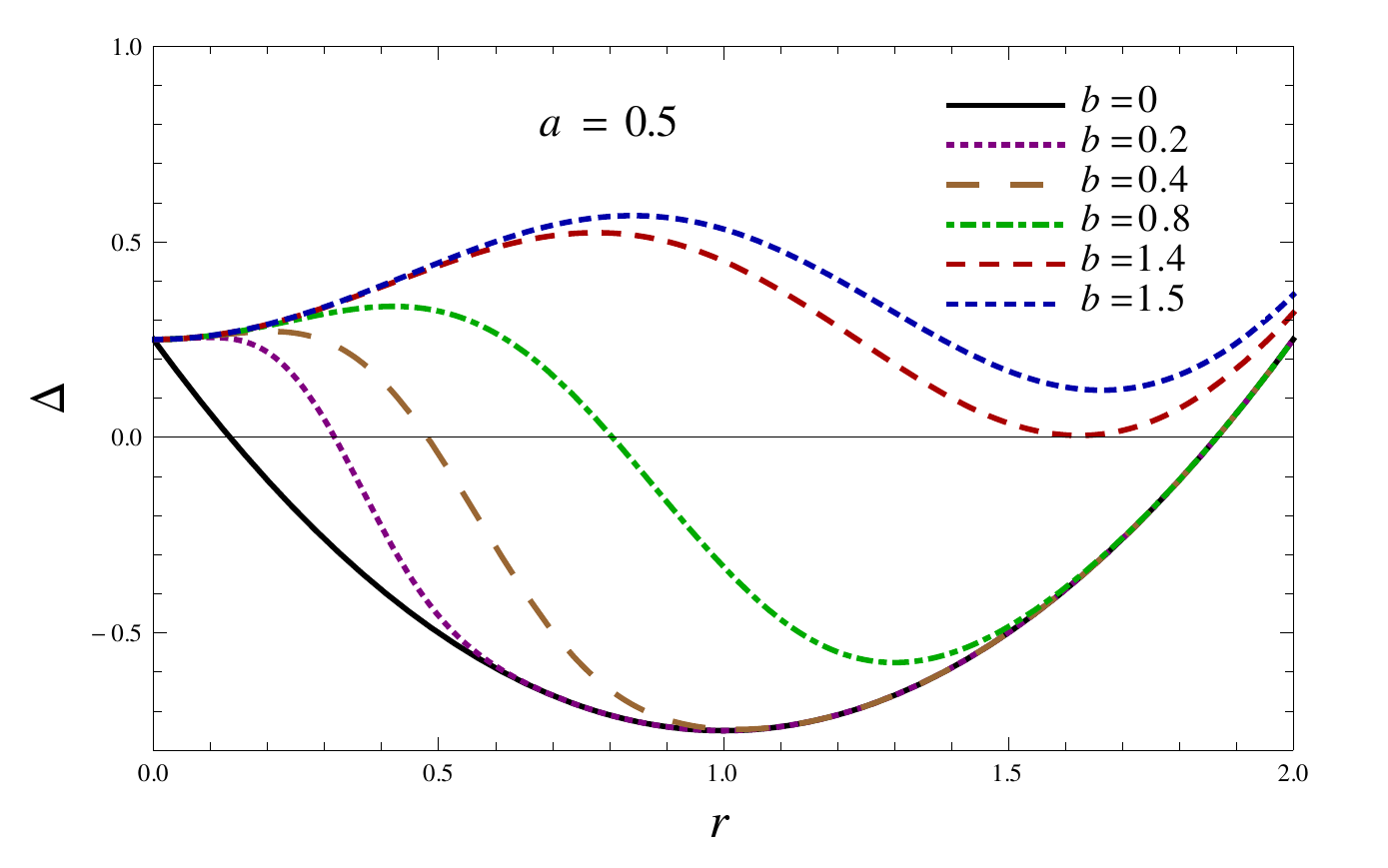}
    \caption{\label{ehf} Plots showing the variation variation of $\Delta$ 
    with radius $r$ for different values of the parameter $b$ and the 
    rotation parameter $a$.}
\end{figure}

Since the spacetime \eqref{metric} has a coordinate singularity at 
$\Delta =0$, which corresponds to the horizon of the rotating regular 
black holes. Actually the event horizon is located at the larger root 
($r_+$) of 
\begin{equation}
    \Delta = r^2 +a^2 -2M(1-e^{-r^3/b^3})r, \label{horz}
\end{equation}
where \eqref{sls} coincides with \eqref{horz} when either $\theta = 0$ or 
$\pi$. Clearly, the radii of horizons depend on the parameter $b$ and they 
are different from the Kerr black hole. We solve \eqref{horz} for horizons 
numerically as well as plot them in Fig.~\ref{ehf} for different values of 
parameters $a$ and $b$. It turns out that the horizons of the rotating 
regular black hole have similar behavior like the Kerr black hole, but 
there exist several extremal black holes corresponding to the various 
values of charge $b$.
\begin{table}
	\begin{center}
		\caption{Radius of inner horizon and outer horizon for different 
		values of parameter $b$ and $a$}\label{table:horizons}
%		\resizebox{\textwidth}{!}{
		\begin{tabular}{| c | c c | c c | c c | c c|}
			\hline 
		&\multicolumn{2}{c}{$a=0.40$}  &\multicolumn{2}{c}{$a=0.45$}
		&\multicolumn{2}{c}{$a=0.50$} &  \multicolumn{2}{c|}{$a=0.55$}\\
			\hline
		$b$ & $r^{-}_{H}$  & $r^{+}_{H}$  & $r^{-}_{H}$ & $r^{+}_{H}$ 
		& $r^{-}_{H}$ & $r^{+}_{H}$ & $r^{-}_{H}$  & $r^{+}_{H}$   \\
			\hline
		0     & 0.28085  & 1.91652  & 0.29917  & 1.89303  & 0.31791 & 1.86603  
		& 0.33734 & 1.83516  \\
		0.40  & 0.43462  & 1.91652  & 0.45736  & 1.89303  & 0.48063 & 1.86603 
		& 0.50460 & 1.83516  \\
			% 			%
		0.80  & 0.74894  & 1.91648  & 0.77631  & 1.89298  & 0.80536 & 1.86594 
		& 0.83625 & 1.83502  \\
			% 			%
		1.10  & 1.02687  & 1.90991  & 1.05876  & 1.88464  & 1.09424 & 1.85500 
		& 1.13402 & 1.82009  \\
			% 			%
		1.20  & 1.13547  & 1.89844  & 1.17071  & 1.87049  & 1.21110 & 1.83687 
		& 1.25840 & 1.79566  \\
			%            %
		1.30  & 1.41374  & 1.81955  & 1.47841  & 1.76301  & 1.35248 & 1.79581 
		& 1.42108 & 1.73422  \\
			\hline 
		\end{tabular}
%		}
	\end{center}
\end{table}

The Fig.~\ref{ehf} and Table~\ref{table:horizons} shows the existence of 
two roots, for a set of values of parameters $a$ and $b$, which corresponds 
to the Cauchy horizon (smaller root) and event horizon (larger root). We 
find that for a given value of $a$, there exists a critical value of 
parameter $b=b_c$, where the two horizons coincide ($r=r_{\pm}$) 
corresponding to the extremal black holes (Fig.~\ref{ehf} and \ref{ergo}). 
When $b<b_c$, we have rotating black hole with two horizons, and for 
$b>b_c$ no black hole will form (Fig.~\ref{ehf} and \ref{ergo}).

\subsection{Ergosphere}
\label{ergo1}
\begin{figure*}[h!]
    \begin{tabular}{c c c c}
        \includegraphics[width=0.245\linewidth]{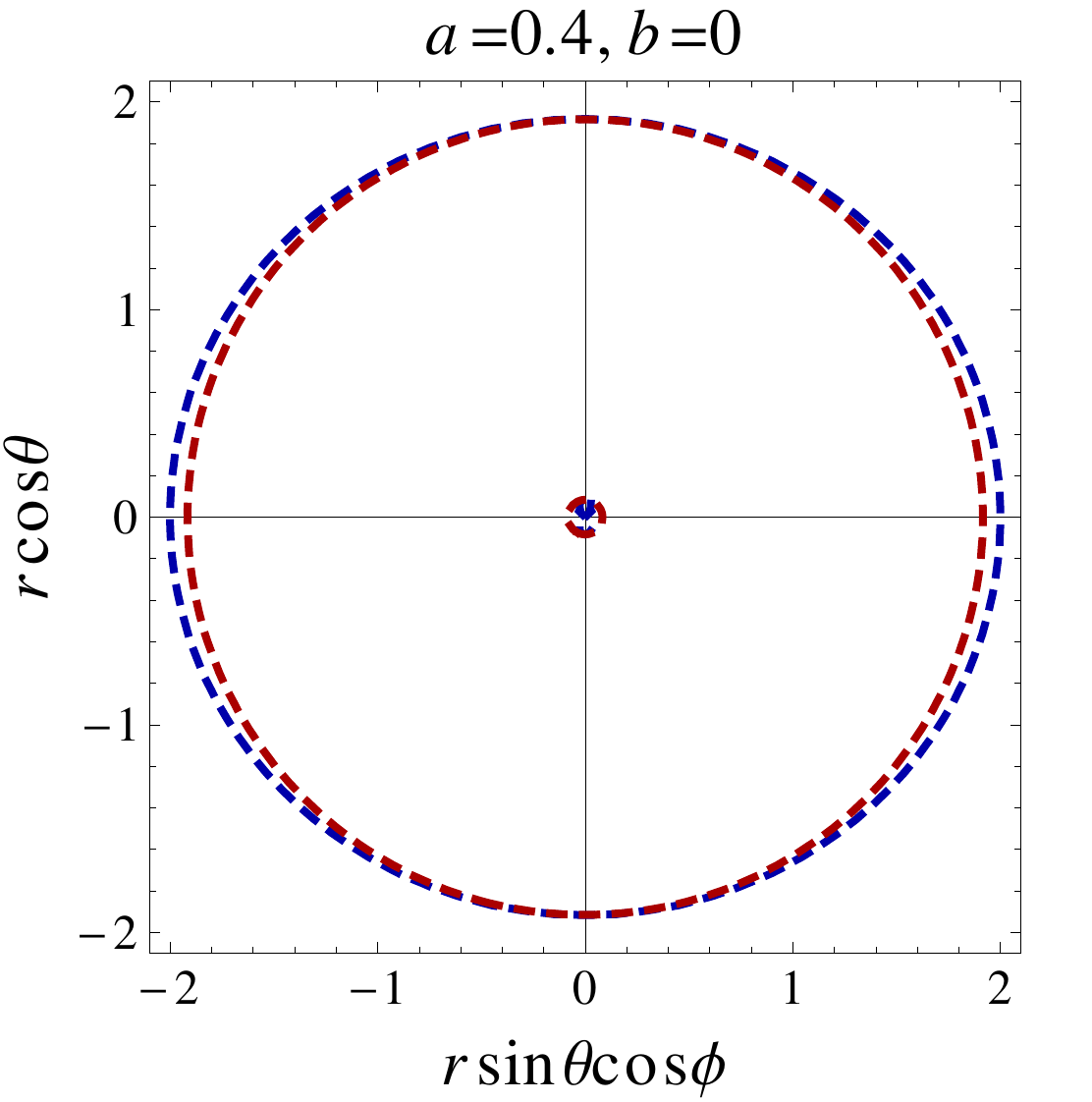}
        \includegraphics[width=0.245\linewidth]{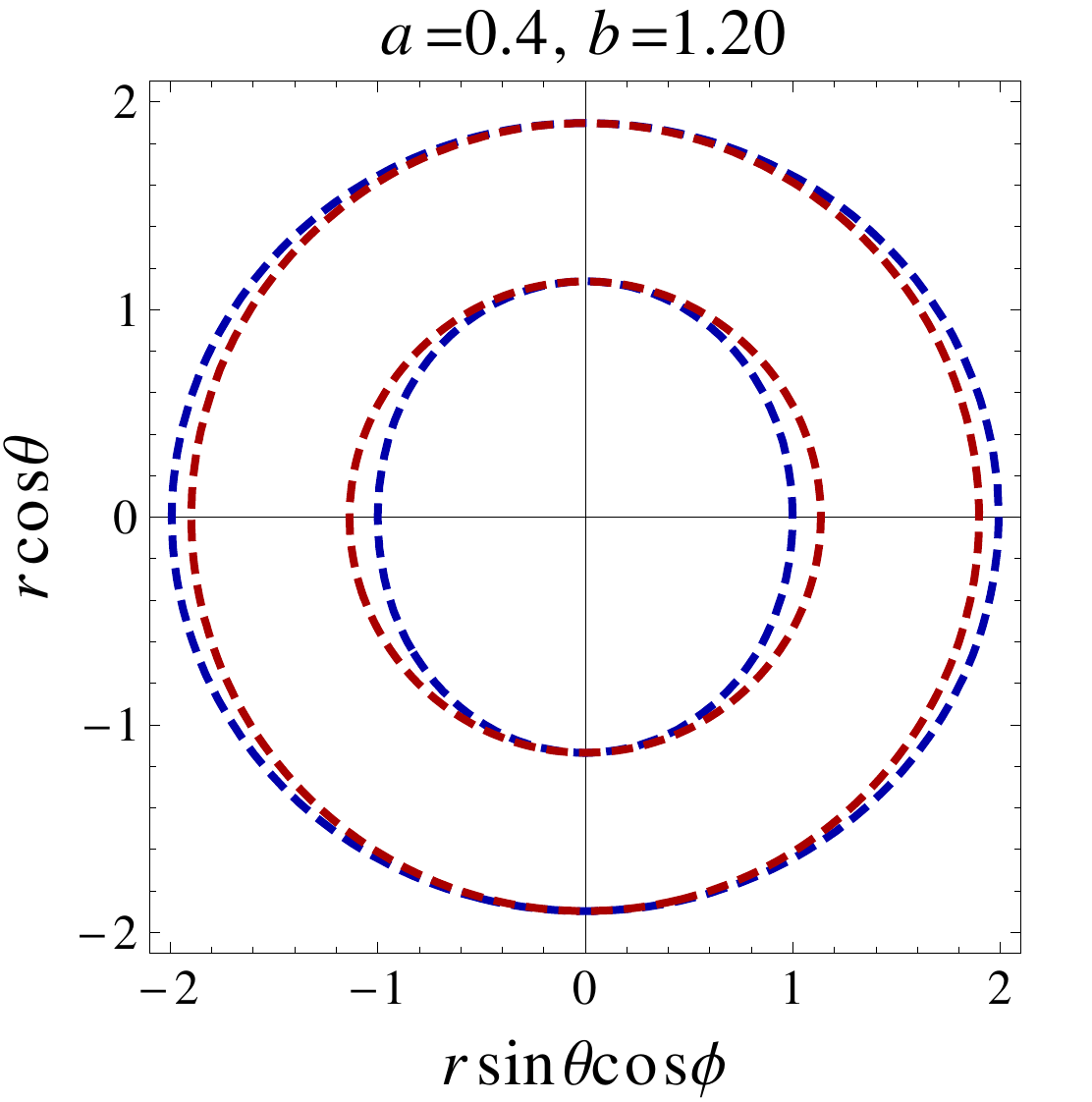}
        \includegraphics[width=0.245\linewidth]{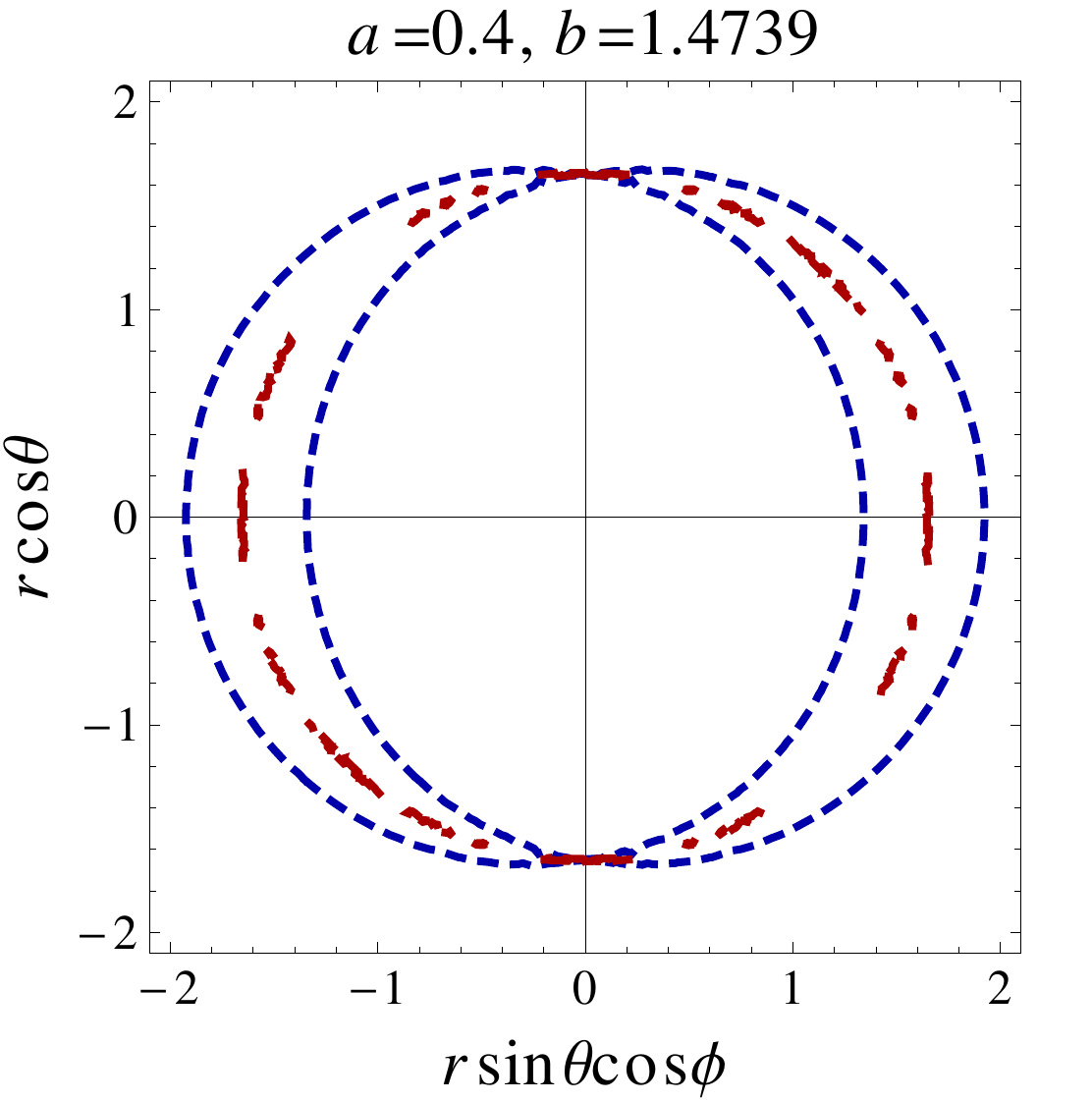}
        \includegraphics[width=0.245\linewidth]{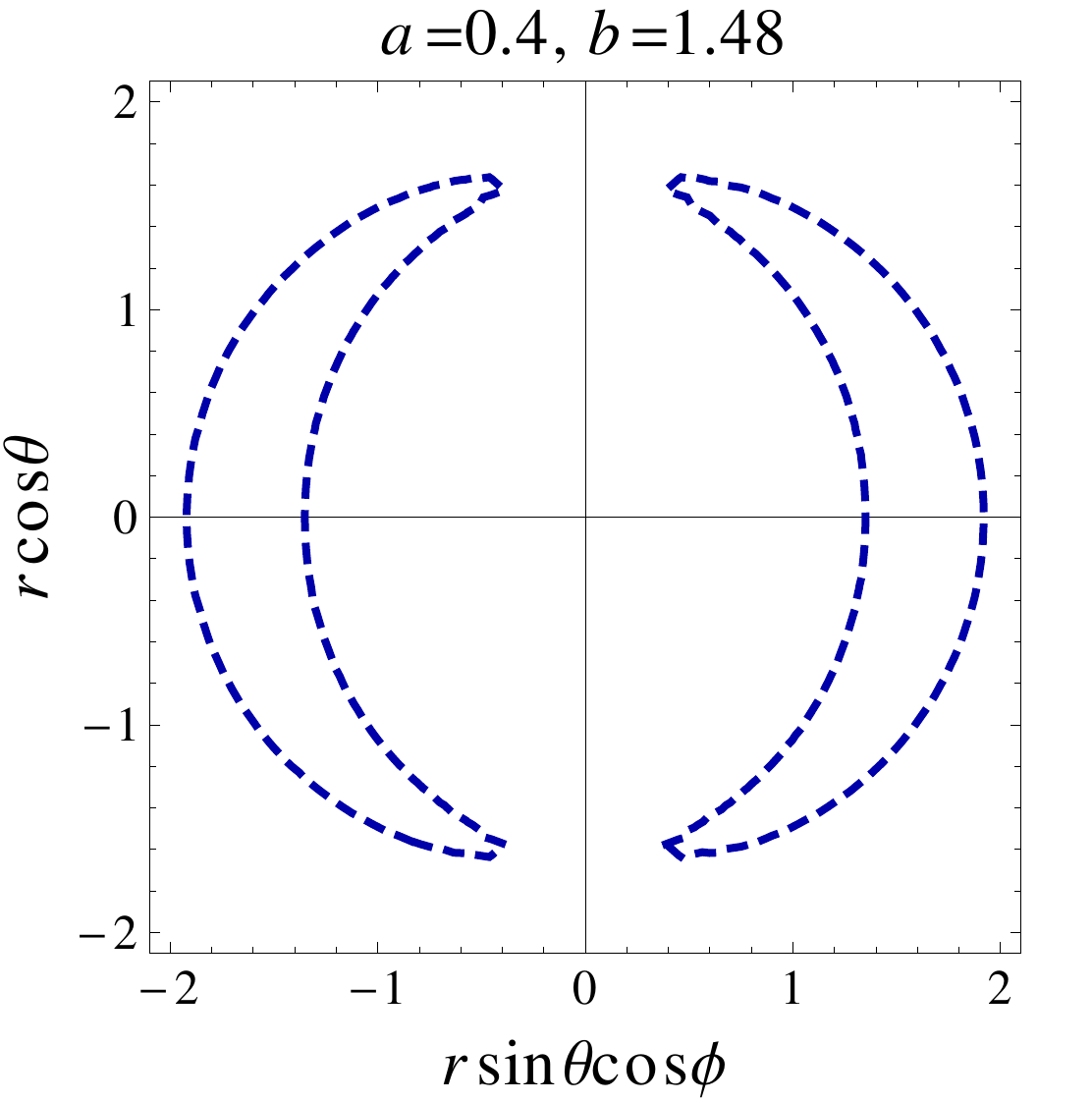}\\
        \includegraphics[width=0.245\linewidth]{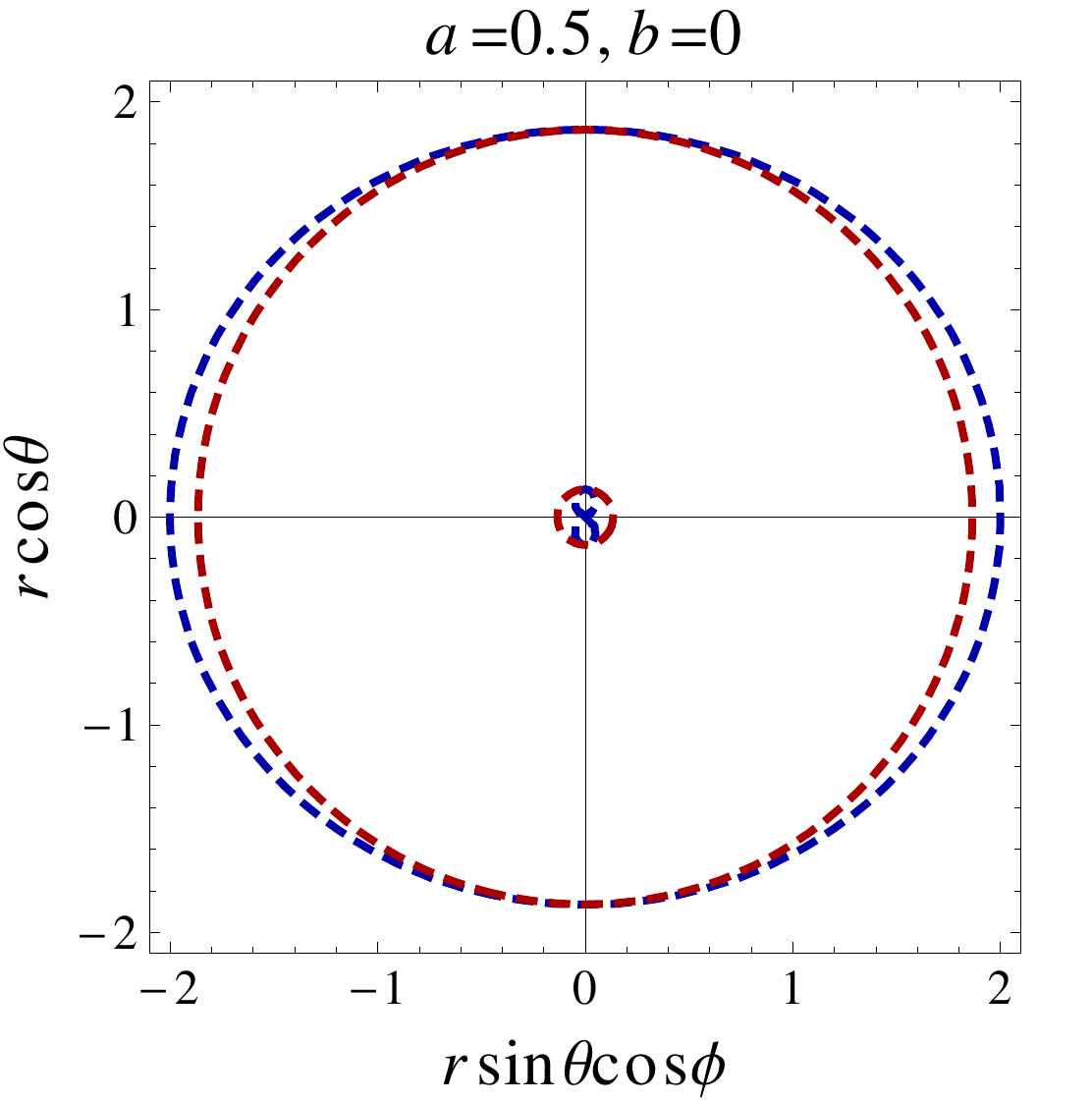}
        \includegraphics[width=0.245\linewidth]{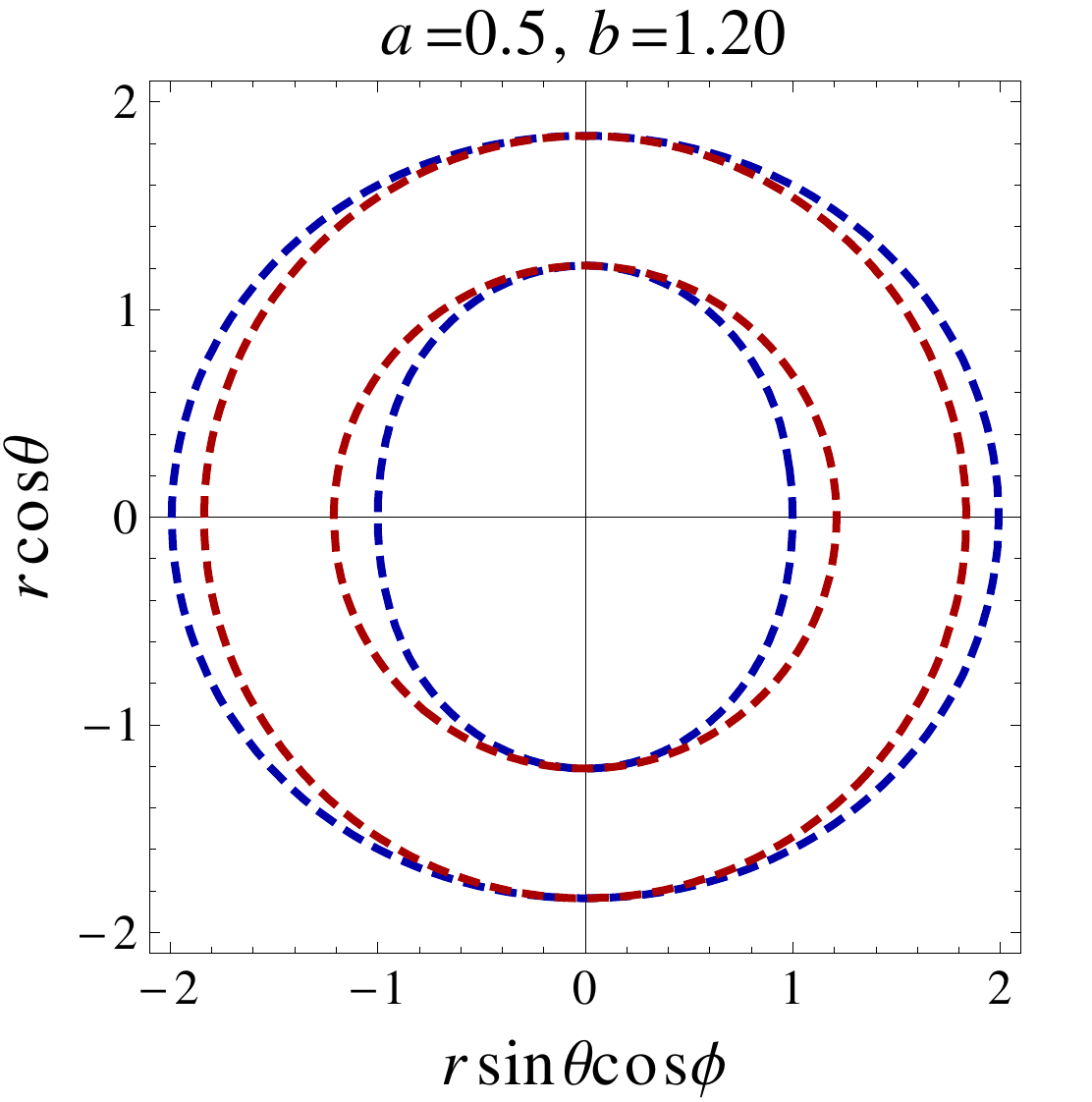}
        \includegraphics[width=0.245\linewidth]{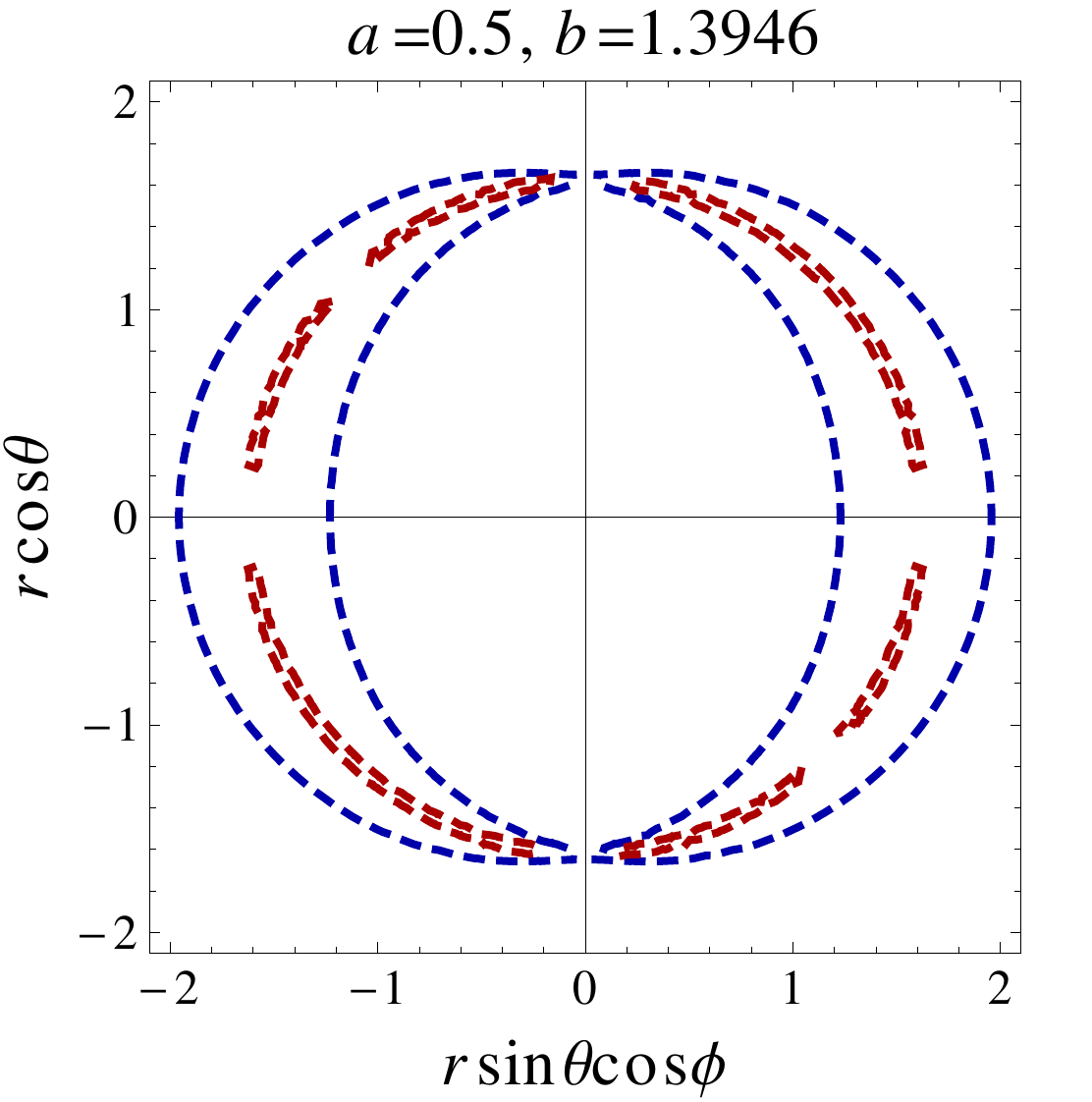}
        \includegraphics[width=0.245\linewidth]{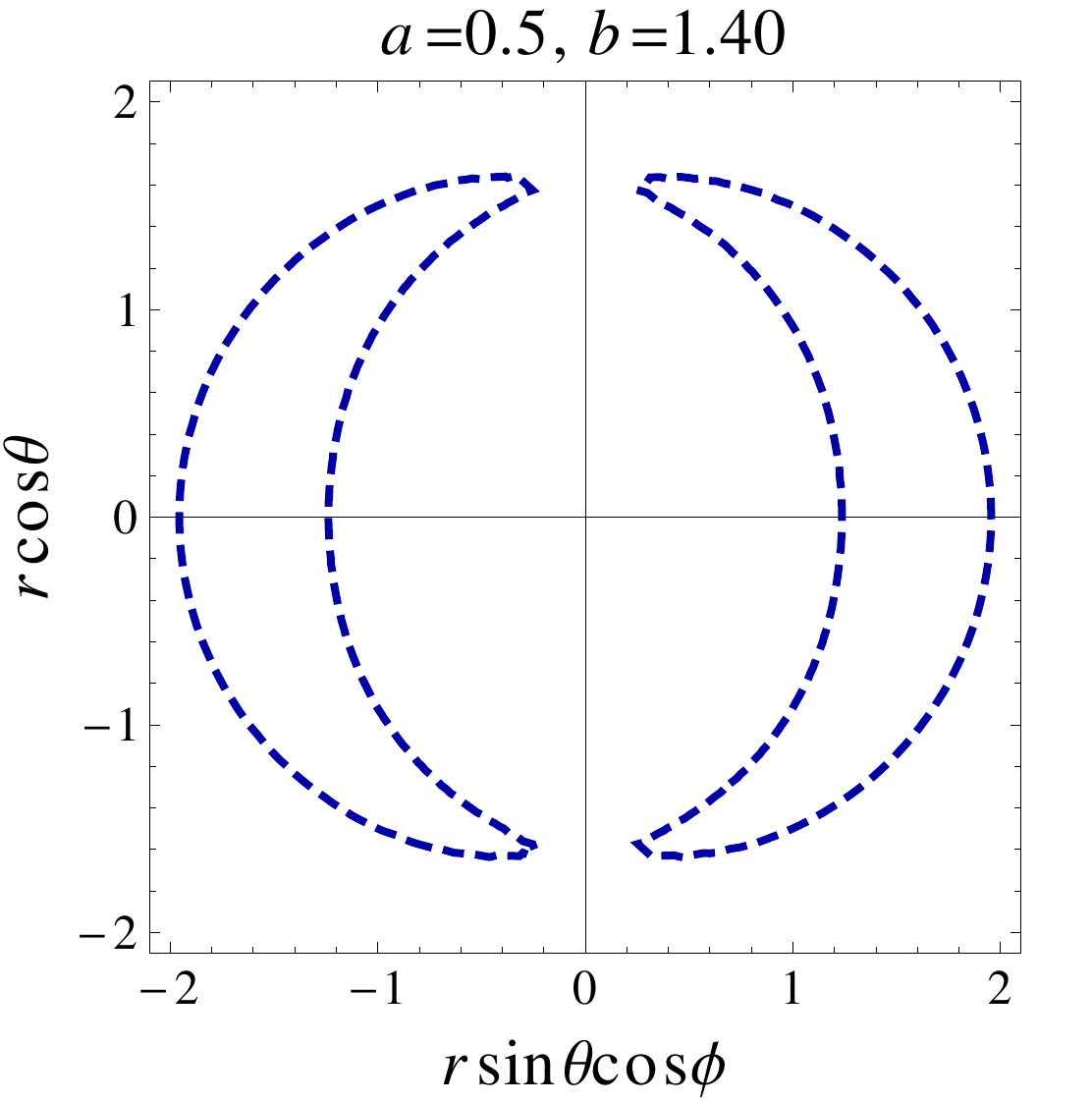}\\
        \includegraphics[width=0.245\linewidth]{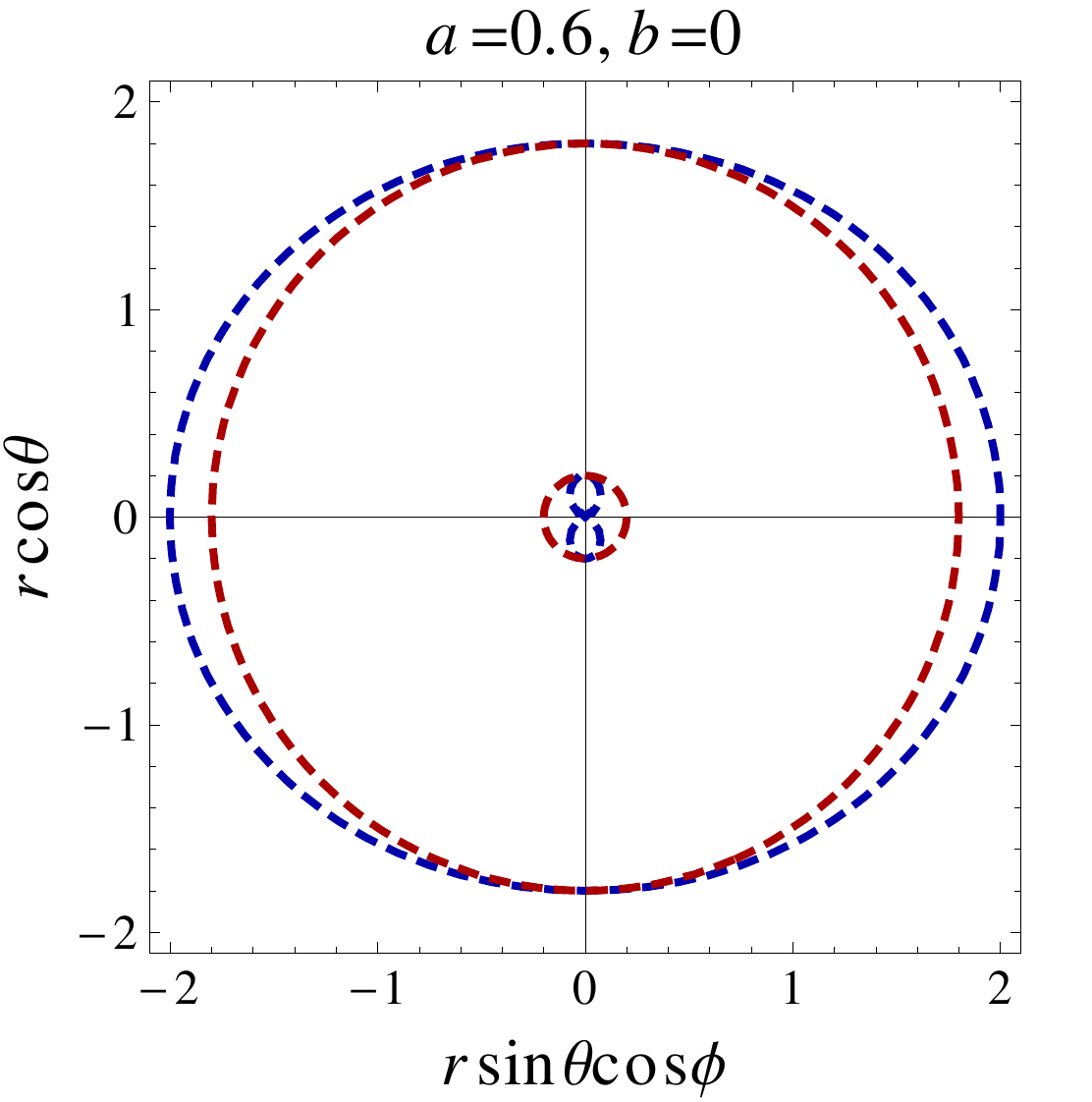}
        \includegraphics[width=0.245\linewidth]{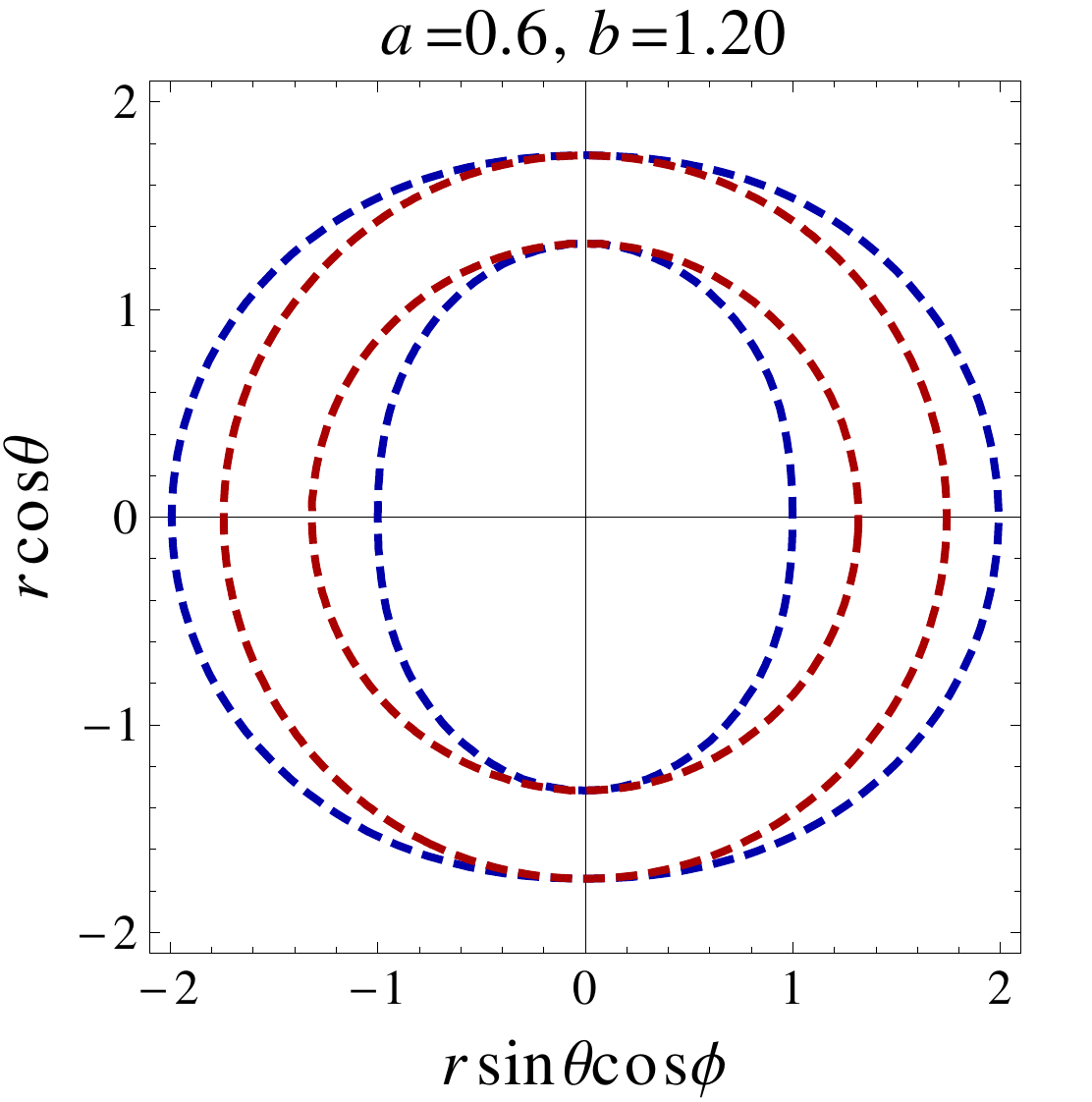}
        \includegraphics[width=0.245\linewidth]{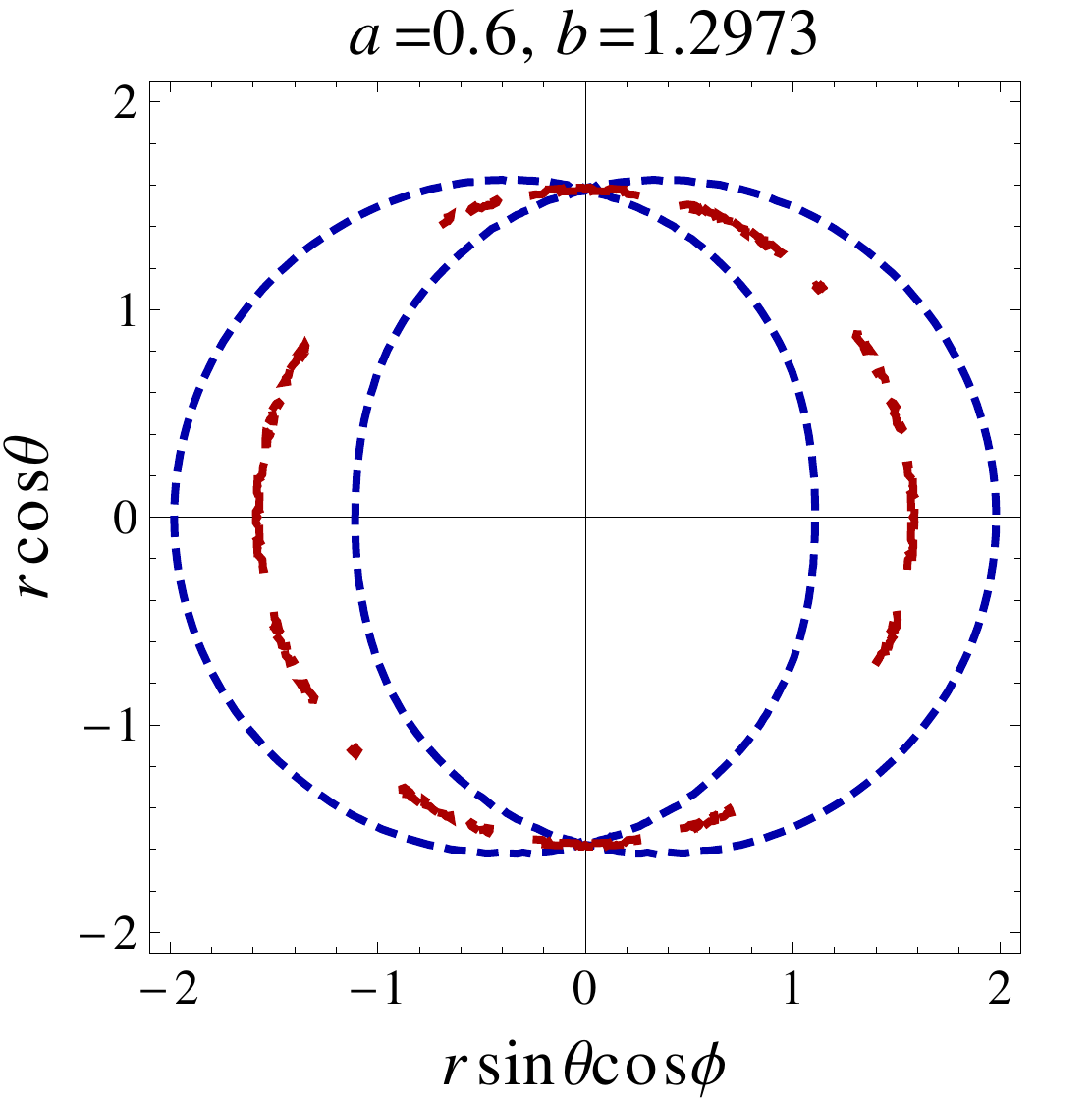}
        \includegraphics[width=0.245\linewidth]{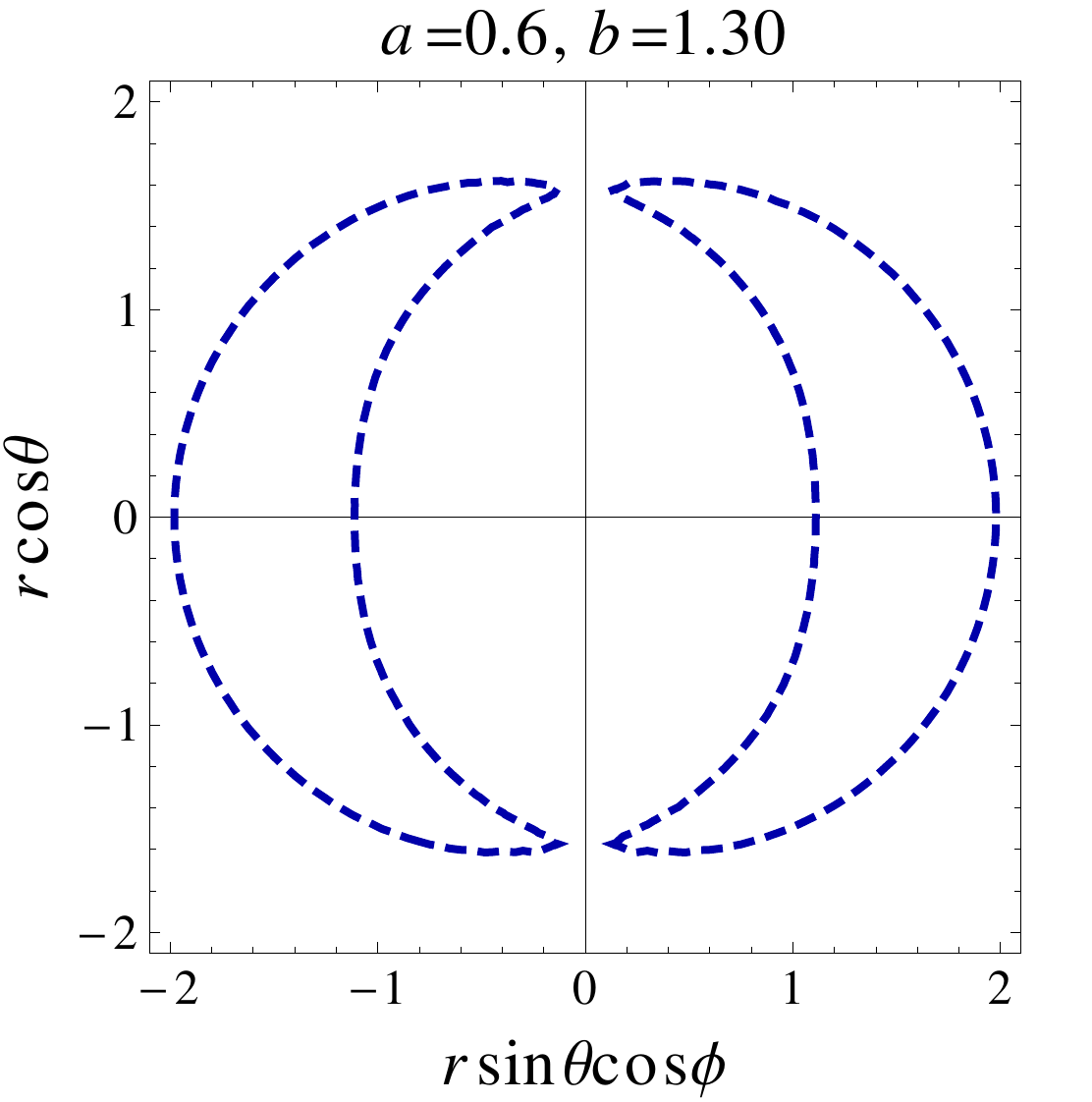}\\
        \includegraphics[width=0.245\linewidth]{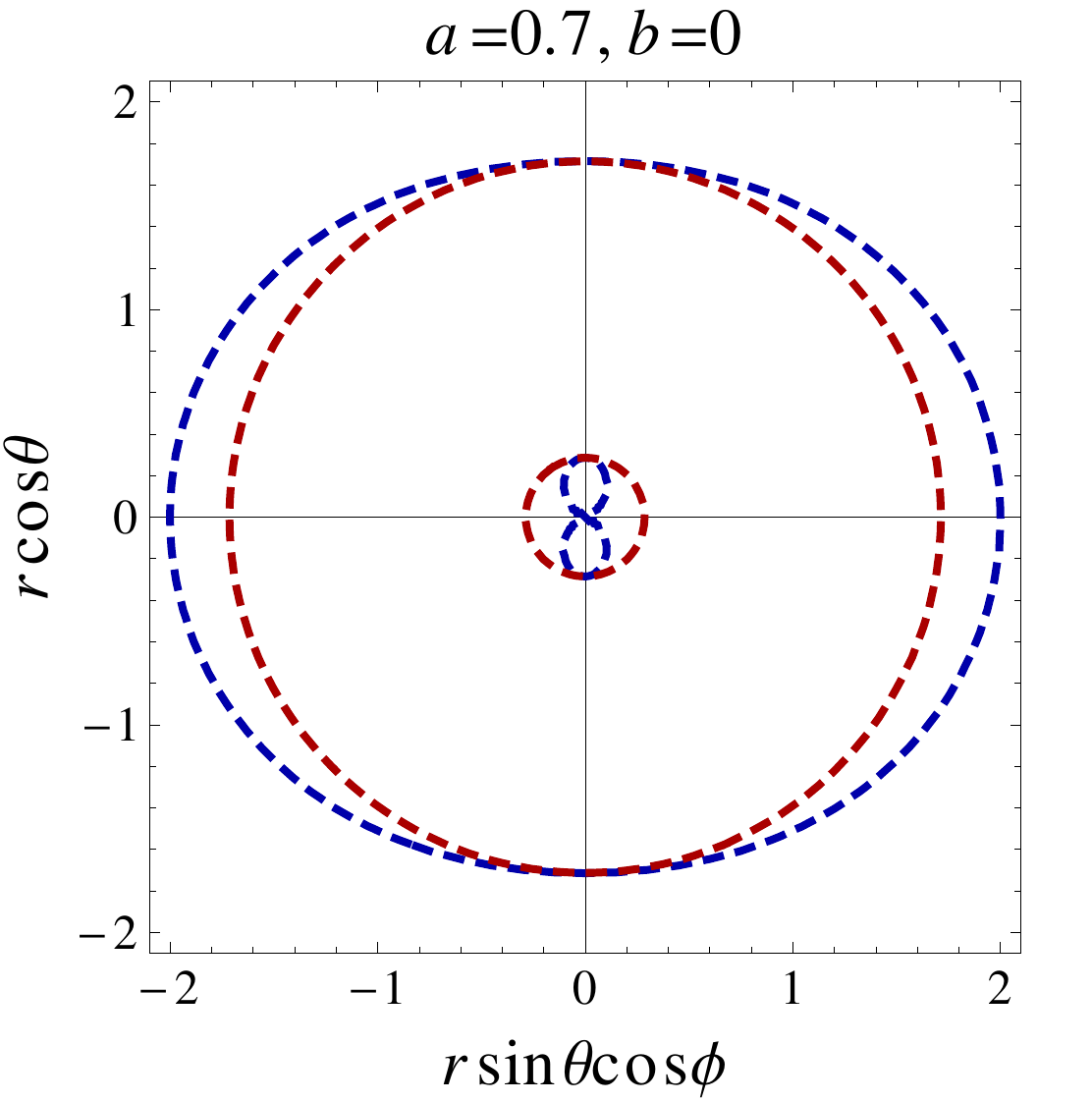}
        \includegraphics[width=0.245\linewidth]{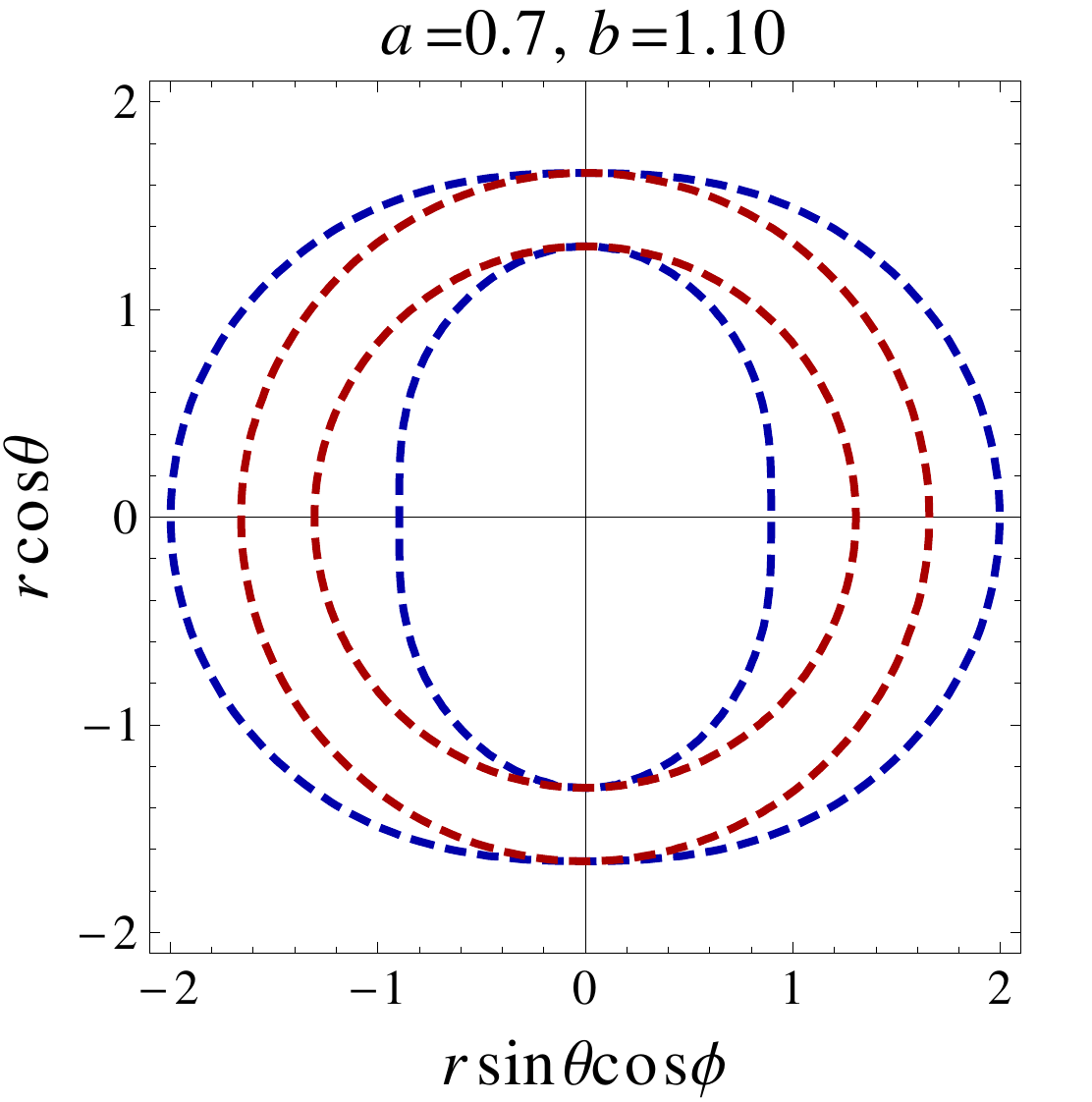}
        \includegraphics[width=0.245\linewidth]{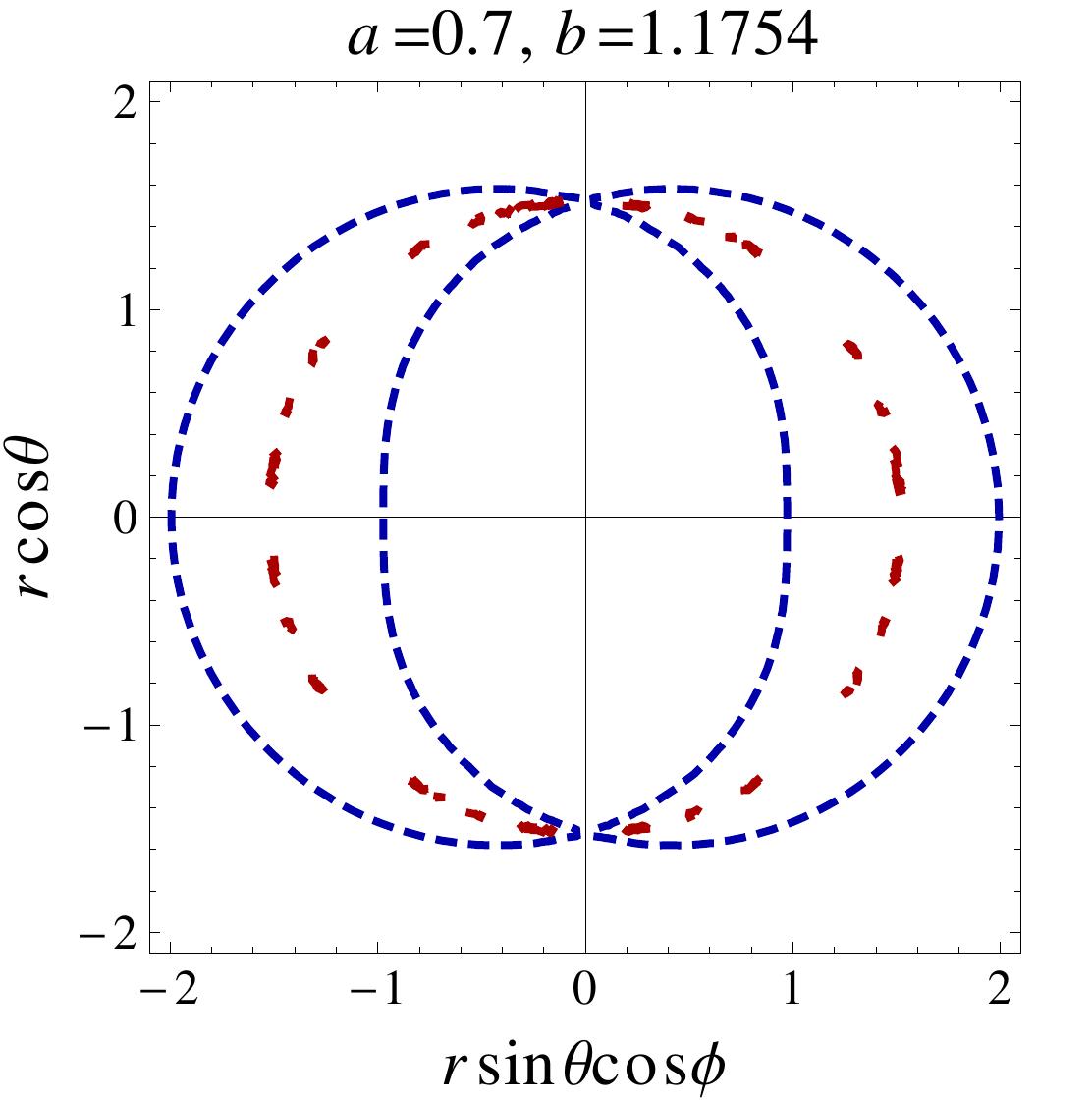}
        \includegraphics[width=0.245\linewidth]{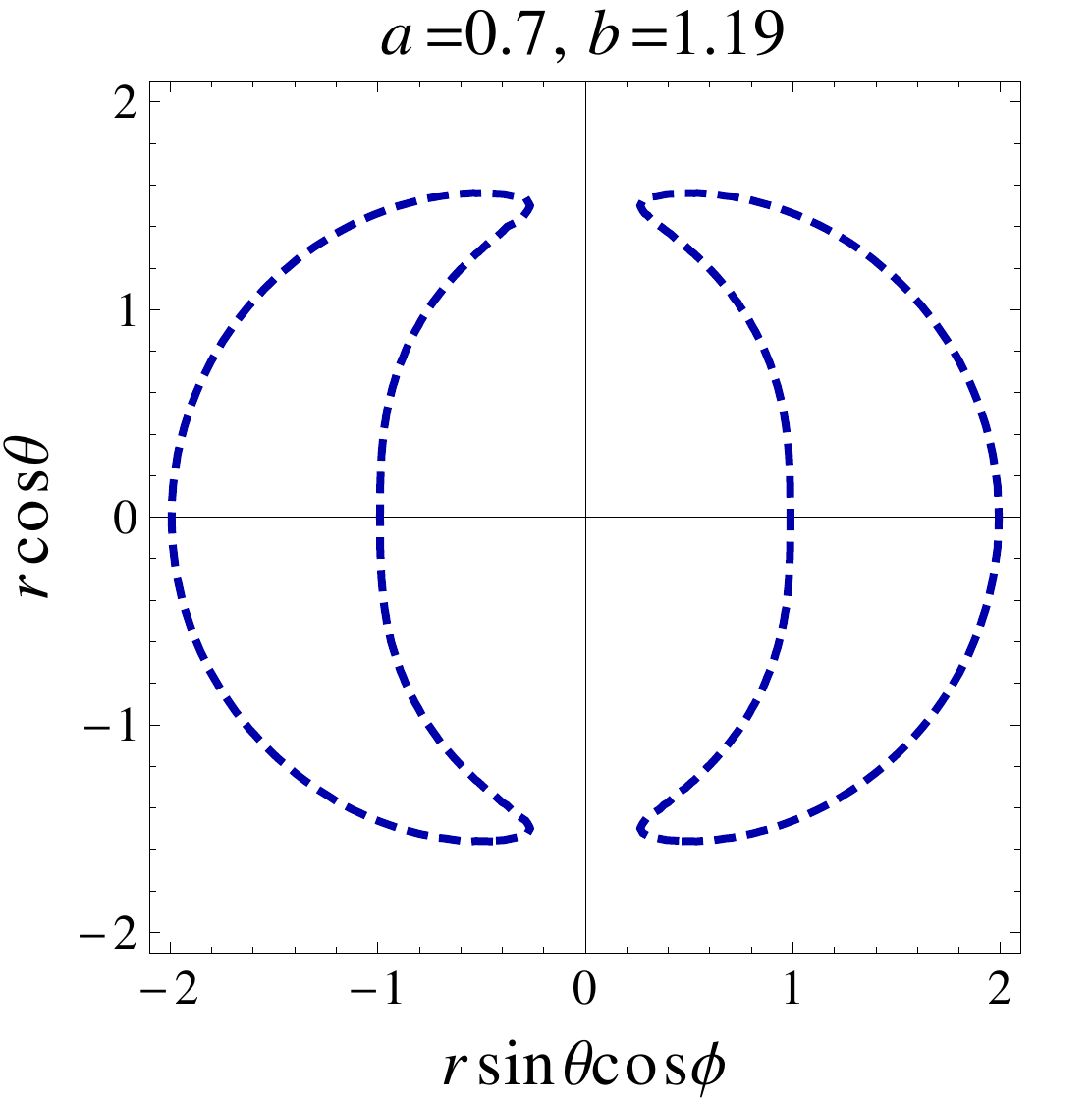}
    \end{tabular}
    \caption{\label{ergo} Plots showing the behavior of the ergoregion 
    in the xz-plane for different values of the parameters $b$ and $a$.}
\end{figure*}
An ergosphere is bounded by the two above discussed surfaces, namely, the 
static limit surface and the event horizon. It lies outside the black hole. 
The ergoregion of the rotating black holes is a region of spacetime where 
the time translation Killing vector $\chi^a$ becomes spacelike, and also 
that every timelike vector acquires rotational ($\phi$) counterpart. Hence, 
a particle can enter into ergoregion and can leave again, but it cannot 
remain stationary without the ergosphere. The ergospheres for the rotating 
regular black hole are shown in the Fig.~\ref{ergo} for different values of 
parameters $b$, which are polar plots of \eqref{sls} and \eqref{horz}. 

Interestingly, we note that ergosphere becomes more prolate with increasing 
values of charge $b$ (see Fig.~\ref{ergo} horizontally) and with increasing 
spin $a$ (see Fig.~\ref{ergo} from top to bottom). It means the faster 
rotating and charged black holes are more prolate thereby increasing the 
area of the ergosphere. Furthermore, we observe from the Fig.~\ref{ergo} 
that we can find a critical parameter $b=b_c$ where two surface shrinks to 
one, and for $b>b_c$, we have no ergoregion. It turns out that energy can 
be extracted from ergoregion via Penrose process \cite{Penrose:1971uk} and 
the magnetic charge shall affect the Penrose process.

\section{Black hole shadow}
\label{shadow}
\begin{figure*}
	\begin{tabular}{c c c c}
		\includegraphics[scale=0.6]{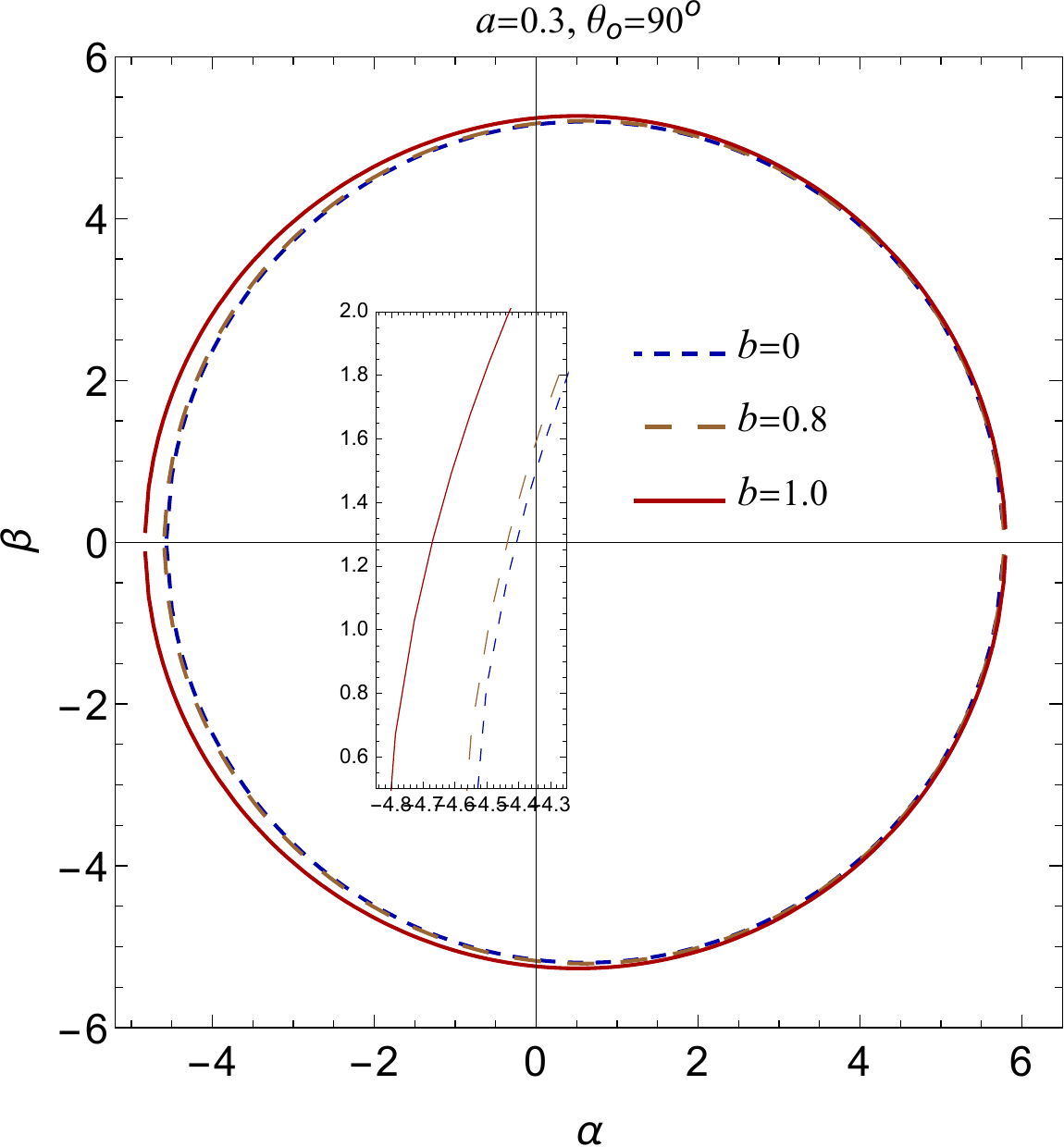} \quad
		\includegraphics[scale=0.6]{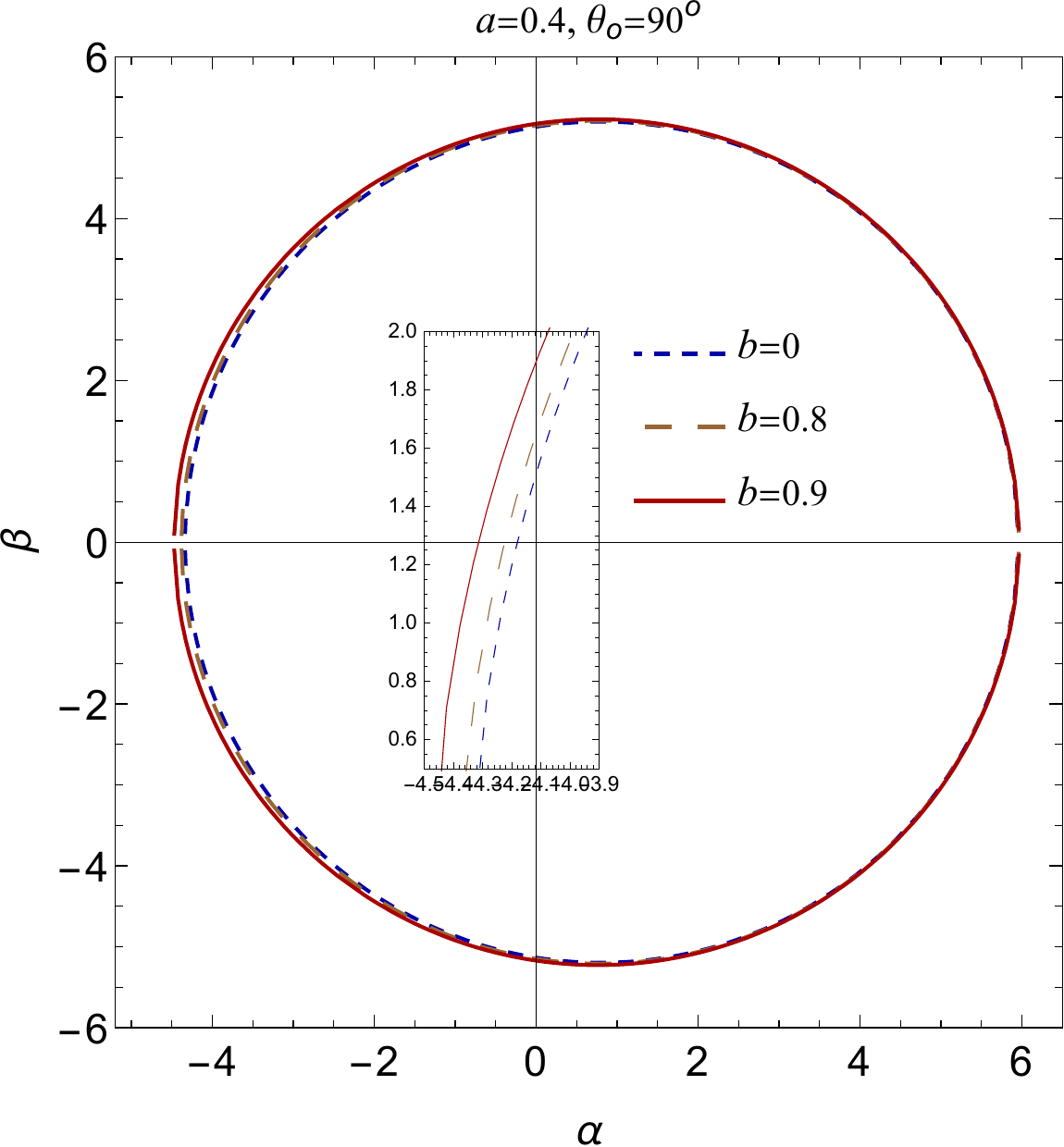}\\
		\includegraphics[scale=0.6]{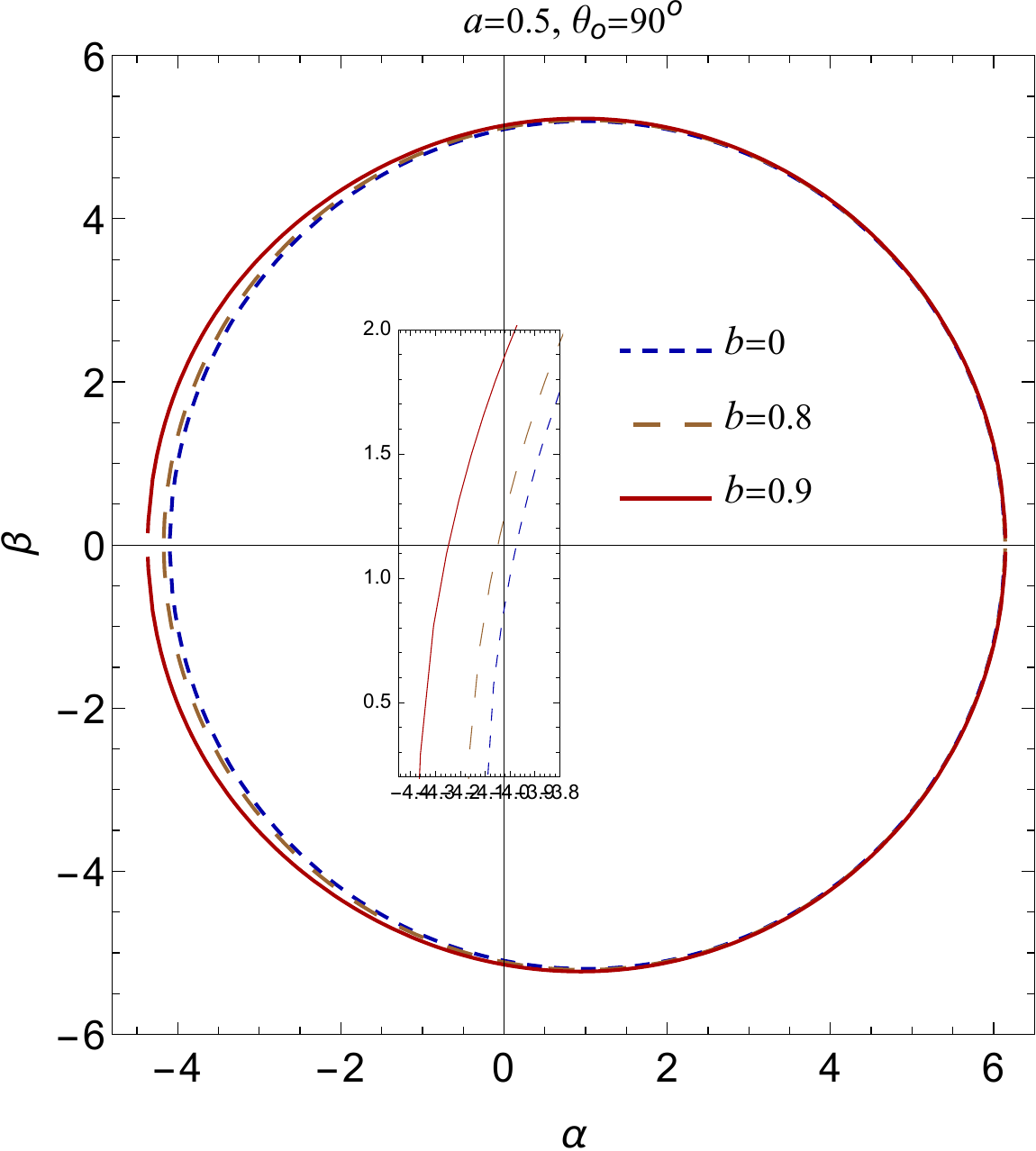} \quad
		\includegraphics[scale=0.6]{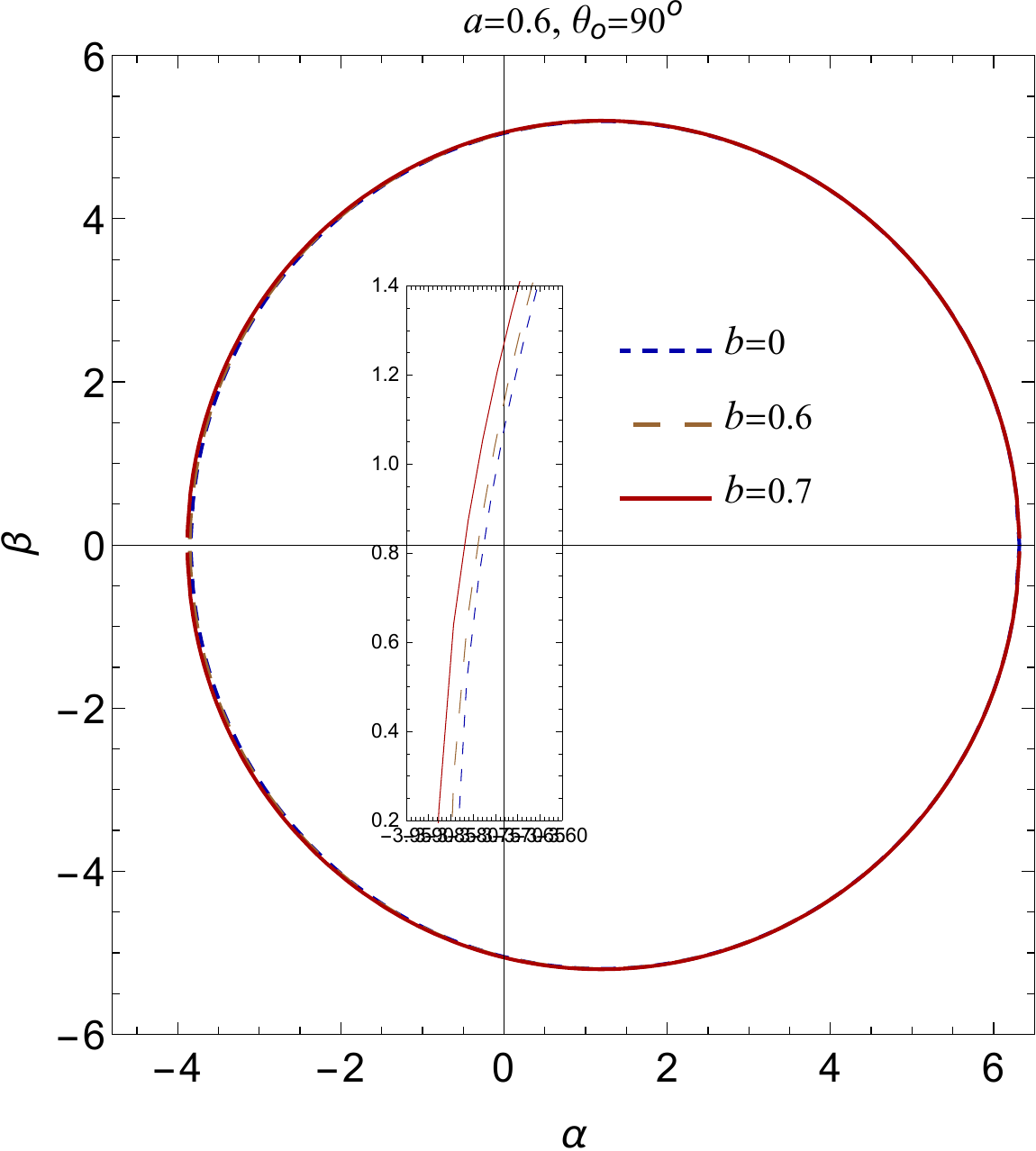}
	\end{tabular}
	\caption{\label{shad-3} Plots showing the shapes of shadow for 
	different values of the parameter $b$.}
\end{figure*}

\begin{figure*}
        \includegraphics[scale=0.64]{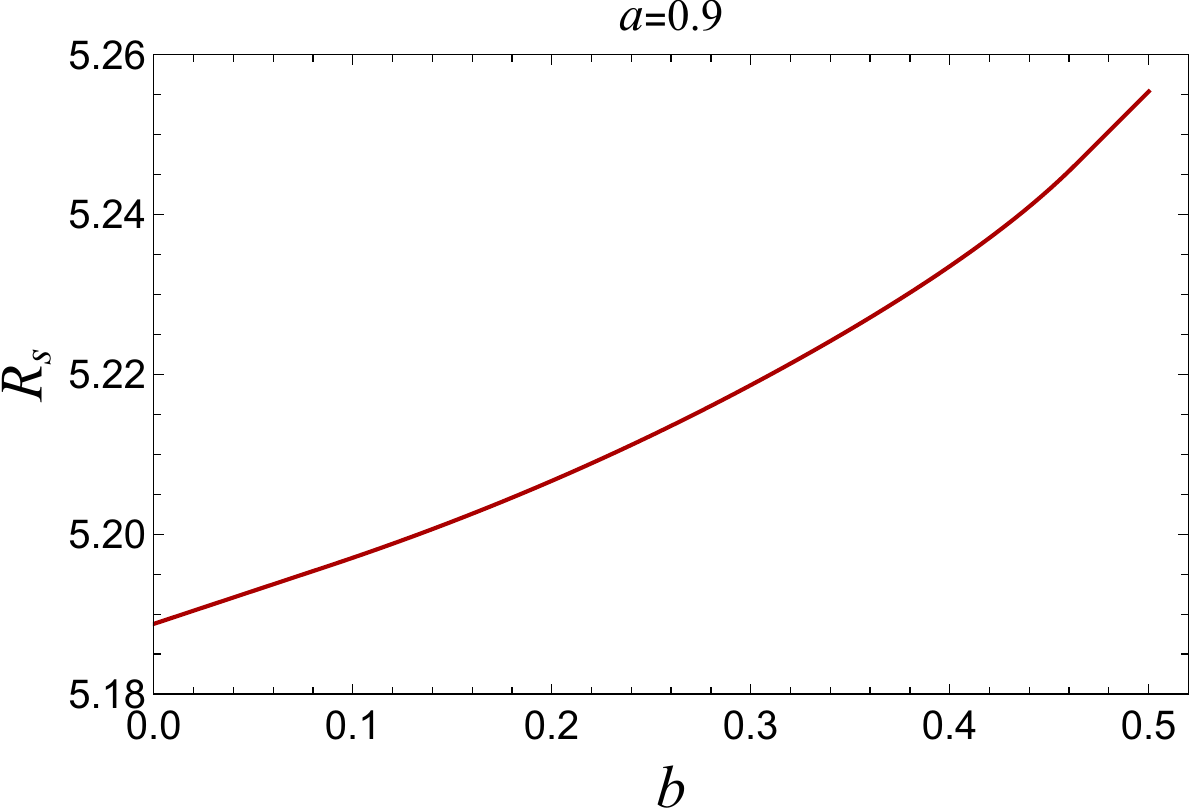} \quad
        \includegraphics[scale=0.64]{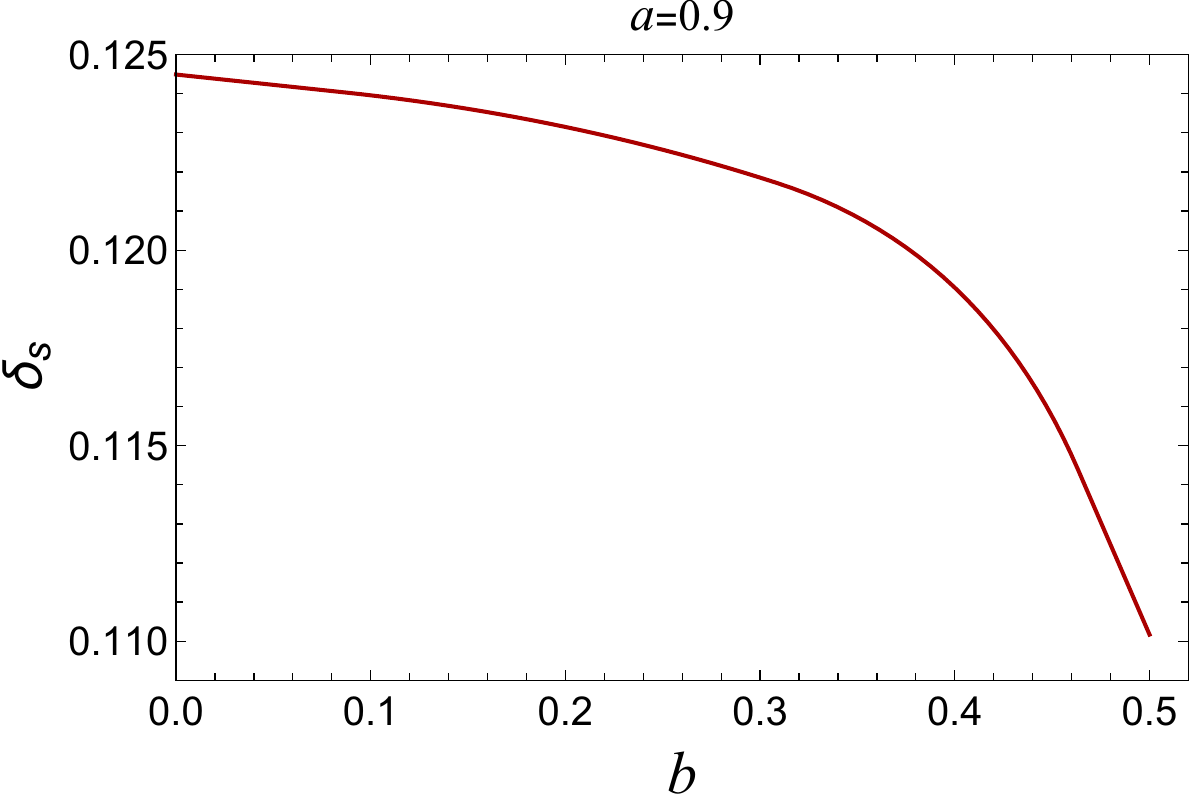}       
   \caption{\label{obs} Plots showing the behavior of observable $R_s$ 
   and $\delta_s$ with parameter $b$. }
\end{figure*}

In this section, we are going to study one of the most interesting 
astrophysical phenomena for the rotating regular black hole known as 
black hole shadow. This can be understood as follows. When a black hole 
is placed in between an observer and the bright background source, the 
photons with a small angular momentum fall into the black hole without 
reaching the observer create a dark spot in the sky, which is called the 
black hole shadow. Thus, to obtain the apparent shape or shadow of the 
black hole and discuss the properties of shadow, we must discuss the 
geodesic of the photon in the background of the rotating regular black 
hole. These geodesics can be determined by the Hamilton-Jacobi formulation 
\cite{Carter:1968rr} given by
\begin{equation}
    \frac{\partial S}{\partial \sigma} = -\frac{1}{2} 
    g^{\mu\nu} \frac{\partial S}{\partial x^{\mu}} 
    \frac{\partial S}{\partial x^{\nu}}, \label{hje}
\end{equation}
which has a solution in the separable form as follows
\begin{equation}
    S = \frac{1}{2} m_0^2 \sigma -Et +L_{z} \phi +S_{r}(r) 
    +S_{\theta}(\theta), \label{hja}
\end{equation}
where $ m_0 $ corresponds to the rest mass of the particle, and $S_{r}(r)$ 
and $ S_{\theta}(\theta)$ are respectively functions of $r$ and $\theta$ 
only. Note that in case of photon the rest mass, $m_0=0$. By following the 
standard procedure of the Hamilton-Jacobi method, we obtain the following 
forms of the geodesic equations, 
\begin{equation}
    \Sigma \frac{d t}{d \sigma} =  -a \left(aE \sin^2 \theta 
    - L_{z}\right) + \frac{(r^2 + a^2) \left[(r^2+a^2)E-aL_{z}\right]}
    {r^2 + a^2- 2 M (1-e^{-r^3/b^3})r}, \label{u^t}
\end{equation}
\begin{equation}
    \Sigma \frac{d \phi}{d \sigma} = -\left(aE - L_{z}\csc^2 \theta \right) 
    + \frac{a \left[(r^2+a^2)E-aL_{z}\right]}
    {r^2 + a^2- 2 M (1-e^{-r^3/b^3})r}, \label{u^phi}
\end{equation}
\begin{equation}
    \Sigma \frac{d r}{d \sigma} = \pm  \sqrt{\mathcal{R}}, \label{u^r}
\end{equation}
\begin{equation}
    \Sigma \frac{d \theta}{d \sigma} = \pm  \sqrt{\Theta}, \label{u^theta}
\end{equation}
where $\mathcal{R}$ and $\Theta$ can be expressed as follows
\begin{eqnarray}
    \mathcal{R} &=& \left[(r^2+a^2)E-aL_{z}\right]^2 
    -\left[r^2 + a^2- 2 M (1-e^{-r^3/b^3})r\right] 
    \left[\mathcal{K}+ (L_{z}-a E)^2 \right], 
    \nonumber \\
    \Theta &=& \mathcal{K} +\cos^2 \theta 
    \left(a^2E^2-L_{z}^2\csc^2 \theta \right). \label{R-Th}
\end{eqnarray}
These geodesic equations represent first-order differential equations with 
respect to the affine parameter $\sigma$ and contain three constants of the 
motion: the energy $E$, the angular momentum $L_z$, and the Carter constant 
$\mathcal{K}$ \cite{Carter:1968rr}. If we substitute $b=0$, in the geodesic 
equations, they reduce to the Kerr spacetime case \cite{Hioki:2009na}. 
Now in order to get the shadow of the rotating regular black hole, we 
introduce the two independent dimensionless quantities or impact 
parameters such that $\xi=L_{z}/E$ and $\eta=\mathcal{K}/E^2$. Equation 
$\mathcal{R}=0$ explores the radial turning points of the photons and the 
spherical photon orbits around the black hole can be determined by the 
following relations \cite{Bardeen:1972fi}
\begin{equation}
    \mathcal{R} = 0 \quad \text{and} \quad \frac{d \mathcal{R}}{dr}=0.
    \label{condition}
\end{equation}
On solving (\ref{condition}) for impact parameters $\xi$ and $\eta$, we 
immediately get their forms as follows
\begin{eqnarray}
    \xi &=& \frac{M \left[a^2 (b^3-3 r^3)-3 r^2 (b^3+r^3)\right]
    -b^3 e^{r^3/b^3} \left[a^2 (r+M)+r^2 (r-3 M)\right]}
    {a M (b^3-3r^3)+a b^3(r-M)e^{r^3/b^3}},
    \nonumber \\
    \eta &=& \frac{4 a^2 b^3 r^3 M e^{r^3/b^3} 
    \left[b^3(e^{r^3/b^3}-1)-3 r^3\right]
    -r^4 \left[b^3 e^{r^3/b^3} (r-3 M)+3 M(b^3+r^3)\right]^2}
    {\left[a M (b^3-3r^3)+a b^3(r-M)e^{r^3/b^3}\right]^2}. \label{xi-eta}
\end{eqnarray}
It turns out that the impact parameters have a dependency on charge $b$. If 
charge $b$ is set to be zero, then Eq.~\eqref{xi-eta} reduces to the Kerr 
black hole case \cite{Hioki:2009na}. This calculation is helpful to 
determine the celestial coordinates for a distant observer in order to get 
the shadow of the rotating regular black hole. The celestial coordinates 
for a distant observer in ($\alpha$, $\beta$)-plane are given 
\cite{Bardeen:1973gb} by
\begin{eqnarray}
    \alpha &=& \lim_{r_{0} \rightarrow \infty} 
    \left(-r_{0}^2 \sin \theta_{0} \frac{d \phi}{dr} \right), 
    \nonumber \\
    \beta &=& \lim_{r_{0} \rightarrow \infty} 
    \left(r_{0}^2 \frac{d \theta}{dr} \right), \label{alp-bet}
\end{eqnarray}
where $r_{0}$ is the distance from the black hole to the observer and 
$\theta_{0}$ represents the inclination angle between the direction to 
observer and the rotation axis of the black hole. By using \eqref{u^t}, 
\eqref{u^phi}, \eqref{u^r}, \eqref{u^theta}, and \eqref{R-Th}, the 
celestial coordinates \eqref{alp-bet} transform into 
\begin{eqnarray}
    \alpha &=& -\xi \csc \theta_0, \nonumber\\
    \beta &=& \pm \sqrt{\eta +a^2 \cos^2 \theta 
    -\xi^2 \cot^2 \theta_0}. \label{alphabeta}
\end{eqnarray}
\begin{figure*}
    \begin{tabular}{c c c c}
        \includegraphics[scale=0.35]{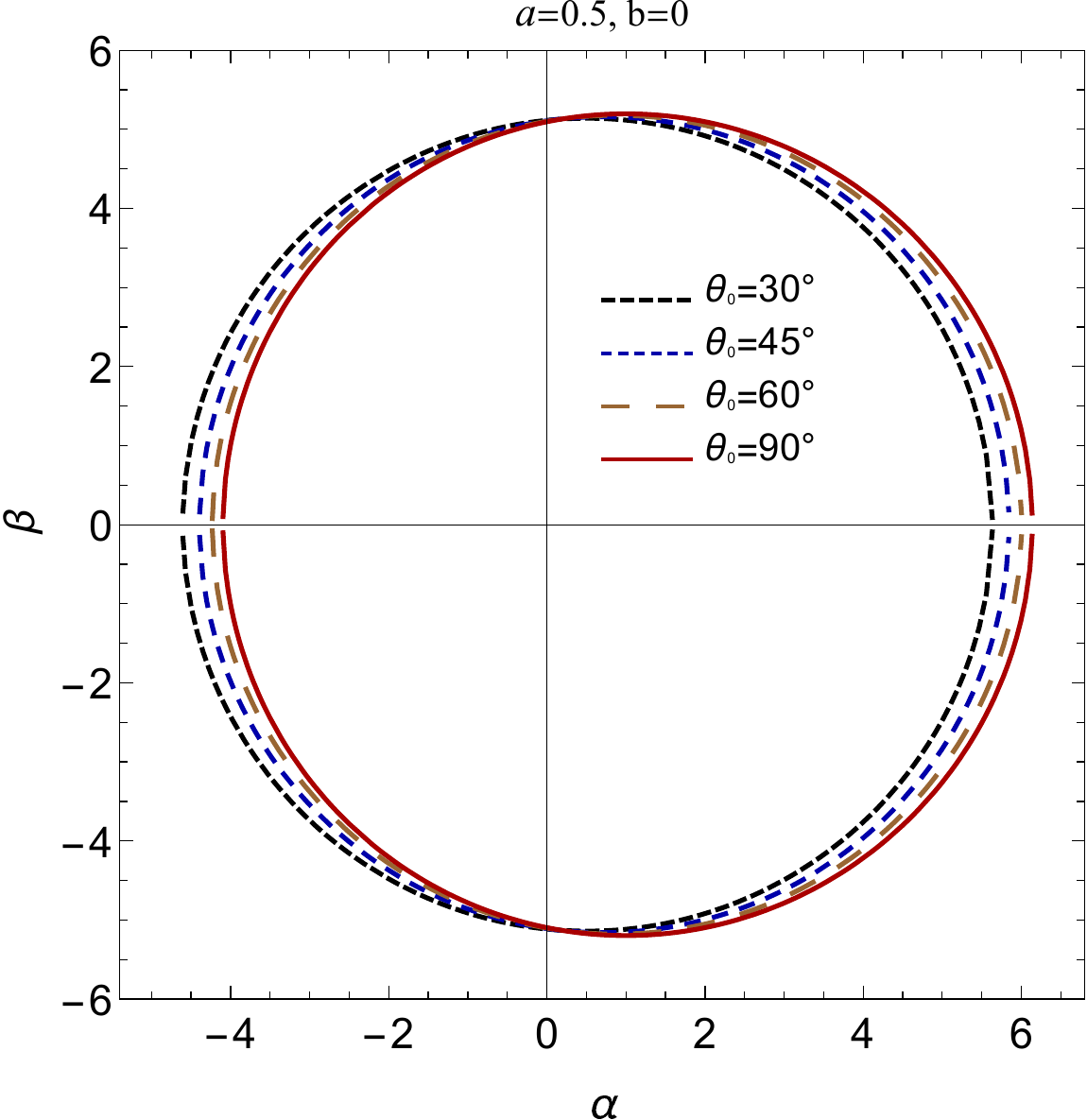}
        \includegraphics[scale=0.35]{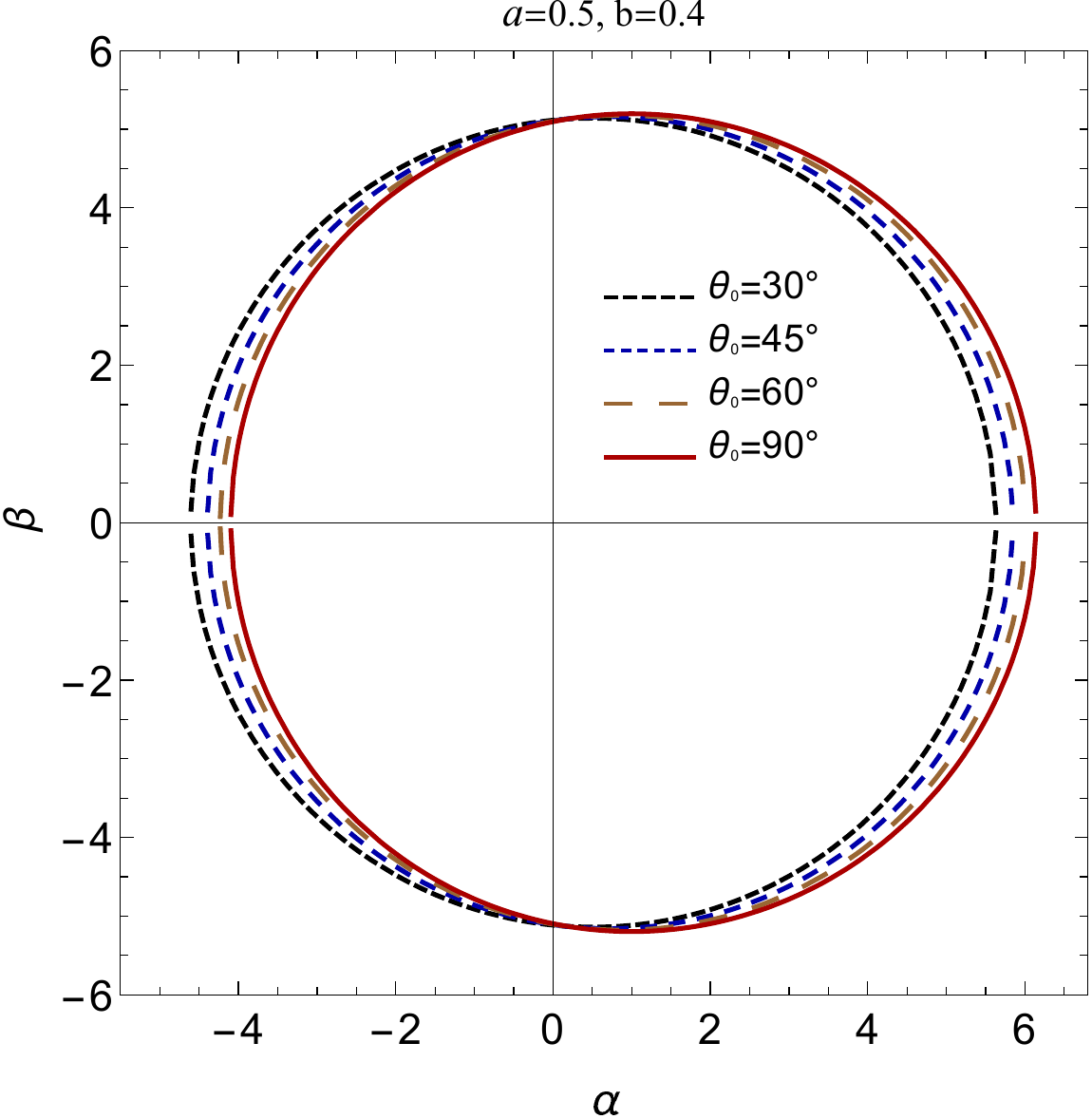}
        \includegraphics[scale=0.35]{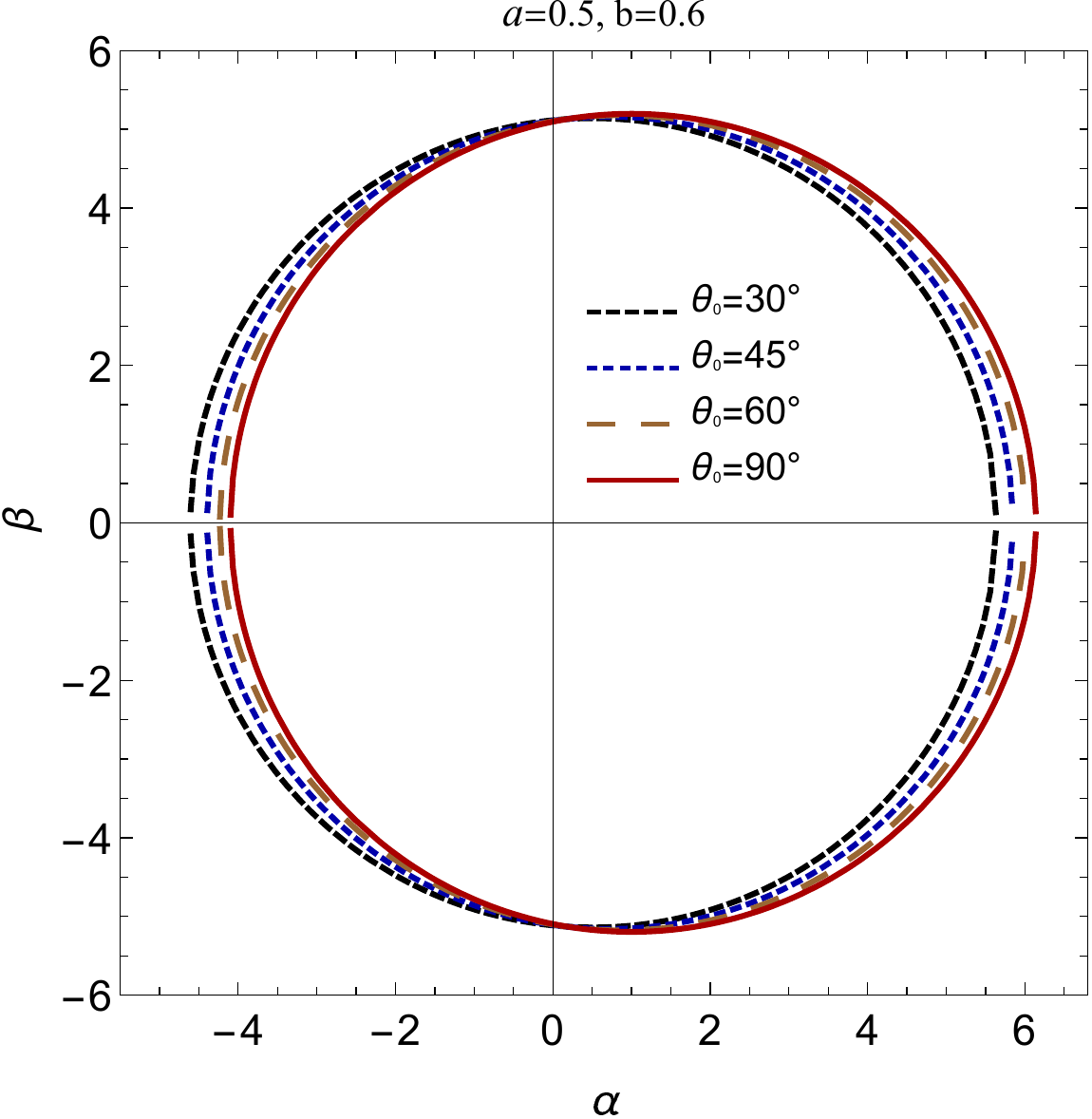}
        \includegraphics[scale=0.35]{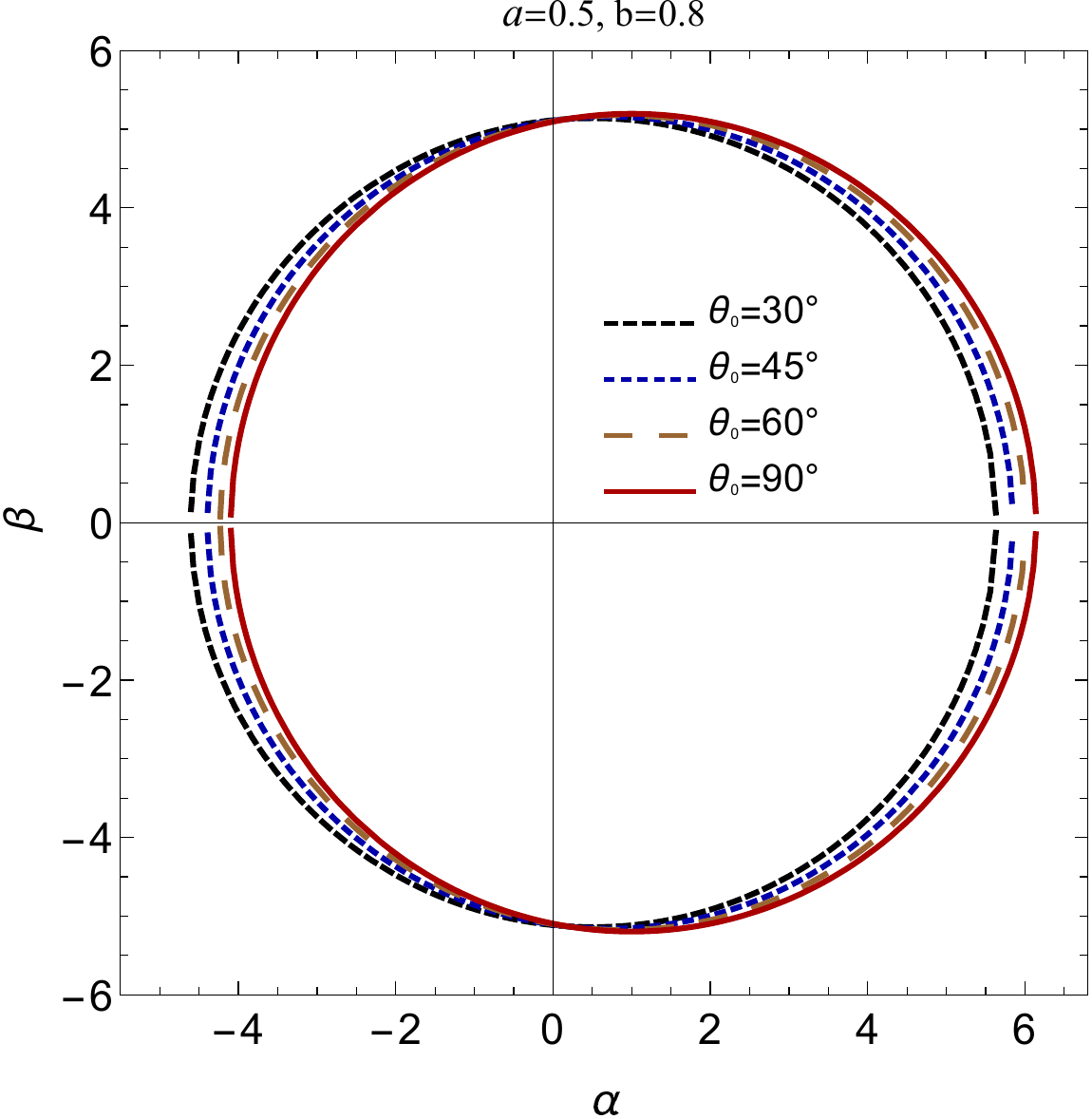}
    \end{tabular}
    \caption{\label{shad-1} Plots showing the shapes of shadow with 
    variation of inclination angle $\theta_0$ for fixed values of $b$ 
    and $a$.}
\end{figure*}
\begin{figure*}
    \begin{tabular}{c c c c}
        \includegraphics[scale=0.35]{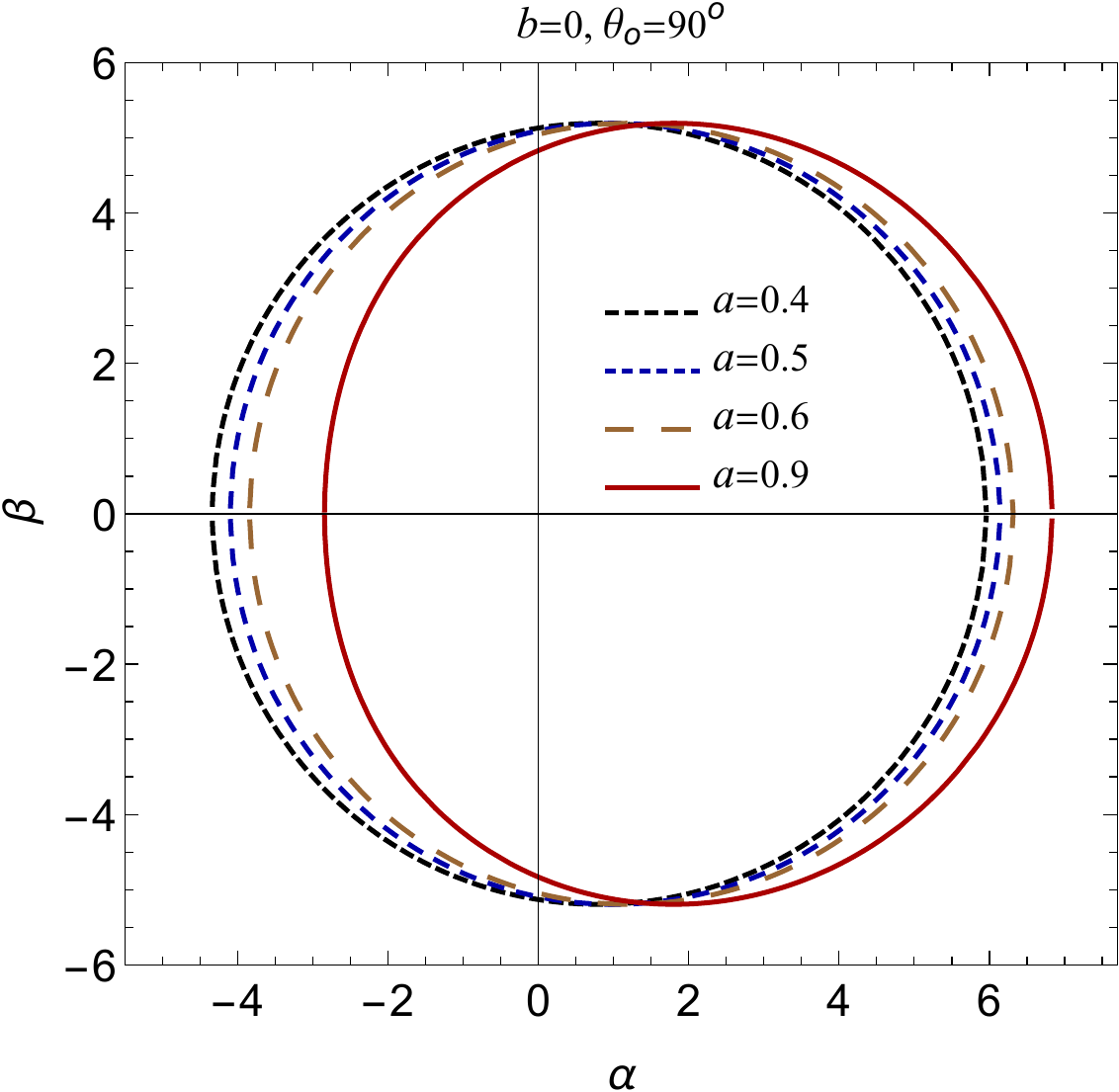}
        \includegraphics[scale=0.35]{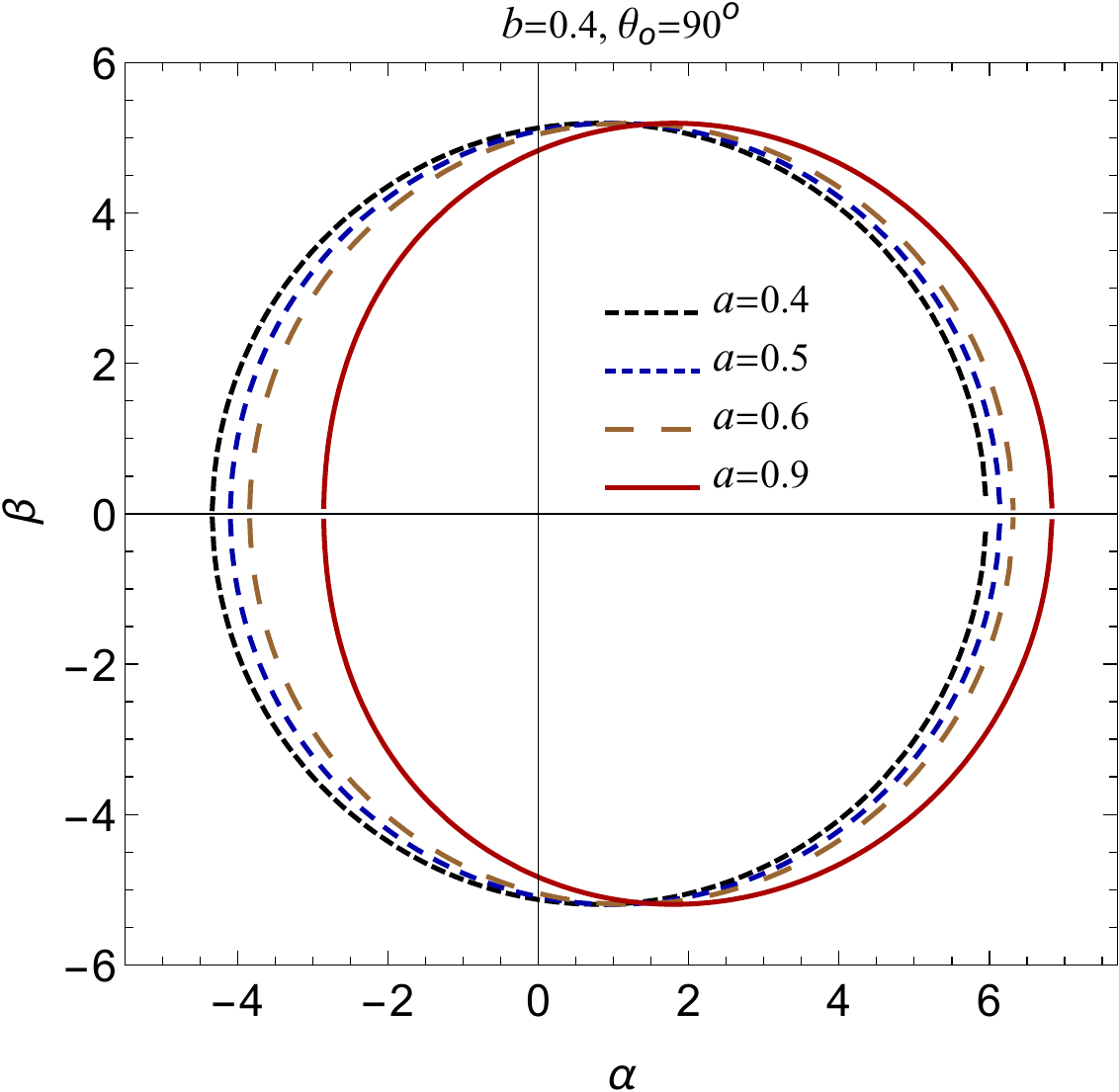}
        \includegraphics[scale=0.35]{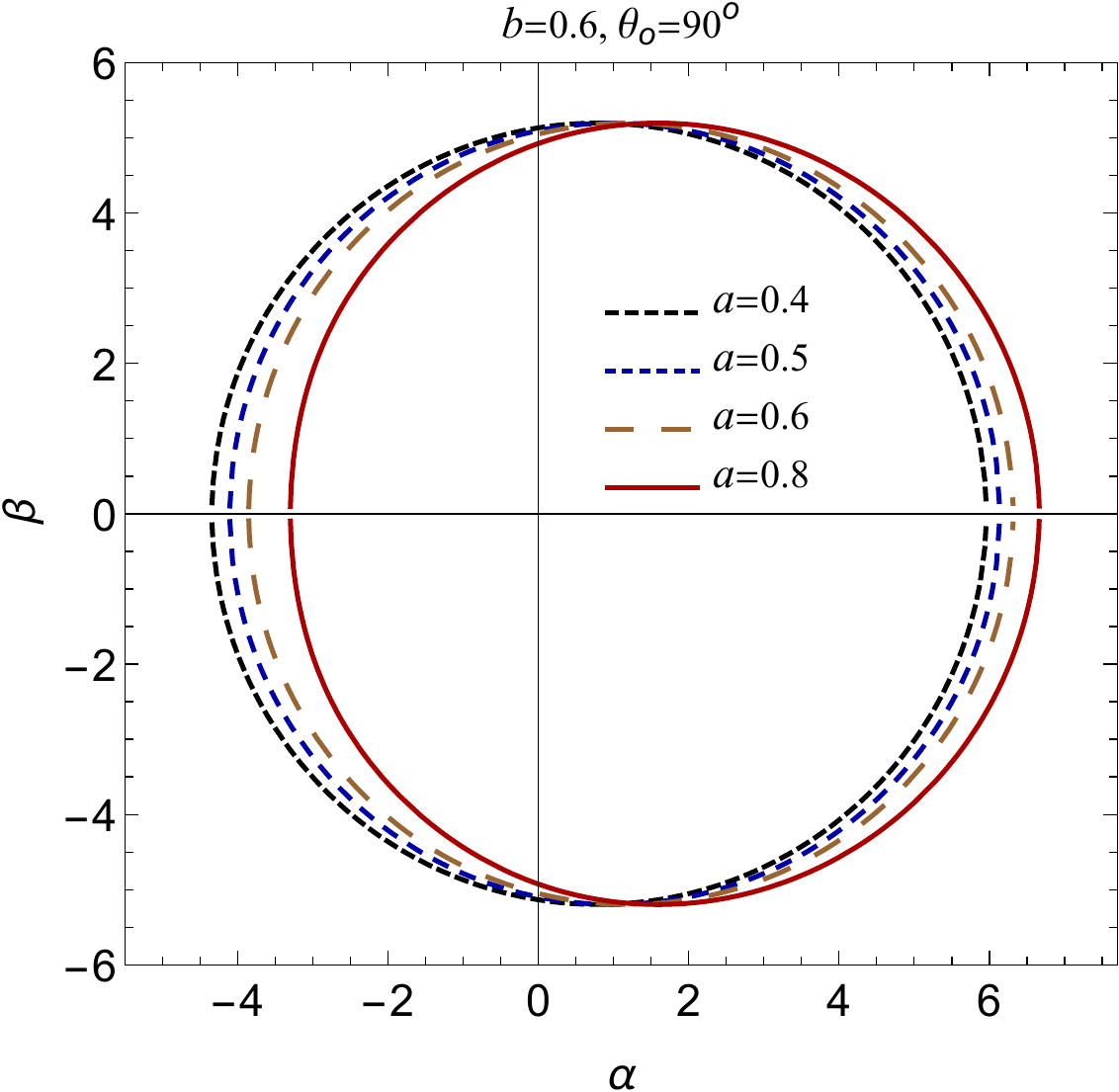}
        \includegraphics[scale=0.35]{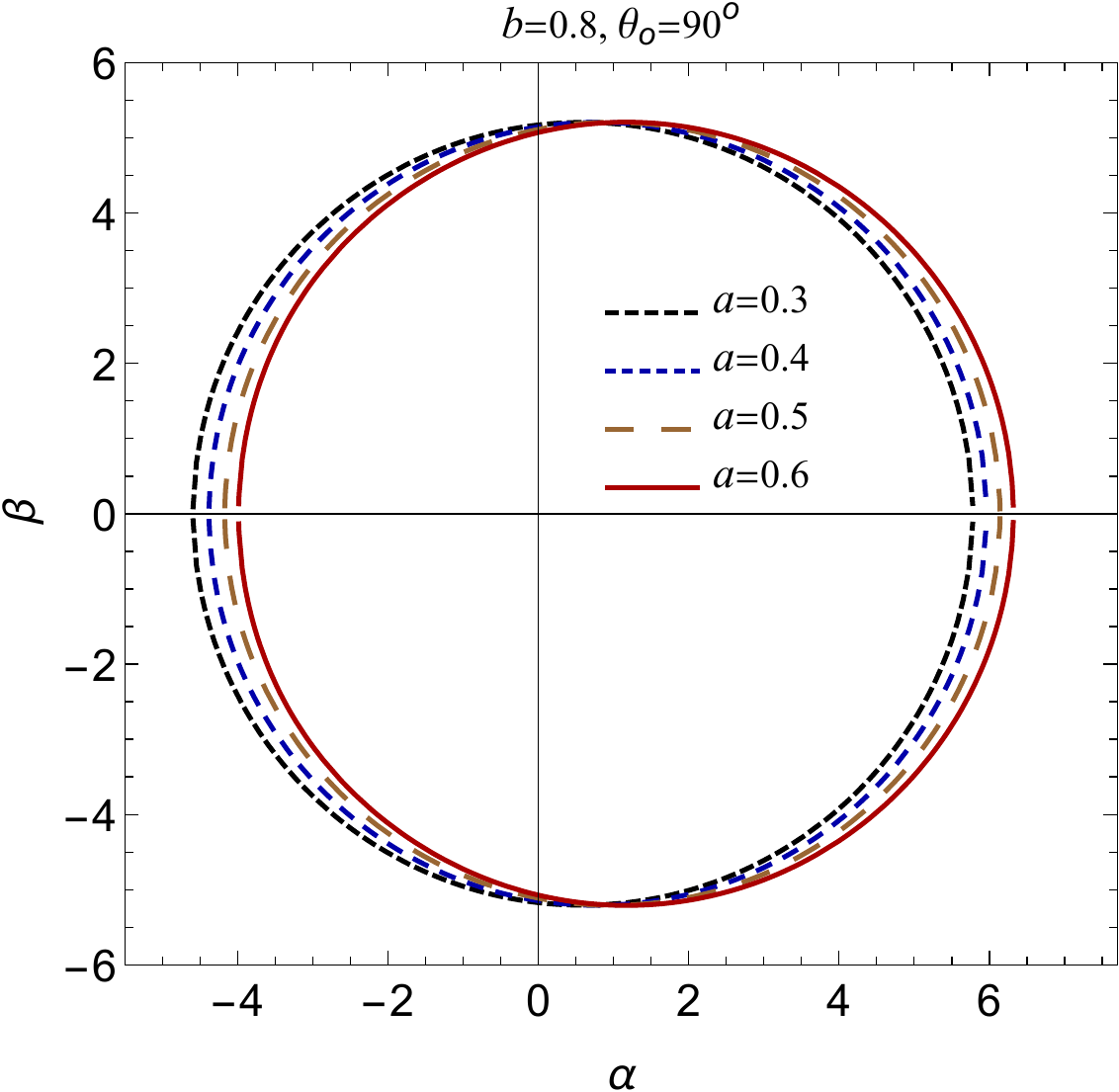}
    \end{tabular}
  \caption{\label{shad-2} Plots showing the shapes of shadow for different 
   values of the parameter $b$ and $a$.}
\end{figure*}
Equation~\eqref{alphabeta} represents a direct relationship between the 
celestial coordinates ($\alpha,\beta$) and the impact parameters 
($\xi,\eta$). Every single photon that approaches the observer determines a 
point on the ($\alpha$, $\beta$)-plane of the image, which can be visible 
with the help of telescope. The size and shape of the shadow depends on the 
parameters of black hole as well as on the inclination angle $\theta_0$. 
The pictorial representation of the rotating regular black hole's shadow 
can be seen from Figs.~\ref{shad-1}, \ref{shad-2} and \ref{shad-3}. We 
obtain several images of the rotating regular black hole's shadow for the 
particular choice of the parameters $a$, $b$, and $\theta_0$. A distortion 
in the shape of the black hole's shadow arises that increases with higher 
values of $a$ as well as $\theta _{0}$. Fig.~\ref{shad-3} depicts 
the effect of parameter $b$ on the shadow of the rotating regular black 
hole, which includes the case of the Kerr black hole ($b=0$) for a 
comparison.
Moreover, the actual size and distortion in shape of the rotating regular 
black hole's shadow can be determined by the observables, namely, $R_s$ and 
$\delta_s$ \cite{Hioki:2009na}. Here $R_s$ and $\delta_s$ correspond to 
the actual size or radius of the shadow and distortion in the shape of the 
shadow, respectively. In order to compute the radius of the shadow, we 
consider a circle passing through the three different points, namely, top, 
bottom, and rightmost corresponding to $(\alpha_t, \beta_t)$, 
$(\alpha_b, \beta_b)$, and $(\alpha_r, 0)$, respectively \cite{Hioki:2009na}. 
We are considering the definition of the observables given by Hioki and 
Maeda \cite{Hioki:2009na}:
\begin{eqnarray}
    R_s &=& \frac{(\alpha_t - \alpha_r)^2 
    +\beta_t^2}{2 |\alpha_t - \alpha_r|}, \\
    \delta_s &=& \frac{(\tilde{\alpha_p} - \alpha_p)}{R_s},
\end{eqnarray}
\begin{figure}
   \includegraphics[scale=0.35]{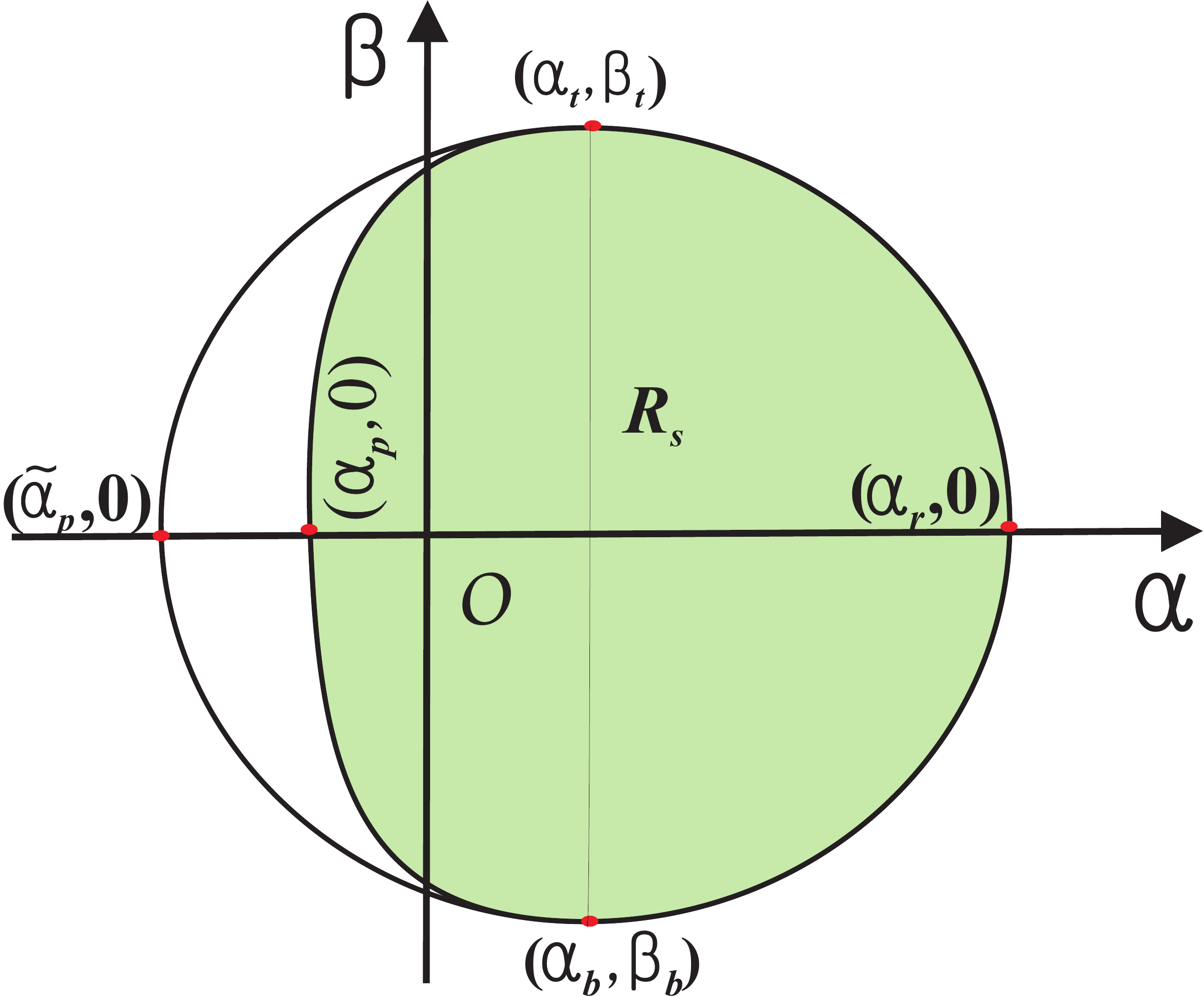}
   \caption{\label{char} Illustration of the observable of the black hole 
   shadow \cite{Amir:2016cen}.}
\end{figure}
where $(\tilde{\alpha_p},0)$ and $(\alpha_p,0)$ are the points where the 
reference circle and the silhouette of the shadow cut the horizontal axis 
at the opposite side of $(\alpha_r,0)$ (cf., Fig.~\ref{char}). This special 
characterization helps us to find the effect of charge $b$ on radius and 
distortion in the shape of the rotating regular black hole's shadow. 
The typical behavior of these observables with charge $b$ can be seen in 
Fig.~\ref{obs}. We examine that an increase in the magnitude of charge $b$ 
increases the radius of the black hole's shadow ($R_s$). On the other hand, 
the distortion in the shape of the black hole's shadow ($\delta_s$) 
decreases with an increase in magnitude of charge $b$ (cf.~Fig.~\ref{obs}).     
When a comparison of our results with the Kerr-Newman and the braneworld 
Kerr spacetimes is taking into account, we discover that the effect of 
nonlinear charge (our case) is opposite that of electric charge $q$ 
(Kerr-Newman) and tidal charge (braneworld Kerr). In our case, the 
radius of the black hole's shadow increases with charge instead of 
being decreased at the same time the distortion decreases with charge 
instead of being increase.

\section{Concluding remarks}
\label{conclusion}
We have shown that the regular black hole metric \eqref{dymn} is an exact 
solution of the Einstein's field equations coupled to the nonlinear 
electrodynamics associated with the Lagrangian \eqref{lf} corresponds to 
a magnetic charge $b$. We have also constructed a rotating counterpart of 
the regular black hole \eqref{dymn} containing an additional parameter $b$, 
which encompasses the Kerr black hole in the particular case when $b=0$, 
and therefore belongs to a family of non-Kerr black holes. The source of 
the rotating regular black hole has also been computed in order to validate 
the solution. We have further discussed the energy conditions of the 
rotating regular black hole. The rotating regular spacetime describes the 
extremal black holes with degenerate horizons for a critical amount of 
charge $b=b_c$, the non-extremal black holes with two distinct horizons 
for $b < b_c$ corresponding to the Cauchy and the event horizon. We have 
made a comprehensive analysis of horizon structure and ergosphere of the 
rotating regular black holes and explicitly brings out the effect of charge 
$b$. The ergosphere is sensitive to the charge $b$, which enlarges the 
ergoregion and becomes more prolate with an increasing magnitude of $b$.   

The immediate target of the Event Horizon Telescope is observations of the 
image by supermassive black hole Sgr A$^*$ and M87$^*$ (shadow) which is a 
major attempt of understanding the nature of these black holes and explain 
the strong field gravity. Motivated from this, we have studied the shapes 
of the shadow from rotating regular black hole to discuss the effect of 
charge $b$ on the Kerr black hole shadow. We have done a detailed analysis 
of particle motion in the rotating regular black hole spacetimes. In order 
to achieve the goal, we have derived necessary analytical expressions to 
obtain the black hole shadow. The shadow is completely characterized by 
the two observables, namely, radius $R_s$ and distortion $\delta_s$. 
It turns out that the radius $R_s$ increases with an increase in the 
magnitude of charge $b$, it results in a larger shadow of the rotating 
regular black hole than the Kerr black hole shadow. The rotating regular 
black hole is less distorted when compared with the Kerr black hole shadow 
as distortion $\delta_s$ decreases with increasing charge $b$. We have also 
compared our results with the Kerr-Newman and braneworld Kerr spacetimes; 
as a consequence, we found that in our case, the charge increases the 
radius of shadow and decreases the distortion in the shadow of the black 
hole. The results obtained in our study may be useful in the light of 
the observational outcome of the Event Horizon Telescope. The black hole 
shadow may help to conclude that whether the Kerr metric is an accurate 
description of astrophysical black holes or there is room for non-Kerr 
black holes like one we have discussed.

\begin{acknowledgements} 
We thank the DST INDO-SA bilateral project for Grant No. 
DST/INT/South Africa/P-06/2016. M.A. acknowledges that this research work 
is supported by the National Research Foundation, South Africa. S.D.M. 
acknowledges that this work is based upon research supported by South 
African Research Chair Initiative of the Department of Science and 
Technology and the National Research Foundation. We would like to 
thank IUCAA, Pune, for hospitality, where a part of this work was done.
\end{acknowledgements}

%%%%%%%%%%%%%%%%%%%%%%%%%%%%%%%%%%%%%%%%%%%%%

\noindent

\begin{thebibliography}{99}
% 
%\cite{Penrose:1969pc}
\bibitem{Penrose:1969pc} 
  R.~Penrose,
  %``Gravitational collapse: The role of general relativity,''
  Riv.\ Nuovo Cim.\  {\bf 1}, 252 (1969) [Gen.\ Rel.\ Grav.\  {\bf 34}, 1141 (2002)].
  
%\cite{Sakharov:1966}
\bibitem{Sakharov:1966}
  A.D.~Sakharov,
  %``Initial stage of an expanding universe and appearence of a nonuniform distribution of matter,''
  Sov.\ Phys.\ JETP, {\bf 22}, 241 (1966).

%\cite{Gliner:1966}
\bibitem{Gliner:1966}
  E.B.~Gliner,
  %`` Algebraic properties of the energy-momentum tensor and vacuum-like states of matter,''
  Sov.\ Phys.\ JETP, {\bf 22}, 378 (1966).

%\cite{Bardeen:1968}
\bibitem{Bardeen:1968}
  J.~Bardeen,
  in {\it Proceedings of GR5} (Tiflis, U.S.S.R., 1968)
     
%\cite{AyonBeato:2000zs}
\bibitem{AyonBeato:2000zs} 
  E.~Ay{\'o}n-Beato and A.~Garc{\'i}a,
  %``The Bardeen model as a nonlinear magnetic monopole,''
  Phys.\ Lett.\ B {\bf 493}, 149 (2000)
  
%\cite{Dymnikova:2004zc}
\bibitem{Dymnikova:2004zc} 
  I.~Dymnikova,
  %``Regular electrically charged structures in nonlinear electrodynamics coupled to general relativity,''
  Class.\ Quantum\ Gravity\  {\bf 21}, 4417 (2004).

%\cite{Bronnikov:2000vy}
\bibitem{Bronnikov:2000vy} 
  K.A.~Bronnikov,
  %``Regular magnetic black holes and monopoles from nonlinear electrodynamics,''
  Phys.\ Rev.\ D {\bf 63}, 044005 (2001). 

%\cite{Shankaranarayanan:2003qm}
\bibitem{Shankaranarayanan:2003qm} 
  S.~Shankaranarayanan and N.~Dadhich,
  %``Nonsingular black holes on the brane,''
  Int.\ J.\ Mod.\ Phys.\ D {\bf 13}, 1095 (2004).

%\cite{Hayward:2005gi}
\bibitem{Hayward:2005gi}
  S.A.~Hayward,
  %``Formation and evaporation of regular black holes,''
  Phys.\ Rev.\ Lett.\  {\bf 96}, 031103 (2006).

%\cite{Ansoldi:2008jw}
\bibitem{Ansoldi:2008jw}
  S.~Ansoldi,
  %``Spherical black holes with regular center: A Review of existing models including a recent realization with Gaussian sources,''
  \url{arXiv:0802.0330}.

%\cite{Culetu:2014lca}
\bibitem{Culetu:2014lca}
  H.~Culetu,
  %``On a regular charged black hole with a nonlinear electric source,''
  Int.\ J.\ Theor.\ Phys.\  {\bf 54}, 2855 (2015).

%\cite{Balart:2014jia}
\bibitem{Balart:2014jia} 
  L.~Balart and E.C.~Vagenas,
  %``Regular black hole metrics and the weak energy condition,''
  Phys.\ Lett.\ B {\bf 730}, 14 (2014).
  
%\cite{Balart:2014cga}
\bibitem{Balart:2014cga} 
  L.~Balart and E.C.~Vagenas,
  %``Regular black holes with a nonlinear electrodynamics source,''
  Phys.\ Rev.\ D {\bf 90}, 124045 (2014).

%\cite{Xiang:2013sza}
\bibitem{Xiang:2013sza} 
  L.~Xiang, Y.~Ling, and Y.G.~Shen,
  %``Singularities and the Finale of Black Hole Evaporation,''
  Int.\ J.\ Mod.\ Phys.\ D {\bf 22}, 1342016 (2013).

%\cite{Bambi:2011mj}
\bibitem{Bambi:2011mj} 
  C.~Bambi,
  %``Testing the Kerr black hole hypothesis,''
  Mod.\ Phys.\ Lett.\ A {\bf 26}, 2453 (2011).

%\cite{Bambi:2014nta}
\bibitem{Bambi:2014nta} 
  C.~Bambi,
  %``Testing the Bardeen metric with the black hole candidate in Cygnus X-1,''
  Phys.\ Lett.\ B {\bf 730}, 59 (2014).
      
%\cite{Bambi:2013ufa}
\bibitem{Bambi:2013ufa}
  C.~Bambi and L.~Modesto,
  %``Rotating regular black holes,''
  Phys.\ Lett.\ B {\bf 721}, 329 (2013).

%\cite{Toshmatov:2014nya}
\bibitem{Toshmatov:2014nya}
  B.~Toshmatov, B.~Ahmedov, A.~Abdujabbarov, and Z.~Stuchl{\'i}k,
  %``Rotating Regular Black Hole Solution,''
  Phys.\ Rev.\ D {\bf 89}, 104017 (2014).
    
\bibitem{Ghosh:2014hea}
  S.G.~Ghosh and S.D.~Maharaj,
  %``Radiating Kerr-like regular black hole,''
  Eur.\ Phys.\ J.\ C {\bf 75}, 7 (2015).

%\cite{Neves:2014aba}
\bibitem{Neves:2014aba} 
  J.C.S.~Neves and A.~Saa,
  %``Regular rotating black holes and the weak energy condition,''
  Phys.\ Lett.\ B {\bf 734}, 44 (2014).

%\cite{Toshmatov:2017zpr}
\bibitem{Toshmatov:2017zpr} 
  B.~Toshmatov, Z.~Stuchl{\'i}k, and B.~Ahmedov,
  %``Generic rotating regular black holes in general relativity coupled to nonlinear electrodynamics,''
  Phys.\ Rev.\ D {\bf 95}, 084037 (2017).
  
%\cite{Newman:1965tw}
\bibitem{Newman:1965tw}
  E.T.~Newman and A.I.~Janis,
  %``Note on the Kerr spinning particle metric,''
  J.\ Math.\ Phys.\  {\bf 6},  915 (1965).
    
%\cite{Azreg-Ainou:2014aqa}
\bibitem{Azreg-Ainou:2014aqa} 
  M.~Azreg-A{\"i}nou,
  %``From static to rotating to conformal static solutions: Rotating imperfect fluid wormholes with(out) electric or magnetic field,''
  Eur.\ Phys.\ J.\ C {\bf 74}, 2865 (2014).

%\cite{Azreg-Ainou:2014nra}
\bibitem{Azreg-Ainou:2014nra} 
  M.~Azreg-A{\"i}nou,
  %``Regular and conformal regular cores for static and rotating solutions,''
  Phys.\ Lett.\ B {\bf 730}, 95 (2014).

%\cite{Azreg-Ainou:2014pra}
\bibitem{Azreg-Ainou:2014pra} 
  M.~Azreg-A{\"i}nou,
  %``Generating rotating regular black hole solutions without complexification,''
  Phys.\ Rev.\ D {\bf 90}, 064041 (2014).

\bibitem{Ghosh:2014pba} 
  S.G.~Ghosh,
  %``A nonsingular rotating black hole,''
  Eur.\ Phys.\ J.\ C {\bf 75}, 532 (2015).
  
%\cite{Ghosh:2014mea}
\bibitem{Ghosh:2014mea}
  S.G.~Ghosh, P.~Sheoran, and M.~Amir,
  %``Rotating Ayon-Beato-Garcia black hole as a particle accelerator,''
  Phys.\ Rev.\ D {\bf 90}, 103006 (2014).

%\cite{Amir:2015pja}
\bibitem{Amir:2015pja} 
  M.~Amir and S.G.~Ghosh,
  %``Rotating Hayward's regular black hole as particle accelerator,''
  J.\ High\ Energy\ Phys. 07 (2015) 015.
  
%\cite{Ghosh:2015pra}
\bibitem{Ghosh:2015pra} 
  S.G.~Ghosh and M.~Amir,
  %``Horizon structure of rotating Bardeen black hole and particle acceleration,''
  Eur.\ Phys.\ J.\ C {\bf 75}, 553 (2015).
  
%\cite{Amir:2016nti}
\bibitem{Amir:2016nti} 
  M.~Amir, F.~Ahmed and S.G.~Ghosh,
  %``Collision of two general particles around a rotating regular Hayward's black holes,''
  Eur.\ Phys.\ J.\ C {\bf 76}, 532 (2016).

%\cite{Ahmed:2018fge}
\bibitem{Ahmed:2018fge} 
  F.~Ahmed, M.~Amir and S.G.~Ghosh,
  %``Particle acceleration of two general particles in the background of rotating Ayón-Beato-García black holes,''
  Astrophys.\ Space Sci.\  {\bf 364}, 10 (2019).

%\cite{Toshmatov:2015wga}
\bibitem{Toshmatov:2015wga} 
  B.~Toshmatov, A.~Abdujabbarov, Z.~Stuchl{\'i}k, and B.~Ahmedov,
  %``Quasinormal modes of test fields around regular black holes,''
  Phys.\ Rev.\ D {\bf 91}, 083008 (2015).
  
%\cite{Toshmatov:2018tyo}
\bibitem{Toshmatov:2018tyo} 
  B.~Toshmatov, Z.~Stuchl{\'i}k, J.~Schee, and B.~Ahmedov,
  %``Electromagnetic perturbations of black holes in general relativity coupled to nonlinear electrodynamics,''
  Phys.\ Rev.\ D {\bf 97}, 084058 (2018).
  
%\cite{Toshmatov:2018ell}
\bibitem{Toshmatov:2018ell}
  B.~Toshmatov, Z.~Stuchl{\'i}k, and B.~Ahmedov,
  %``Electromagnetic perturbations of black holes in general relativity coupled to nonlinear electrodynamics: Polar perturbations,''
  Phys.\ Rev.\ D {\bf 98},  085021 (2018) .
  
%\cite{Toshmatov:2019gxg}
\bibitem{Toshmatov:2019gxg}
  B.~Toshmatov, Z.~Stuchl{\'i}k, B.~Ahmedov, and D.~Malafarina,
  %``Relaxations of perturbations of spacetimes in general relativity coupled to nonlinear electrodynamics,''
  Phys.\ Rev.\ D {\bf 99},  064043 (2019).
  
\bibitem{Chandrasekhar92}
  S.~ Chandrasekhar, 
  {\it The Mathematical Theory of Black Holes} 
  (Oxford University Press, New York, 1992).
  
%\cite{Falcke:1999pj}
\bibitem{Falcke:1999pj} 
  H.~Falcke, F.~Melia, and E.~Agol,
  %``Viewing the shadow of the black hole at the galactic center,''
  Astrophys.\ J.\  {\bf 528}, L13 (2000).
  
  %\cite{Takahashi04}
\bibitem{Takahashi04} 
  R.~Takahashi,
  %``Shapes and positions of black hole shadows in accretion disks and spin parameters of black holes,''
  J.\ Korean Phys.\ Soc.\  {\bf 45}, S1808 (2004) [Astrophys.\ J.\  {\bf 611}, 996 (2004)].

%\cite{Takahashi:2005hy}
\bibitem{Takahashi:2005hy} 
  R.~Takahashi,
  %``Black hole shadows of charged spinning black holes,''
  Publ.\ Astron.\ Soc.\ Jap.\  {\bf 57}, 273 (2005).

%\cite{Hioki:2009na}
\bibitem{Hioki:2009na} 
  K.~Hioki and K.i.~Maeda,
  %``Measurement of the Kerr Spin Parameter by Observation of a Compact Object's Shadow,''
  Phys.\ Rev.\ D {\bf 80}, 024042 (2009).
    
%\cite{Bambi:2008jg}
\bibitem{Bambi:2008jg} 
  C.~Bambi and K.~Freese,
  %``Apparent shape of super-spinning black holes,''
  Phys.\ Rev.\ D {\bf 79}, 043002 (2009).
    
%\cite{Bambi:2010hf}
\bibitem{Bambi:2010hf} 
  C.~Bambi and N.~Yoshida,
  %``Shape and position of the shadow in the $\delta = 2$ Tomimatsu-Sato space-time,''
  Class.\ Quant.\ Grav.\  {\bf 27}, 205006 (2010).

%\cite{Amarilla:2010zq}
\bibitem{Amarilla:2010zq} 
  L.~Amarilla, E.F.~Eiroa, and G.~Giribet,
  %``Null geodesics and shadow of a rotating black hole in extended Chern-Simons modified gravity,''
  Phys.\ Rev.\ D {\bf 81}, 124045 (2010).
    
%\cite{Amarilla:2011fxx}
\bibitem{Amarilla:2011fxx} 
  L.~Amarilla and E.F.~Eiroa,
  %``Shadow of a rotating braneworld black hole,''
  Phys.\ Rev.\ D {\bf 85}, 064019 (2012).

%\cite{Yumoto:2012kz}
\bibitem{Yumoto:2012kz} 
  A.~Yumoto, D.~Nitta, T.~Chiba, and N.~Sugiyama,
  %``Shadows of Multi-Black Holes: Analytic Exploration,''
  Phys.\ Rev.\ D {\bf 86}, 103001 (2012).

\bibitem{Zakharov} 
A.F.~Zakharov, F.~De Paolis, G.~Ingrosso, and A.A.~Nucita, 
New Astronomy Reviews {\bf 56} (2012) 64-73.
    
%\cite{Amarilla:2013sj}
\bibitem{Amarilla:2013sj} 
  L.~Amarilla and E.F.~Eiroa,
  %``Shadow of a Kaluza-Klein rotating dilaton black hole,''
  Phys.\ Rev.\ D {\bf 87}, 044057 (2013).
    
%\cite{Atamurotov:2013dpa}
\bibitem{Atamurotov:2013dpa} 
  F.~Atamurotov, A.~Abdujabbarov, and B.~Ahmedov,
  %``Shadow of rotating HoÅ™ava-Lifshitz black hole,''
  Astrophys.\ Space Sci.\  {\bf 348}, 179 (2013).
  
%\cite{Atamurotov:2013sca}
\bibitem{Atamurotov:2013sca} 
  F.~Atamurotov, A.~Abdujabbarov, and B.~Ahmedov,
  %``Shadow of rotating non-Kerr black hole,''
  Phys.\ Rev.\ D {\bf 88}, 064004 (2013).

%\cite{Wei:2013kza}
\bibitem{Wei:2013kza} 
  S.W.~Wei and Y.X.~Liu,
  %``Observing the shadow of Einstein-Maxwell-Dilaton-Axion black hole,''
  J.\ Cosmol.\ Astropart.\ Phys.\ 11 (2013) 063.
  
%\cite{Abdujabbarov:2012bnn}
\bibitem{Abdujabbarov:2012bnn} 
  A.~Abdujabbarov, F.~Atamurotov, Y.~Kucukakca, B.~Ahmedov, and U.~Camci,
  %``Shadow of Kerr-Taub-NUT black hole,''
  Astrophys.\ Space\ Sci.\  {\bf 344}, 429 (2013).

%\cite{Johannsen:2015qca}
\bibitem{Johannsen:2015qca} 
  T.~Johannsen,
  %``Photon Rings around Kerr and Kerr-like Black Holes,''
  Astrophys.\ J.\  {\bf 777}, 170 (2013).
      
%\cite{Grenzebach:2014fha}
\bibitem{Grenzebach:2014fha} 
  A.~Grenzebach, V.~Perlick, and C.~L{\"a}mmerzahl,
  %``Photon Regions and Shadows of Kerr-Newman-NUT Black Holes with a Cosmological Constant,''
  Phys.\ Rev.\ D {\bf 89}, 124004 (2014).
  
%\cite{Li:2013jra}
\bibitem{Li:2013jra} 
  Z.~Li and C.~Bambi,
  %``Measuring the Kerr spin parameter of regular black holes from their shadow,''
  J.\ Cosmol.\ Astropart.\ Phys.\ 01 (2014) 041.

%\cite{Papnoi:2014}
\bibitem{Papnoi:2014} 
  U.~Papnoi, F.~Atamurotov, S.~G.~Ghosh, and B.~Ahmedov,
  %``Shadow of five-dimensional rotating Myers-Perry black hole,''
  Phys.\ Rev.\ D {\bf 90}, 024073 (2014).
    
%\cite{Abdujabbarov:2015xqa}
\bibitem{Abdujabbarov:2015xqa} 
  A.A.~Abdujabbarov, L.~Rezzolla, and B.J.~Ahmedov,
  %``A coordinate-independent characterization of a black hole shadow,''
  Mon.\ Not.\ Roy.\ Astron.\ Soc.\  {\bf 454}, 2423 (2015).
  
%\cite{Goddi:2016jrs}
\bibitem{Goddi:2016jrs} 
  C.~Goddi {\it et al.},
  %``BlackHoleCam: fundamental physics of the Galactic center,''
  Int.\ J.\ Mod.\ Phys.\ D {\bf 26}, 1730001 (2016).
    
%\cite{Abdujabbarov:2016hnw}
\bibitem{Abdujabbarov:2016hnw} 
  A.~Abdujabbarov, M.~Amir, B.~Ahmedov, and S.G.~Ghosh,
  %``Shadow of rotating regular black holes,''
  Phys.\ Rev.\ D {\bf 93}, 104004 (2016).
  
%\cite{Amir:2016cen}
\bibitem{Amir:2016cen} 
  M.~Amir and S.G.~Ghosh,
  %``Shapes of rotating nonsingular black hole shadows,''
  Phys.\ Rev.\ D {\bf 94}, 024054 (2016).
    
%\cite{Younsi:2016azx}
\bibitem{Younsi:2016azx} 
  Z.~Younsi, A.~Zhidenko, L.~Rezzolla, R.~Konoplya, and Y.~Mizuno,
  %``New method for shadow calculations: Application to parametrized axisymmetric black holes,''
  Phys.\ Rev.\ D {\bf 94}, 084025 (2016).
  
%\cite{Grenzebach:2016}
\bibitem{Grenzebach:2016}
A.~Grenzebach, 
\textit{The Shadow of Black Holes: An Analytic Description},
Springer Briefs in Physics, Springer, Heidelberg (2016).

%\cite{Amir:2017slq}
\bibitem{Amir:2017slq} 
  M.~Amir, B.P.~Singh, and S.G.~Ghosh,
  %``Shadows of rotating five-dimensional charged EMCS black holes,''
  Eur.\ Phys.\ J.\ C {\bf 78}, 399 (2018).
  
%\cite{Kumar:2017vuh}
\bibitem{Kumar:2017vuh} 
  R.~Kumar, B.P.~Singh, M.S.~Ali, and S.G.~Ghosh,
  %``Rotating black hole shadow in Rastall theory,''
  \url{arXiv:1712.09793}.

%\cite{Singh:2017vfr}
\bibitem{Singh:2017vfr} 
  B.P.~Singh and S.G.~Ghosh,
  %``Shadow of Schwarzschild–Tangherlini black holes,''
  Ann.\ Phys.\  {\bf 395}, 127 (2018).
  
%\cite{Nedkova:2013msa}
\bibitem{Nedkova:2013msa}
  P.G.~Nedkova, V.K.~Tinchev, and S.S.~Yazadjiev,
  %``Shadow of a rotating traversable wormhole,''
  Phys.\ Rev.\ D {\bf 88}, 124019 (2013).

%\cite{Bambi:2013nla}
\bibitem{Bambi:2013nla} 
  C.~Bambi,
  %``Can the supermassive objects at the centers of galaxies be traversable wormholes? The first test of strong gravity for mm/sub-mm very long baseline interferometry facilities,''
  Phys.\ Rev.\ D {\bf 87}, 107501 (2013).
   
%\cite{Azreg-Ainou:2014dwa}
\bibitem{AzregAinou:2014dwa} 
  M.~Azreg-A{\"i}nou,
  %``Confined-exotic-matter wormholes with no gluing effects—Imaging supermassive wormholes and black holes,''
  J.\ Cosmol.\ Aastro.\ Phys.\ 1507 (2015) 037.

%\cite{Ohgami:2015nra}
\bibitem{Ohgami:2015nra} 
  T.~Ohgami and N.~Sakai,
  %``Wormhole shadows,''
  Phys.\ Rev.\ D {\bf 91}, 124020 (2015).

%\cite{Abdujabbarov:2016efm}
\bibitem{Abdujabbarov:2016efm} 
  A.~Abdujabbarov, B.~Juraev, B.~Ahmedov, and Z.~Stuchlík,
  %``Shadow of rotating wormhole in plasma environment,''
  Astrophys.\ Space Sci.\  {\bf 361}, 226 (2016).

%\cite{Shaikh:2018kfv}
\bibitem{Shaikh:2018kfv} 
  R.~Shaikh,
  %``Shadows of rotating wormholes,''
  Phys.\ Rev.\ D {\bf 98}, 024044 (2018).
  
%\cite{Ohgami:2016iqm}
\bibitem{Ohgami:2016iqm} 
  T.~Ohgami and N.~Sakai,
  %``Wormhole shadows in rotating dust,''
  Phys.\ Rev.\ D {\bf 94}, 064071 (2016).
  
%\cite{Amir:2018szm}
\bibitem{Amir:2018szm} 
  M.~Amir, A.~Banerjee and S.D.~Maharaj,
  %``Shadow of charged wormholes in Einstein–Maxwell–dilaton theory,''
  Annals Phys.\  {\bf 400}, 198 (2019).
  
%\cite{Amir:2018pcu}
\bibitem{Amir:2018pcu} 
  M.~Amir, K.~Jusufi, A.~Banerjee and S.~Hansraj,
  %``Shadow images of Kerr-like wormholes,''
  Class.\ Quantum\ Gravity\  {\bf 36},  215007 (2019).

%\cite{Akiyama:2019cqa}
\bibitem{Akiyama:2019cqa} 
  K.~Akiyama {\it et al.} [Event Horizon Telescope Collaboration],
  %``First M87 Event Horizon Telescope Results. I. The Shadow of the Supermassive Black Hole,''
  Astrophys.\ J.\  {\bf 875}, L1 (2019).
  
%\cite{Akiyama:2019fyp}
\bibitem{Akiyama:2019fyp} 
  K.~Akiyama {\it et al.} [Event Horizon Telescope Collaboration],
  %``First M87 Event Horizon Telescope Results. V. Physical Origin of the Asymmetric Ring,''
  Astrophys.\ J.\  {\bf 875}, L5 (2019).

%\cite{Akiyama:2019eap}
\bibitem{Akiyama:2019eap} 
  K.~Akiyama {\it et al.} [Event Horizon Telescope Collaboration],
  %``First M87 Event Horizon Telescope Results. VI. The Shadow and Mass of the Central Black Hole,''
  Astrophys.\ J.\  {\bf 875}, L6 (2019).
    
%\cite{Dymnikova:1992ux}
\bibitem{Dymnikova:1992ux} 
  I.~Dymnikova,
  %``Vacuum nonsingular black hole,''
  Gen.\ Relativ.\ Grav.\  {\bf 24}, 235 (1992).

%\cite{Fernando:2016ksb}
\bibitem{Fernando:2016ksb} 
  S.~Fernando,
  %``Bardeen–de Sitter black holes,''
  Int.\ J.\ Mod.\ Phys.\ D {\bf 26}, 1750071 (2017).
  
%%\cite{Bronnikov:2017sgg}
%\bibitem{Bronnikov:2017sgg} 
%  K.~A.~Bronnikov,
%  %``Nonlinear electrodynamics, regular black holes and wormholes,''
%  Int.\ J.\ Mod.\ Phys.\ D {\bf 27}, 1841005 (2018).
    
%\cite{Bardeen:1972fi}
\bibitem{Bardeen:1972fi}
  J.M.~Bardeen, W.H.~Press, and S.A.~Teukolsky,
  %``Rotating black holes: Locally nonrotating frames, energy extraction, and scalar synchrotron radiation,''
  Astrophys.\ J.\  {\bf 178}, 347 (1972).

%\cite{Carmeli:1975kg}
\bibitem{Carmeli:1975kg} 
  M.~Carmeli and M.~Kaye,
  %``Gravitational Field of a Radiating Rotating Body,''
  Annals Phys.\  {\bf 103}, 97 (1977).
    
%\cite{Penrose:1971uk}
\bibitem{Penrose:1971uk} 
  R.~Penrose and R.M.~Floyd,
  %``Extraction of rotational energy from a black hole,''
  Nature (London) {\bf 229}, 177 (1971). 
 
%\cite{Carter:1968rr}
\bibitem{Carter:1968rr} 
  B.~Carter, 
  Phys.\ Rev.\ {\bf 174}, 1559 (1968).  
  
\bibitem {Bardeen:1973gb} 
  J.M.~Bardeen, 
  in {\it Black holes, in Proceedings of the Les Houches Summer School, Session 215239}, edited by C.~ De Witt and B.S.~ De Witt (Gordon and Breach, New York, 1973).


\end{thebibliography}
\end{document}